\documentclass[english]{article}

\newcommand{\elvrm }{}

\catcode`\@=11 

\def\marginnote#1{}

\def\numberbysection{\@addtoreset{equation}{section}
	\def\theequation{\thesection.\arabic{equation}}} 

\catcode`@=12
\relax 

\begin{document}

\title{\Large A challenge in enumerative combinatorics:
The graph of contributions\footnote{
In honor of F.Y. Wu on the occasion of his 70th birthday. 
Statphys-Taiwan 2002. The second
APCTP and sixth Taiwan International Symposium on Statistical
Physics.} 
 of Professor Fa-Yueh WU}

\author{J-M.Maillard
\footnote{
E-mail: maillard@lpthe.jussieu.fr}}

\date{\today}

\maketitle

% \vskip .1cm
\centerline{LPTHE, Universit\'e de Paris VI,}
\vskip .1cm
 \centerline{Tour 16, $1^{\rm er}$ \'etage,
bo\^\i te 126, 4 Place Jussieu, }
\vskip .1cm
\centerline{ F--75252 Paris Cedex 05, France.}

\vskip .5cm

\begin{abstract}

We will try to sketch Professor F. Y. Wu's contributions
in lattice statistical mechanics, solid state physics,
graph theory, enumerative combinatorics
and so many other domains of physics and mathematics.
We will recall  F. Y. Wu's most important
and well-known classic results and we will also  sketch 
his most recent researches dedicated to the 
connections of lattice statistical mechanical models with deep
problems in pure mathematics. Since it is hard to 
provide an exhaustive list 
of all his contributions,
to give some representation of F. Y.  Wu's 
"mental  connectivity" we will concentrate
on the interrelations between the various results he has obtained
in so many different domains of physics and mathematics. 
Along the way we will also try to 
understand   Wu's motivations and his favorite
concepts, tools and ideas.

\end{abstract}

\vskip .5cm 
{PAR--LPTHE 02--29} \hfill 

\noindent {\bf AMS Classification scheme numbers}: 82A68, 82A69, 14E05, 14J50,
16A46, 16A24, 11D41 


\newpage
% \tableofcontents \newpage

\tableofcontents \newpage

\section{Introduction}

The publish-or-perish period of science 
could soon be seen as a golden age: our brave new world
now celebrates the triumph of Enron's financial 
and accounting creativity.
Sadly science is now also, increasingly,
considered from an accountant's viewpoint. In this
 respect, if one takes this ``modern'' 
point of view Professor F. Y. Wu's
contribution\footnote{
Professor F. Y. Wu is presently the Matthews University Distinguished Professor
at Northeastern University.
He is a fellow of the American Physical Society and a permanent member of
 the Chinese Physical Society (Taipei).
His research has been supported by the National Science Foundation
 since 1968, a rare accomplishment by
itself in an environment of  declining research support in the U.S., and
he currently serves as the editor of three professional
journals: the Physica A, International Journal of Modern
Physics B and the Modern Physics Letters B.
} is clearly a very good return on investment:
he has given more than 270 talks in meetings or conferences,
 published over 200 papers and
monographs in refereed journals, and 
 had many students. He has also published in, or
 is the editor\footnote{For instance, 
%Exactly Soluble Models in Statistical Mechanics:
%Historical Perspectives and Current Status, Proceedings
%Northeastern University, Boston 30 - 31 March 1996
%edited by C King and  F Y Wu.
Ref. \cite{w169bis} contains the proceedings of the conference on "Exactly 
Soluble Models in Statistical Mechanics:
Historical Perspectives and Current Status", held at Northeastern 
University in March 1996 - the first ever international conference to deal
exclusively with this topic. The proceedings reflect
the broad range of interest in exactly soluble models as well as 
the diverse fields in physics and mathematics
 that they connect. }
of, many 
books~\cite{w18bis,w28,w66bis,w116bis,w131bis,w149bis,w162bis,w168,w169,w185} 
 
Professor  Wu was trained  in theoretical condensed
 matter physics~\cite{w1,w2,w17,w18,w24,w32,w103}, but he is now
 seen as a mathematical physicist who is a leading
expert in mathematical modeling of phase transition phenomena
 occurring in complex systems. 
  Wu's research includes
 both theoretical studies and practical
applications\footnote{He has 
considered, for instance, the modeling of physical adsorption and applied
it to describe processes used in chemical and 
environmental engineering~\cite{w141,w165}.
He has even published one experimental 
paper on slow neutron detectors~\cite{w3}.
}. Among his recent researches he has studied
connections of statistical mechanical models with deep
problems in pure mathematics. This includes the generation of
knot and link invariants from soluble models of statistical
mechanics and the study of the long-standing unsolved
mathematical problem of multidimensional partitions of
integers in number theory using a Potts model approach.

Professor Wu's contributions to lattice statistical mechanics have been
mostly in
the area of exactly solvable lattice models.  
%%%% More recently he
%%%% has also worked  on duality 
%%%% relations for  Potts  correlation functions~\cite{w172,w175,w178,w182}. 
While integrable models
 have continued to occupy a prominent place in his work (such as the exact solution 
 of two- and three-dimensional spin models and  interacting
  dimer systems), his 
 work has ranged over a wide variety of problems including exact lattice
 statistics in two and three dimensions, graph theory
and combinatorics to mention just a few. 
%%% Professor Wu is widely known for his works in statistical mechanics
%%%  and more specifically, lattice statistical mechanics.  But his work in
 His work in
many-body theory~\cite{w1,w2,w5,w6,w13,w19,w25,w33,w62}, especially those
 on liquid helium~\cite{w0,w1,w4,w22,w23}, has also been
influential for many years. 
%%% His early papers on the  formulation of the correlated-basis function
%%% approach~\cite{w5,w13,w19,w25,w33}
%%%  already showed  inklings
%%% of his  mathematical skills.

F. Y. Wu joined the faculty of Northeastern University to work with Elliott Lieb in
1967, and in 1968 they published a joint paper\footnote{
 This paper has
become prominent in the theory of high-$T_c$ superconductors. P. W. Anderson 
even attributed to this paper as ``predicting'' the existence of
quarks in his {\em Physics Today} (October, 1997) article on 
the centennial of the discovery 
of electrons. 
} on the ground
state of the Hubbard model~\cite{w9} which has since become a classic.
% (absence of
% Mott transition in an exact solution of 
% the short-range one-band model in one dimension). 
The Baxter-Wu
model~\cite{w42,w46} is 
also, clearly, an important milestone in
the history of integrable lattice models. 

F. Y. Wu has published several very important reviews
of lattice statistical mechanics.
First,  Lieb and  Wu  wrote a monograph in 1970 
on vertex models which has become the fundamental reference in the field for
decades~\cite{w28}.  Wu's 1982 review 
on the Potts model is another classic~\cite{w84}.  At more than one
hundred citations per year
 ever since it was published, it is one of the
most cited papers in physics\footnote{There was once a study
   published in 1984 (E. Garfield, Current Comments {\bf 48},
   3 (1984)) on citations in physics for the year of 1982.
   It reports that in 1982, the year this Potts review was
   published, it was the fifth most-cited paper among
   papers published in all physics.}. In 1992  F.Y. Wu published yet
 another extremely well-received review
on knot theory and its connection with 
lattice statistical 
mechanics~\cite{w147}.
% (see also~\cite{w143,w144}).
In addition, in 1981, F. Y. Wu and Z. R. Yang  published
a series of expository papers on critical phenomena written in 
Chinese~\cite{w79} - \cite{w83}.
This  review is
 well-known to  Chinese researchers.

\vskip .3cm 
\subsection{The choice of presentation: A challenge in enumerative combinatorics}

 An intriguing aspect of 
lattice statistics is that seemingly totally
different problems are sometimes related to each other, and that the
solution of one problem can often lead to solving other outstanding 
unsolved problems. At first sight, most of the work of F. Y. Wu could be said to 
correspond to {\em exact results in lattice statistical mechanics}, but because
of the relations between seemingly totally
different problems it can equivalently be seen, and sometimes
be explicitly presented, as exact results in
various domains of mathematical physics or mathematics:
sometimes exact results in  {\em graph theory}, sometimes in
 {\em enumerative
combinatorics},  sometimes in {\em knot
theory}, sometimes in {\em number theory},
% (partition of integers~\cite{w163}), 
etc.   Wu's ``intellectual walk'' goes from vertex models to
circle theorems or duality relations, from dimers to Ising models and
back, from percolations or animal problems to Potts models,
from Potts models to  the Whitney-Tutte Polynomials, to polychromatic
polynomials or to knot theory, from 
results, or conjectures, on  critical manifolds\footnote{
The critical manifolds deduced or conjectured by F. Y. Wu
are mostly algebraic varieties and not simple differentiable
or analytical manifolds.} to Yang-Baxter
integrability, perhaps on the way revisiting duality or Lee-Yang zeros, etc., etc.   
The simple listing of Professor Wu's 
results and contributions,
and the inter-relations between these results and the associated
concepts and tools, is by itself a  {\em challenge 
in enumerative combinatorics}.  

Actually it is impossible to describe  Wu's contributions
{\em linearly}, in a sequence of sections in a review paper
like this, or even with a website-like 
``{\em tree} organization'' of paragraphs.
F. Y. Wu's contributions really correspond to a quite large ``graph'' of concepts,
results, tools and models, with many ``intellectual loops''.
The only possible ``linear'' and exhaustive description of  Wu's contributions
is his list of publications.  
\vskip 4mm

{\noindent
{\em We have therefore chosen to give his exhaustive
list of publications at the end of this paper. No other references are
given.}}

\vskip 4mm
 We have chosen to
keep the notation F.Y. Wu used 
in his publications\footnote{The price  paid is, for instance, that the spin edge
Boltzmann weights will sometimes be
denoted 
$\, e^{K_1}, \,e^{K_2}, $$\, e^{K_3},$$
 \, e^{K_4}$,  or  $\, a,\,  b, \,$$ c,\,  d$,
or $\, x_1, \, x_2,$$ \,x_3, \, x_4$, and the 
vertex Boltzmann weights $\, \omega_1, \,$
$\omega_2, \, \cdots $
or $\, a,\,  b, \,$$ c,\,  d, \, $
 $\, a',\,  b', \,$$ c',\,  d' $. This corresponds
to the spectrum of notations used in the 
lattice statistical mechanics literature. These different notations
were often introduced when one faces large polynomial expressions
and that the $\, e^{K_i}$ or $\, e^{-\beta\cdot J_i}$ notations
for Boltzmann weights would be painful.}, 
and {\em not to normalize
them}, so that the reader who wants to 
see more and goes back to the cited publications 
will immediately recover the equations and notations.

\vskip 4mm

 Obviously, we will not try to provide an exhaustive description
of  Wu's contributions but, rather, to  provide some considered
well-suited specific ``morceaux choisis\footnote{I apologize, in advance, for 
the fact that these ``morceaux choisis'' are obviously
biased by my personal taste for effective birational 
 algebraic geometry 
in lattice statistical mechanics.
}'',
{\em comments} on some of his results, some hints
of the kind of concepts he likes to work with, and trying to 
explain why his results are important, fruitful and
stimulating for anyone who works in lattice statistical mechanics
or in mathematical physics.

\section{Even before vertex models: the exact 
solution of the Hubbard model}
\label{Hubbard}

%We will see in section (\ref{vertexmodels}) below
%a large set of results from
%F. Y. Wu's monograph with E. H.  Lieb
%on {\em vertex models}, in particular the {\em six-vertex
%model}. Looking at the six-vertex model 
%amounts to considering the {\em Bethe ansatz}
%explicitly or implicitly. Actually, among the exactly soluble models
%(the bread-and-butter of F. Y. Wu)  was one
%that, for a long time, was a ``sleeper'', namely, Bethe's 1931 solution
%of the ground state energy and elementary excitations 
%of the one-dimensional quantum-mechanical 
%spin-${1\over 2}$ Heisenberg model of antiferromagnetism.

 Elliott H. Lieb and F. Y. Wu published in 1968 a 
 joint paper on the ground
state of the Hubbard model~\cite{w9}
% (absence of
% Mott transition in an exact solution of 
% the short-range one-band model in one dimension) 
which has since
become a classic, and served as a
cornerstone in the theory of high-$T_c$ superconductors. 
An important question there corresponds to
the spin-charge decoupling which is exact
and explicit in one-dimensional models: 
is the spin-charge decoupling a 
characteristic of one dimension?
Is it possible that some ``trace'' of  spin-charge decoupling 
remains for quantum two-dimensional
models which are supposedly  related to  high-$T_c$ superconductors?

Let us describe briefly 
the classic Lieb-Wu solution
of the Hubbard model.
One assumes that the electrons can hop between the Wannier states of
neighboring lattice sites and that each site is capable
of accommodating two electrons of opposite spins with an interaction
energy $\, U \, > \, 0$. The corresponding Hamiltonian reads:
\begin{eqnarray}
\label{Hubbar}
H \, = \, \, T \, \sum_{<ij>} \sum_{\sigma} \,
 c_{i\; \sigma}^{\dag}\; c_{j\; \sigma}\; \, + \, U \, \sum_{i} \; 
 c_{i\; \uparrow }^{\dag}\; c_{i\; \uparrow}\;
 c_{i\; \downarrow }^{\dag}\; c_{i\; \downarrow}\; \nonumber 
\end{eqnarray}
where $\, c_{i\; \sigma}^{\dag}\;$ and   $\,  c_{i\; \sigma}\, $ are
the creation and annihilation
operators for an electron of spin $\, \sigma\, $ in the Wannier state
at the $\, i-th$ lattice site and the first sum 
is taken over nearest neighbor sites.
Denoting $\, f(x_1, \, x_2, \, \cdots\, , x_M; \, x_{M+1} ,
 \,\cdots\, ,  x_N) \, $ the amplitude of the wavefunction
for which the down spins are located at sites $\, x_1, \, x_2, \,
\cdots\,, x_M\, $ and the up spins are located at sites $\, x_{M+1} ,
 \,\cdots\, ,  x_N$. The eigenvalue equation $\, H \psi \, = \, E \,
\psi\, $ leads to:
\begin{eqnarray}
\label{Gaudin}
&&-\sum_{i=1}^{N} \sum_{s=\pm 1} f(x_1, \, x_2, \, \cdots\, , x_{i} \,
 +s,  \,\cdots\, ,  x_N) \,\qquad \qquad  \\
&&\quad + \, U \, \sum_{i<j} 
\delta(x_{i}\, -x_{j}) \,  f(x_1, \, x_2, \, \,\cdots\, ,  x_N) \, 
= \,\, \, E\, f(x_1, \, x_2, \,\,\cdots\, ,  x_N) \nonumber \nonumber 
\end{eqnarray}
 where $\, f(x_1, \, x_2, \, \,\cdots\, ,  x_N) \,$ 
is antisymmetric in
the first $\, M$ and the last $\, N-M$ variables (separately). 
Let $\mu_+$ (resp. $\mu_-$) denote the chemical potential 
of adding (resp. removing) one electron.
In the half-filled band one has $\, \mu_{+} \, = \, \, U\, -\mu_{-}$,
and the calculation of $\, \mu_{-}\, $ can
 be done in closed form with the result:
\begin{equation}
\label{mu}
\mu_{-}= 2  -4 \, \int_{0}^{\infty} {{J_1(\omega) \cdot
d\omega } \over {\omega \cdot (1+{\rm exp}\,({\omega\, U/2}))}} \,  \, \,
 \end{equation}
% \begin{eqnarray}
% \label{mu}
% &&\mu_{-}\, -2 \, \, = \, \, -4 \, \int_{0}^{\infty} {{J_1(\omega) \cdot
% d\omega } \over {\omega \cdot (1+exp(\omega\, U/2))}} \,  \, \,
% \qquad \nonumber  \\
% && \qquad \qquad  \, = \, \, -4 \, \sum_{n=1}^{\infty} (-1)^n 
% \cdot \Bigl((1+ {{n^2\, U^2} \over {4}})^{1/2}\,
%  -{{n} \over {2}} \, U \Bigr) \nonumber 
% \end{eqnarray}
where $\, J_1\, $ is the Bessel function.
It can be established from (\ref{mu}) 
and $\, \mu_{+} \, = \, \, U\,
-\mu_{-}$ that $\, \mu_{+} \, >\, \mu_{-}$
for $\, U \, > \, 0$. In other words, the ground state for a
half-filled band is insulating for any 
nonzero $\, U$, and conducting for $\, U\, =\, 0$.  Equivalently,
there is no Mott transition for
 nonzero  $\, U$, {\em i.e.}, the ground state 
is analytic in  $\, U$ on the real axis except at the origin. 

%The ground-state energy, wave functions and the chemical potentials
%were obtained~\cite{w9} and it was found that the ground state exhibits no
%conductor-insulator transition as the correlation strength 
%is increased~\cite{w9}. 

\section{Vertex models}
\label{vertexmodels}
The distinction between vertex models and spin models
is traditional in lattice statistical mechanics,
but there are ``bridges'' between
these two sets of lattice models~\cite{w73}. 
Roughly speaking one can say that  
 F. Y. Wu  first obtained results
 on vertex models~\cite{w11,w12} (five-vertex
models~\cite{w7,w8}, 
free-fermion vertex models~\cite{w47}, dimer models 
seen as vertex models, ...) and then obtained results
on spin models (Ising model with 
second-neighbor Interactions~\cite{w10},
the Baxter-Wu model~\cite{w42,w46},
Potts model, ...), introducing more and more 
graph theoretical approaches up to looping the loop
with  knot theory
which is, in fact, closely related to vertex models
and Potts models! As far as vertex models are
concerned, we will first sketch the approach given in 
his monograph with   Lieb (section (\ref{Liebmonograph})),
in a second step we will sketch his free-fermion results
(section (\ref{asymmetric})) closely followed by his dimer
 results (section (\ref{dimers})),
and, then, we will discuss some miscellaneous results
he obtained on five-, six- and eight-vertex
models (section (\ref{miscel})).

\subsection{Two-dimensional Ferroelectric Models}
\label{Liebmonograph}

Elliott  Lieb and F. Y. Wu  wrote a monograph on vertex models in 1970,
entitled  ``Two-dimensional Ferroelectric Models'',
which has become a fundamental reference in the field for
decades~\cite{w28}. This monograph gives
the best introduction to the sixteen-vertex model, which is
a fundamental model in lattice statistical mechanics.
But unfortunately it is not   known  well enough even to many specialists
of lattice models that it contains the most general eight-vertex model,
most of the (Yang-Baxter) integrable vertex models (the 
symmetric eight-vertex model, various free-fermion models,
 the asymmetric free-fermion model, 
the asymmetric and symmetric six-vertex, the five-vertex models,
three-coloring of square maps, and others) and
 also fundamental  {\em non-integrable} models
such as, for instance, the Ising model in
 a magnetic field. In particular the monograph mentions
explicitly the weak-graph duality (see section (\ref{duali}) below)
on the  sixteen-vertex model (see page 457 of~\cite{w28}):
\begin{eqnarray}
\label{weakgraph}
&&
\omega_1^{*} \, = \, \, {{1} \over {4}} \cdot \sum_{i=1}^{16}
\omega_{i}, \qquad \quad 
\omega_2^{*} \, = \, \, {{1} \over {4}} \cdot \Bigl(\sum_{i=1}^{8}
\omega_{i} \, - \sum_{i=9}^{16}
\omega_{i}\Bigr) \qquad  \\
&&\omega_3^{*} \, = \, \, {{1} \over {4}} \cdot \Bigl(\sum_{i=1}^{4}
\omega_{i} \, - \sum_{i=5}^{8}\omega_{i}\,
+(\omega_{10}+\omega_{12}+\omega_{14}+\omega_{16})\,\qquad  \nonumber \\
&&\qquad \qquad  \qquad  -(\omega_9+\omega_{11}+\omega_{13}+\omega_{15})
\omega_{i}\Bigr) , \,\,\,  \cdots \qquad  \nonumber 
\end{eqnarray}

The 154 pages of this monograph are 
still, by today's standard, an extremely valuable
document for any specialist of lattice models.
Beyond the taxonomy of ferro and ferrielectric models
(ice model, KDP~\cite{w7,w16}, modified KDP~\cite{w38}, F
 model~\cite{w11}, modified F model~\cite{w35,w70,w75}, F model
with a staggered field, ...),
% the equivalence with 
%sing models with two-, three- and four-body interactions, 
this monograph remains extremely modern and 
valuable from a technical viewpoint.

Among the exactly soluble models
(the bread-and-butter of F. Y. Wu)  was one
that, for a long time, was a ``sleeper'', namely, Bethe's 1931 solution
of the ground state energy and elementary excitations 
of the one-dimensional quantum-mechanical 
spin-${1\over 2}$ Heisenberg model of antiferromagnetism.
We will see  below
a large set of results from
the Lieb-Wu monograph 
on {\em vertex models}, in particular the six-vertex
model. 
%Looking at the six-vertex model 
%amounts to considering the {\em Bethe ansatz}
%explicitly or implicitly. 

The monograph gives an extremely
 lucid exposition of the Bethe ansatz for the
six-vertex model. The  Bethe ansatz is analyzed and explained
 in the most general framework
(with horizontal and vertical fields) and it is
a must-read to  anyone who wants to work
seriously on the coordinate Bethe ansatz. It is certainly
much more interesting and deeper  than so many subsequent
papers that have revisited,
at nauseum, the Bethe ansatz 
%in the simple case 
of the 
symmetric six-vertex model, re-styling this simple 
Bethe ansatz with a conformal resp. quantum group, resp. knot theory, 
resp. ... framework.
The analysis of the conditions
for the transfer matrix $\, T$ of the most general sixteen-vertex model
to have a non-trivial ``linear operator'' (1D quantum Hamiltonian)
 that commutes\footnote{Which is
the most obvious manifestation of the Yang-Baxter integrability.} 
with $\, T$ (pages 367 to 373) are probably one of the first pages
any student who wants to study integrable lattice models should 
read.

The monograph makes crystal clear of the fact that the
Bethe ansatz is related to the conservation of certain charge.
This is seen from the fact that most of the analyses
 (from page 374 to page 444)
relies on the use (page 363 equation (81))
of the variable $\, y\, = \, 1-2\;n/N$, which in the spin language is the average 
$\, z$-component of the spin per vertical bond, namely, 
$\, y \, = \, <S_z>/N$ for a  square lattice of size
 $\, N \times M\, $, 
where $\, n$ denotes the number of 
down arrows  and $\, N$ the number
of vertical bonds in a row.

We use the same notation as in Lieb-Wu. In particular let
us introduce
the horizontal and vertical fields 
  $\, H\, $ and $\, V\, $ respectively.
The partition function per site
%, $\, \ln(Z)$, reads 
in the thermodynamic limit is:
\begin{eqnarray}
\label{z(y)V(y)}
%\ln(Z) \, = \,\, \, \, 
\lim_{N \, \rightarrow\,  \infty} {{1} \over {N}} \cdot
 \ln(\Lambda) 
\, \,  \,  = \, \,  \,  \, \, \max_{-1 \le y \le +1} [z(y)\, + \, V \cdot y] 
\end{eqnarray}
where $\, \Lambda\, $ denotes the largest
eigenvalue of the transfer matrix.
The monograph details a large set of situations. 
Let us  consider here the regime 
\begin{equation}
 \Delta \, \equiv \, (\omega_1\, \omega_2\, +\omega_3\, \omega_4\, 
-\omega_5\, \omega_6) \,/2 \sqrt {\omega_1\, \omega_2\, \omega_3\, \omega_4} \,  <\, -1,
\nonumber
\end{equation}
and introduce  the variable:
% $\, \theta_0 $:
\begin{eqnarray}
&&e^{\theta_0} \, = \, \, {{1\, + \eta \; e^{\lambda}} \over
{e^{\lambda}\, +\;\eta}}, \qquad \quad
 0 \le \,\theta_0\,  \le \,  \lambda \qquad \quad \hbox{where:} \nonumber \\
&&\quad \, \eta \, = \, e^{|K_1|} \, = \, \Bigl({{\omega_1 \, \omega_2 } \over
{\omega_3 \, \omega_4 }}  \Bigr)^{1/2} \qquad \hbox{or:} \qquad
 \Bigl({{\omega_3 \, \omega_4 } \over
{\omega_1 \, \omega_2 }}  \Bigr)^{1/2}. \qquad \nonumber 
\end{eqnarray}
When $\, \Delta <-1$, $\, z(y)\, $ reads:
\begin{eqnarray}
&&z(y) \,\, = \, \, \, -K_2 \, + \, \max(0, \, -K_1) \, + {{1} \over
{4\;\pi}} \cdot \int_{-b}^{+b} \, R(\alpha) \cdot C(\alpha) \cdot
d\alpha \qquad \qquad  \nonumber \\
&&\quad \hbox{where:}\quad \quad  \qquad C(\alpha) \, = \, \, ln\Bigl( 
{{ \cosh(2\; \lambda -\, \theta_0)\, - \, \cos(\alpha) } 
\over {\cosh(\theta_0)\, - \, cos(\alpha)  }}
\Bigr) \nonumber
\end{eqnarray}
and the (normalized) density\footnote{One has $\,R(\alpha)\cdot
d\alpha \, = \, \, 2\; \pi \, \rho(p) \cdot dp$.}
$\, R(\alpha)\, $ 
satisfies the Bethe-ansatz integral equation with the kernel $\, K(\alpha)$:
\begin{eqnarray}
\label{kernel}
&&R(\alpha)\, = \,\, \,\, {{\sinh(\lambda) }
 \over { \cosh(\lambda)\, - \,\cos(\alpha)
}}\, - \, \int_{-b}^{+b} K(\alpha-\beta) \cdot R(\beta)
\, d\beta  \\
&& \qquad  \hbox{with:} \quad  \qquad  
2\; \pi \cdot K(\alpha-\beta) \, = \, \,
 {{\sinh(2\;\lambda) } \over { \cosh(2\; \lambda)\, - \,\cos(\alpha-\beta)
}}. \nonumber 
\end{eqnarray}
The integral equation (\ref{kernel}) is nothing but the
well-known Yang-Yang Bethe ansatz  integral equation on the density 
$\, \rho(q)\, $:
\begin{eqnarray}
1 \, = \,\, \,2 \; \pi \cdot \rho(p) \, - \, 
\int_{-Q}^{+Q} \, {{ d\theta(p,q) } \over { dp}} \rho(q) \cdot dq
\qquad \hbox{with:} \qquad Q \, = \, {{ \pi \cdot (1-y) } \over {2}}.
\nonumber 
\end{eqnarray}
\vskip .3cm
The range $\, b$ of the new variable $\, \alpha$ 
in the integral relation (\ref{kernel}) can be deduced from the definition
of the  density $\, R(\alpha)\, $:
\begin{eqnarray}
 \pi \cdot (1-y) \,\, = \,\, \,\, \int_{-b}^{+b} R(\alpha)\, d\alpha \,\, = \,
 \,\, \int_{-Q}^{+Q} \, \rho(q) \cdot dq \, .\nonumber 
\end{eqnarray}
When $\, y\,= \, 0$ the integral attains its maximum range
and one can solve (\ref{kernel}) by a Fourier series of a Fourier
transform. One thus  gets $\, R(\alpha)\, $ as a simple 
% $\, dn\, $
dn \, elliptic function.
  Not surprisingly one can also
calculate all the derivatives of $\, z(y)\, $ 
at $\, y\,= \, 0\, $.
% $\, z'(0)$, $\, z^{'''}(0),\,$ etc ..
One can thus expand $\, z(y) \, $ namely, write $\, z(y) \, = \, z(0) \, -
z'(0)\cdot y \, + \, z^{'''}(0)\cdot y^3/6 \, + \, \cdots\,.  $ 
%in order to get  an expansion of $\, z(y) \, $ at small $\, y$.
In  first order in $\, y$ one obtains $\, z'(0) \, = \, -
\Xi(\lambda\, - \theta_0)$ where the
 function $\, \Xi$ is related to the
Jacobian elliptic function\,  nd:
%$\, nd$:
\begin{eqnarray}
\Xi(\phi) \,\, = \,\,\, \ln\Bigl( {{\cosh((\lambda\, \,
+ \,\phi)/2)} \over {\cosh((\lambda\,\, -\, \phi)/2)}}
\Bigr)\, -\, {{\phi} \over {2}} \,\, - \sum_{n\, =1}^{\infty} {{ (-1)^n
\cdot e^{-2\;n\, \lambda} \cdot \sinh(n\;\phi)} 
\over {n \cdot \cosh(n\; \lambda)}} \nonumber 
\end{eqnarray}
The function $\, \Xi(\phi)\, $ 
also satisfies the nice involutive functional relations\footnote{In
agreement with the inversion relations on the model.}:
\begin{eqnarray}
\label{functioinvol}
\Xi(\phi) \, = \, \,- \Xi(-\phi), \qquad 
\Xi(\lambda\, +\,\phi) \, = \, \,
 \Xi(\lambda\,-\,\phi), \qquad 
\Xi(4\; \lambda\, +\,\phi) \, = \,\,  \Xi(\phi). \nonumber
\end{eqnarray}

Let us consider the thermodynamic properties of the model
when $\, H\, = \, 0$ and $\, V \ne 0$. From (\ref{z(y)V(y)}) one  
sees that the thermodynamic properties 
depend on the optimal choice of $\, y$ given by:
\begin{eqnarray}
z'(y) \, = \, \, - \, V \nonumber 
\end{eqnarray}

When lowering the temperature
the slope of $\, z(y)$ corresponding to the transition
sticks at $\, y\, \simeq \, 0$, and one thus
has (see page 425 of~\cite{w28}) an antiferroelectric transition
occurring at $\, T_c(V)$ given by:
\begin{eqnarray}
\label{transc}
V \,\, = \,\, \,\Xi(\lambda \, - \, \theta_0).
\end{eqnarray}

This gives a beautiful example of a 
{\em transcendental critical manifold}
which reduces, in some domain of the 
parameters (low temperatures), to
 a transcendental equation (\ref{transc})
and {\em not to an algebraic one} as one is used to
see in exactly solvable models. One thus has 
a {\em transcendental critical
manifold} for a vertex model for which one can actually 
write down the exact Bethe ansatz (see equation (\ref{kernel})).
Writing a closed simple formula for the solution
is not possible, but one can certainly 
find numerical solutions on a computer.
Should we say that the model is exactly solvable 
but not ``computable''? We will revisit these questions
of algebraicity
 of the critical manifold
versus integrability in other sections of this
paper with other critical manifolds
conjectures, or results, of F. Y. Wu (see 
for instance sections (\ref{invduakago}), (\ref{invduacrit}) below). For those 
who have a ``naive'' point of view
on the character of critical manifolds\footnote{With, for instance,
 a prejudice of
algebraicity of the critical manifolds of ``solvable''
models: all examples known in the literature are
polynomial expressions in well-suited variables $\, e^{K_i}$.
These include, for instance, the critical varieties of the anisotropic
Ising, or Potts, models on square, triangular lattices, or the 
critical varieties of the Baxter model. For non-integrable  
models the common wisdom
is, probably, that critical manifolds are always 
analytic, or may be differentiable, and 
the algebraicity of the critical manifolds
is ruled out by the non-integrability. This is 
also a naive point of view: see (\ref{critBax}) in 
section (\ref{invduacrit}).}, the example (\ref{transc}) 
shows that a model having a Bethe ansatz
can have a {\em transcendental} critical manifold.  

The Lieb-Wu review provides 
wonderful pieces of analytical work (analysis 
in one complex variable, see for instance pages 410-411
and the analysis of the analytic structure of the F model
or the temperature Riemann structure for the free
energy of the F model). One finds a festival
of one complex variable analytical tools (Maclaurin 
formula, tools for the evaluations of asymptotic behaviors, path
integration, etc.).
   
Many more results can be found in the monograph
(the three-color problem, the hard 
square model, the F model on the triangular
lattice, three coloring of the edges of the hexagonal lattice ...).
Let us mention, in particular, the 
six-vertex model with {\em site-dependent} weights (which can be
considered as the first example of a $\, Z$-invariant model).
Let us introduce $\, \omega_j(I,J) \, $ where $\, j \, = \, 1, \cdots , \,
6$, are the six possible Boltzmann factors
% ($\omega_7(I,J) \,= \,\omega_8(I,J)$)
 of the vertex in row $\, I$ and column $\, J$, and
let us require that
the algebraic invariant $\, \Delta$ be independent of $\,I,J$:
\begin{eqnarray}
\label{invariantnonconstant}
\Delta \, = \, \, 
{{ \omega_1(I,J) \,\omega_2(I,J) \, + \,  \omega_3(I,J)
\,\omega_4(I,J)\, - \, \omega_5(I,J) \,\omega_6(I,J)} \over { 2
\cdot ( \,  
\omega_1(I,J) \,\omega_2(I,J)\omega_3(I,J)
\,\omega_4(I,J))^{1/2}}}.
\end{eqnarray}
Up to
a  multiplicative factor $\,\kappa_{I,J}$\,,
a (rational) parametrization of these invariance
 conditions (\ref{invariantnonconstant}) is:
\begin{eqnarray}
\label{parainhom}
&&\omega_1(I,J) \,= \, \,  (1\, -\, t\cdot p_{I,J})
\cdot \alpha_{I,J} \, \beta_{I,J} , \quad \nonumber \\
&&\omega_2(I,J) \,= \, \,  (1\, -\, t\cdot p_{I,J})
\cdot {{ 1} \over {\alpha_{I,J} \, \beta_{I,J}}},\\
&&\omega_3(I,J) \,= \, \,  (p_{I,J}\, -\, t)
\cdot  {{ \alpha_{I,J}} \over { \beta_{I,J}}},  \quad 
\omega_4(I,J) \,= \, \,  (p_{I,J}\, -\, t)
\cdot {{ \beta_{I,J}} \over {\alpha_{I,J} }}, \nonumber \\
&&\omega_5(I,J) \,= \, \,   ({{1} \over {t}}\, -\,
t) \cdot  p_{I,J} 
\cdot  \gamma_{I,J}, \quad 
\omega_6(I,J) \,= \, \,   (1\, -\, t^2)
\cdot  {{1} \over {\gamma_{I,J}}}\nonumber
\end{eqnarray}
The Baxter's $\, Z$-invariance condition for integrability requires that
the $\, p_{I,J}$'s are actually products of a (spectral) parameter 
depending on the row and another parameter 
depending on the column: $\, p_{I,J} \, = 
\, \, \rho_I \cdot \sigma_J$. 
We will see 
%in section (\ref{correspond}),
 in section (\ref{Potts}) when sketching 
the correspondence between
the standard scalar Potts model and a staggered asymmetric
six-vertex model, that these product conditions,
$\, p_{I,J} \, = \, \, \rho_I\, \sigma_J$,  actually correspond in the
case of the checkerboard Potts model to
 criticality,
% and integrability conditions 
%of the  model ($u\;v\;w\;z\, = \, 1$ see (\ref{cricheck}))
or to a vanishing conditions of a staggering field $\, H_{\rm stag}\, $
in a Lee-Yang zeros viewpoint: $\, |z| \, = \, e^{H_{\rm stag}} \,
= \, 1$. 
 
Provided that  the $\, p_{I,J} \, = 
\, \, \rho_I \cdot \sigma_J$ integrability conditions are satisfied,
the  partition function, with parametrization
 (\ref{parainhom}), can be expressed as a multiplicative closed formula:
\begin{eqnarray}
\label{multiplicative}
&&Z \, = \, \,\, 2 \, \prod_{I\, = \, 1}^{M}\; \prod_{I\, = \, 1}^{M}
{{\sqrt{\omega_5(I,\, J) \,\omega_6(I,\, J)}
} \over {1-t^2}} \cdot F(\rho_I\, \sigma_J) \cdot F\Bigl( 
{{1 } \over { \rho_I\, \sigma_J}}
\Bigr), \nonumber \\
&& \qquad \quad \quad  \hbox{where:} \quad \qquad \quad 
F(z) \, = \, \, \prod_{m\, = \, 1}^{\infty} {{ 1-t^{4m-1}\;z } \over
{1-t^{4m+1}\; z}} 
\end{eqnarray}

\subsection{Vertex models: free fermions}
\label{subfree}
Another classic work of  F. Y. Wu is his 1970 paper with
C. Fan in which they coined the term the {\em free-fermion model}~\cite{w14}. This work 
was later extended to its checkerboard version
during one of Wu's visit to Taiwan~\cite{w47,w49}. In the following
we shall arrange the homogeneous vertex weights in a  matrix 
$\, R$, whose 
 size and form vary according to the number of
edge states and the coordination
number of the lattice. Typical examples we will consider are the 2D
square and triangular lattices shown below:
\setlength{\unitlength}{0.0075in}
\begin{eqnarray*}
\begin{picture}(120,120)(20,700)
\thinlines
\put( -40,755){\makebox(0,0)[lb]{\raisebox{0pt}[0pt][0pt]{\elvrm $
R^{ij}_{uv} =  $}}} 
\put( 80,820){\line( 0,-1){120}}
\put( 20,760){\line( 1, 0){120}}
\put(125,765){\makebox(0,0)[lb]{\raisebox{0pt}[0pt][0pt]{\elvrm $u$}}}
\put( 85,710){\makebox(0,0)[lb]{\raisebox{0pt}[0pt][0pt]{\elvrm $j$}}}
\put( 85,810){\makebox(0,0)[lb]{\raisebox{0pt}[0pt][0pt]{\elvrm $v$}}}
\put( 20,765){\makebox(0,0)[lb]{\raisebox{0pt}[0pt][0pt]{\elvrm $i$}}}
\end{picture}
\qquad & \qquad
\begin{picture}(120,120)(20,700)
\label{fig1}
\thinlines
\put( 50,730){\line( 1, 1){ 60}}
\put( 110,730){\line( -1,1){60}}
\put( 20,760){\line( 1, 0){120}}
\put(110,795){\makebox(0,0)[lb]{\raisebox{0pt}[0pt][0pt]{\elvrm $v$}}}
\put( 45,710){\makebox(0,0)[lb]{\raisebox{0pt}[0pt][0pt]{\elvrm $j$}}}
\put(125,765){\makebox(0,0)[lb]{\raisebox{0pt}[0pt][0pt]{\elvrm $u$}}}
\put( 50,795){\makebox(0,0)[lb]{\raisebox{0pt}[0pt][0pt]{\elvrm $w$}}}
\put( 110,710){\makebox(0,0)[lb]{\raisebox{0pt}[0pt][0pt]{\elvrm $k$}}}
\put( 20,765){\makebox(0,0)[lb]{\raisebox{0pt}[0pt][0pt]{\elvrm $i$}}}
\put( 15,755){\makebox(0,0)[rb]{\raisebox{0pt}[0pt][0pt]{\elvrm
$R^{ijk}_{uvw}=$}}}
\end{picture}
\qquad & \qquad \\
\mbox{2D square}\qquad & \qquad \mbox{2D triangular}\qquad & \qquad
\end{eqnarray*}

\subsubsection{Free-fermion asymmetric eight-vertex model}
\label{asymmetric}

C. Fan and F. Y. Wu obtained many free-fermion
results~\cite{w10,w14}. The free energy 
of the most general free-fermion model on a square lattice 
evaluated by  Fan and  Wu reads:
\begin{eqnarray}
\label{freeFan}
&&f \, 
%= \, \lim_{N \rightarrow \infty} \, {{1} \over {N}} \cdot \ln(Z) \,
= \, \, {{1} \over {16 \, \pi^2 }} \int_{0}^{2\;\pi}\int_{0}^{2\;\pi}
d\theta \; d\phi \, \ln\Big(2\;a \, + \, 2\; b \, \cos\theta \, +\, \nonumber \\
&&\qquad \qquad \qquad \qquad +\, 2\; c
\, \cos\phi  \, + \, 2\; d \, \cos(\theta-\phi) \, + \, \, 2\; e \,
\cos(\theta+\phi)\Big), \nonumber 
\end{eqnarray}
where:
\begin{eqnarray}
&&a \, = \, {{1} \over {2}} \cdot
(\omega_1^2+\omega_2^2+\omega_3^2+\omega_4^2), \quad  \quad \quad b \, = \,
\omega_1\, \omega_3\, - \omega_2\, \omega_4, \quad\quad  \nonumber \\
&& c \, = \,
\omega_1\, \omega_4\, - \omega_2\, \omega_3,\quad\quad 
d \, = \, \omega_3\, \omega_4\, - \omega_7\, \omega_8, \quad \quad \quad
e \, = \, \, \omega_3\, \omega_4\, - \omega_5\, \omega_6, \nonumber
\end{eqnarray}
provided that the free fermion condition:
\begin{eqnarray}
\label{freefermioncond}
\omega_1 \, \omega_2 \, + \, \omega_3 \, \omega_4 \, \, = \, \, \, 
\omega_5 \, \omega_6 \, + \, \omega_7 \, \omega_8 \,
\end{eqnarray}
is satisfied. 
%Note that
%without any lost of generality one can restrict oneself to:
%$\, \omega_5 \,= \omega_6\, $ and $\,  \omega_7 \,= \omega_8\, $
%using gauge transformations. 
%{\em This is not the case with} 
%$\, \omega_1 \,$ and $\,  \omega_2\, $ on one side and
%$\, \omega_3 \,$ and $\,  \omega_4$ on the other side.
 
Let us revisit some of their results from an 
inversion relation viewpoint.
%  introducing 
% other variables well-suited to the asymmetric eight-vertex model
% namely: $\, a \,  = \omega_1$, $ a' \, = \omega_2$, 
% $\, b \,  = \omega_3$, $ \, b' \, = \omega_4$, 
% $\, c \, = \omega_5$, $\, c' \, = \omega_6$, 
% $\, d\, = \, \omega_7$, $\, d'\, = \, \omega_8$.
Renaming the vertex weights as
  $\, a \,  = \omega_1$, $ a' \, = \omega_2$, 
 $\, b \,  = \omega_3$, $ \, b' \, = \omega_4$, 
 $\, c \, = \omega_5$, $\, c' \, = \omega_6$, 
 $\, d\, = \, \omega_7$, $\, d'\, = \, \omega_8$,
the  matrix $\, R$ of the eight-vertex model is then:
\begin{equation}
\label{8v}
R = \pmatrix {  a & 0 & 0 & d' \cr
                0 & b & c' & 0 \cr
                0 & c & b' & 0 \cr
                d & 0 & 0 & a' }.
\end{equation}
A matrix of the form (\ref{8v}) can be brought, by a similarity
transformation, to a block-diagonal
form:
\begin{eqnarray}
R = \pmatrix{ R_1 & 0 \cr 0 & R_2 }, \qquad\mbox{ with }
\qquad  R_1 = \pmatrix{ a & d' \cr d & a' } \quad \mbox{and}
\quad R_2 = \pmatrix{ b & c' \cr c & b' } \nonumber 
\end{eqnarray}
If one introduces $\, \delta_1= \, a a' - d d' \, $ 
and  $\, \delta_2 = \, b b' - c c'$, the
determinants of the two blocks, then the
 (homogeneous) matrix inverse $\, I$ 
%written polynomially
(namely  $R \rightarrow {\rm det}(R) \cdot R^{-1}$) 
 reads:
\begin{eqnarray}
\label{Ihomo}
&&(a, \, a', \, d, \, d') \,\quad   \rightarrow \quad 
(a' \cdot \delta_2,\,  a \cdot \delta_2,\, -d \cdot \delta_2,\,  -d'
\cdot \delta_2) \\
&&(b, \, b', \,c, \, c') \,\quad    \rightarrow \quad 
(b' \cdot \delta_1, \, 
b \cdot \delta_1, \, 
 -c \cdot \delta_1, \,  -c' \cdot \delta_1) \nonumber 
\end{eqnarray}
%The free-fermion condition~\cite{w14} reads:
%\begin{equation}
%\label{ff8}
%a a' - d d' + b b' - c c' \, =\, \, \, 0.
%\end{equation}

It is straightforward to see that the free-fermion condition
 (\ref{freefermioncond}) is
$\; \delta_1 = - \delta_2 \;$  which has 
the effect of {\em linearizing} the
%(homogeneous) 
inversion (\ref{Ihomo}) into an involution given by:
 \begin{eqnarray}
\label{Ilin}
a \leftrightarrow a' , \quad\quad b \leftrightarrow -b',\quad\quad
(d, \, d') \rightarrow \, \, (-d, \, d') , \quad\quad
b \leftrightarrow -b', \quad\quad
(c, \, c') \rightarrow \, \, (c, \, c')\nonumber 
\end{eqnarray}
The group generated by the two inversion relations of the
model is then realized by permutations of the entries  mixing
with sign changes, and its orbits are thus {\em finite}.
The {\em finiteness} condition of the group is a common feature of all
free-fermion models.

\subsubsection{Free-fermion for the 32-vertex model on a triangular lattice}
\label{Freefermtriangul}

We next consider the free-fermion conditions of J. E. Sacco and  F. Y. Wu~\cite{w50} 
for the 32-vertex model on a
triangular lattice.
 Using the same notation as in~\cite{w50}, we have:
\begin{equation}
\label{rsawu}
R = \left [\begin {array}{cccccccc} {\it f_{0}}&0&0&{\it f_{23}}&0&{\it
f_{13}}&{
\it f_{12}}&0\\0&{\it f_{36}}&{\it f_{26}}&0&{\it f_{16}}&0&0&{\it
{\bar f}_{45}}\\0&{\it
f_{35}}&{\it f_{25}}&0&{\it f_{15}}&0&0&{\it {\bar f}_{46}}\\{\it
f_{56}}&0&0&
{\it {\bar f}_{14}}&0&{
\it {\bar f}_{24}}&{\it {\bar f}_{34}}&0\\0&{\it f_{34}}&{\it
f_{24}}&0&
{\it f_{14}}&0&0&{\it {\bar f}_{56}}
\\{\it f_{46}}&0&0&{\it {\bar f}_{15}}&0&{\it {\bar f}_{25}}&{\it {\bar
f}_{35}}&
0\\{\it f_{45}}&0&0&{\it
{\bar f}_{16}}&0&{\it {\bar f}_{26}}&{\it {\bar f}_{36}}&0\\0&{\it
{\bar f}_{12}}&
{\it {\bar f}_{13}}&0&{\it {\bar f}_{23}}&0&0&{
\it {\bar f}_{0}}\end {array}\right ].\nonumber 
\end{equation}
By permuting  rows and columns, this matrix can be brought into
the block diagonal form:
\begin{equation}
\label{blocks} \nonumber
R= \pmatrix{ R_1 & 0 \cr 0 & R_2 }, \qquad \qquad  \mbox{ with:} \nonumber 
\end{equation}
\begin{equation} \nonumber
R_1 = \left [\begin {array}{cccc} {\it f_{0}}&{\it f_{13}}
&{\it f_{12}}&{\it f_{23}}\\
{\it f_{46}}&{\it {\bar f}_{25}}&
{\it {\bar f}_{35}}&{\it {\bar f}_{15}}\\
{\it f_{45}}&{\it {\bar f}_{26}}&{\it {\bar f}_{36}}&
{\it {\bar f}_{16}}\\
{\it f_{56}}&{\it {\bar f}_{24}}&{\it {\bar f}_{34}}&
{\it {\bar f}_{14}}\end {array}\right ]\, ,
 \qquad
R_2 = \left [\begin {array}{cccc} {\it f_{14}}&{\it f_{34}}&{\it
f_{24}}&{\it {\bar f}_{56}}\\{
\it f_{16}}&{\it f_{36}}&{\it f_{26}}&{\it {\bar f}_{45}}\\{\it
f_{15}}&{\it f_{35}}&{\it f_{25}}&
{\it {\bar f}_{46}}\\{\it {\bar f}_{23}}&{\it {\bar f}_{12}}&{\it {\bar
f}_{13}}&{\it {\bar f}_{0}}\end {array}\right ] \nonumber
\end{equation}
The inverse $\, I$  written polynomially (homogeneous matrix inverse) 
is  now a transformation of
degree $7$.
If one introduces the two determinants $\, \Delta_1 \, = \, \det(R_1) $
and  $\Delta_2 = {\rm det}(R_2) $, then each term in the expression of $\, I(R)$
is a product of a degree three minor, taken within a block, times the
 determinant of the other block. This inverse $\, I$ clearly singles
out one of the three directions of the triangular lattice.
 These three involutions do not commute
and generate a quite large infinite discrete group   
$\Gamma_{\rm triang}$ (see also section (\ref{invduacrit}) below). 

%Denoting $\bar{f}_{12}\, =\, f_{3456}$ and so on, 
The free-fermion conditions
of Sacco and Wu~\cite{w50} read:
\begin{eqnarray}
\label{ff32a}
& f_0 f_{ijkl}& = \, \,  f_{ij} f_{kl} - f_{ik} f_{jl} + f_{il} f_{jk},
\qquad \quad \forall \; i,j,k,l \, = \, 1,\, \dots,\,  6 \nonumber \\
\label{ff32b}
& f_0 \bar{f}_0 & =\, \,   f_{12} \bar{f}_{12} - f_{13} \bar{f}_{13} + f_{14}
\bar{f}_{14} -
f_{15} \bar{f}_{15} + f_{16} \bar{f}_{16} 
\end{eqnarray}
which we denote by $\, {\cal V}$.
What is remarkable is that, not only
is the rational  variety $\, {\cal V}$
%, defined by (\ref{ff32b}), 
globally invariant under
$\Gamma_{\rm triang}$, but {\em again} the realization of this
(generically very large infinite discrete
 group) $\, \Gamma_{\rm triang}$ on this
variety becomes {\em finite}.
This comes about from
the degeneration of $\, I$ into a mixture of 
sign changes  and
permutations of the
entries, as in the case in the preceding subsection.

\vskip .3cm
{\em Remark}: The ordinary {\em matrix product of
 three matrices} (\ref{rsawu}) {\em solutions} of (\ref{ff32a}) 
is {\em another 
solution} !
In other words, if $\, R_{\alpha}, \; R_{\beta}, \; R_{\gamma} \; \in {\cal V}$, then
$ R_{\alpha} \cdot R_{\beta} \cdot R_{\gamma} \; \in {\cal V} $, 
while $\, R_{\alpha} \cdot R_{\beta} \notin {\cal V
}$.
 This was also the case for 
%solutions 
involutions of (\ref{freefermioncond}) in the case
of the square lattice, but
 the mechanism is more subtle here
as conditions  (\ref{ff32b}) imply $\, \Delta_1=\, \Delta_2$.

\subsection{Dimers and spanning trees}
\label{dimers}
Before Fan and Wu's free-fermion vertex models
the Onsager solution of the two-dimensional Ising
model is clearly the first free-fermion model
ever solved.
There were also several approaches to the two-dimensional Ising model that
did not use the transfer matrix formalism; the most interesting one is
perhaps the mapping of the problem onto a dimer-covering problem 
on a slightly more complicated lattice. 
The dimer problem was first
solved by Temperley-Fisher and Kasteleyn. Kasteleyn found out 
how to treat the most general planar graph. 

The dimer problem has a
life of its own and has since found many followup works, 
not only in statistical mechanics, but also 
 in combinatorial theory.  In this regard
 Wu has provided a large number of new
 results~\cite{w34,w164,w173,w183,w195,w196},
including applications to condensed matter physics as well as
in pure combinatorial analysis. 
In addition, Wu has  obtained new results on the  spanning
tree problem~\cite{w187,w189}, a problem intimately related
to the dimer problem through a bijection due to Temperley.
In the following we shall describe some of the contributions in this area.

%The dimer model on the honeycomb lattice was first solved by
%Kasteleyn. This was generalized by 
%H. Y. Huang,  F. Y. Wu, H. Kunz, and D. Kim,
%o the case where the dimers have
%nearest-neighbor interaction~\cite{w164} 
%(an exact solution of the five-vertex model). 
%This model relates to
% a degenerate case of the six-vertex model,
%requiring a special Bethe ansatz analysis. The resulting phase diagram
%of this five-vertex model is quite complicated. This work is also used
%in F. Y. Wu's papers on the three-dimensional 
%dimer models~\cite{w167} (see section (\ref{3ddimers})).
%(Spanning Trees on Hypercubic Lattices and Non-orientable Surfaces).

\subsubsection{Revisiting dimers: the honeycomb lattice}
The dimer model on the honeycomb lattice was first solved by
Kasteleyn, but he never published the solution except hinting at
the existence of a transition. 
This deficiency was made up by  Wu in a 1968 paper~\cite{w8}
in which he presented details of the analysis for the honeycomb
lattice, and applied the results to describe the physics
of a modified KDP model.
  
%(Spanning Trees on Hypercubic Lattices and Non-orientable Surfaces).

\subsubsection{Revisiting dimers: Interacting dimers in 2 and 3 dimensions}
\label{3ddimers}

Almost 30 years after the publication of the solution of the dimers on
the honeycomb lattice~\cite{w8},  Wu and co-workers made two important
extensions of the earlier Kasteleyn solution.  In the first,
 H. Y. Huang,  F. Y. Wu, H. Kunz, and D. Kim~\cite{w164}
considered  the case where the dimers have
nearest-neighbor interaction. 
%(an exact solution of the five-vertex model). 
This model turns out  to be identical to the most general five-vertex model, 
 a degenerate case of the six-vertex model which
requires a special Bethe ansatz analysis. The resulting phase diagram
of this five-vertex model is very complicated and the analysis extremely
lengthy.

In the second work H. Y. Huang, V. Popkov and F. Y. Wu~\cite{w167,w173}
introduced, and solved, a three-dimensional 
model  consisting of layered honeycomb dimer lattices 
as described in the preceding subsection, but with 
a specific layer-layer interaction. 
Again, the phase diagram is very complicated.
 It is noted that this model is 
the only solvable three-dimensional lattice model
 with {\em physical} Boltzmann weights
(the Baxter solution of the 3D Zamolodchikov
 model has negative weights).
However, the layered dimer model, while
 having strictly positive weights, describes dimer configurations 
in which dimers are confined in planes. As a consequence the critical  
behavior is essentially two-dimensional.

\subsubsection{Revisiting dimers: a continuous-line model}
\label{contline}
F. Y. Wu and H. Y. Huang~\cite{w150} have further used a dimer mapping to solve a 
continuous-line
lattice model in three and higher dimensions.  They have also applied it
to model a type-II superconductor~\cite{w152}.
  In three dimensions the model is a special case of an $\, O(n)$ model 
on a finite $\, L_1 \times L_2 \times L_3$ cubic
lattice with periodic boundary conditions with the partition function:
\begin{eqnarray}
\label{Z3D}
Z(n)\,  =\, \, \, \sum_{closed \, \, polygons} n^l \cdot z^b, \nonumber 
\end{eqnarray}
where the summation is taken over all closed non-intersecting 
polygonal configurations, $\, l\, $ is the number of 
polygons, and $\, b\, $ is the number of edges of
 each configuration. They considered the $\, n \, = \, -1\, $ special case
which they showed to be in one-to-one 
correspondence with a dimer problem whose partition function 
can be evaluated as a {\em Pfaffian}. The
 result  for a finite
lattices is:
\begin{eqnarray}
\label{resuZ3D}
Z(-1)\,  =\, \,\,  
\prod_{n_1\, =1}^{L_1} \prod_{n_2\, =1}^{L_2}\prod_{n_3\, =1}^{L_3}
\, \left| 1\, + \, \sum_{i=1}^{3}\, z \cdot e^{2\; \pi \; n_i/N_i}
\right |.  \nonumber 
\end{eqnarray}
In the thermodynamic limit, this leads to the per-site free energy:
 \begin{eqnarray}
\label{thermresuZ3D}
f\,  =\,\,  \, {{1} \over {(2 \; \pi)^3}} \cdot
\int_{0}^{2 \; \pi} d\theta_1\; \int_{0}^{2 \; \pi} d\theta_2
\; \int_{0}^{2 \; \pi} d\theta_3 \, \, 
 \ln \left|1\, + \, \sum_{i=1}^{3}\, z \cdot e^{2\;
 \pi \; n_i/N_i}\right |.  \nonumber 
\end{eqnarray}
The   phase diagram is rich and  quite non-trivial.

However, it must be said that this exactly soluble three-dimensional
 \, O$(-1)$ model describes line configurations running only
in a {\em preferred directions} and, secondly,\footnote{But we are used
 to this after R. J. Baxter's solution of the 3D Zamolodchikov model.}
the Boltzmann weights can be negative. 

\subsubsection{Revisiting dimers: nonorientable surfaces}
% and other Klein bottles. }
\label{Klein}
More recently W. T. Lu and  F. Y. Wu initiated studies on dimers and Ising models
 on {\em nonorientable} surfaces~\cite{w183,w192,w196}.
% (dimer statistics on a M\"obius
% strip and the Klein bottle; Ising Model for non-orientable surfaces;
% closed-packed dimers on non-orientable surfaces).
%One very nice  recent result is contained om
% a paper with W. T. Lu~\cite{w196}.
For   dimers on an $\, M \times N\, $ 
net, embedded on non-orientable surfaces, 
 they solved {\em both} the M\"obius strip and 
the Klein bottle problems for all sizes $\, M \, $ and  $ \, N$ and obtained the 
% Recalling the bijection between dimer
% and spanning tree configurations (Temperley, Kenyon,
% Propp and Wilson) one can also equivalently 
% consider spanning tree problems on non-orientable surfaces~\cite{w187} 
% (Spanning Trees on Hypercubic Lattices and Non-orientable Surfaces).
dimer generating function $\, Z_{M, N}\,$ as:
 \begin{eqnarray}
&&{{Z_{M, N}} \over { z^{M\,N/2}}} \,  \, \, \,  \nonumber \\
&&  \, = \, \, \,
 {\rm Re}\Bigl(
(1-i) \cdot \prod_{m=1}^{(M+1)/2} \prod_{n=1}^{N} \Bigl(
2\; i \, (-1)^{M/2+m+1} \, \sin\Bigl( {{(4\;n-1) \, \pi}
 \over {2\; N}}\Bigr)
\, + 2\, X_m
\Bigr)\Bigr) \nonumber 
\end{eqnarray}
where \, Re\,  denotes the real
part, $\, z_v\, $ and $\, z_h\, $ are the dimer weights in the
vertical and horizontal directions respectively, 
 and  $\,  X_m$ is given for
the M\"obius strip and the Klein bottle respectively by:
\begin{eqnarray}
X_m \, = \, \, {{z_v} \over {z_h}} \cdot \cos\Bigl( {{m\; \pi} \over
{M+1}} \Bigr), \qquad \qquad 
X_m \, = \, \, {{z_v} \over {z_h}} \cdot \cos\Bigl( {{(2\; m\;-1) \pi} \over
{M}} \Bigr) \nonumber 
\end{eqnarray}
%\vskip .6cm 
In  paper~\cite{w196} they also obtained  an {\em extension
of the Stanley-Propp reciprocity theorem} for dimers\footnote{A subject matter 
of pertinent interest to mathematicians.}.
Inspired by this work, there is now much
 activity in this area.
There is also very much current interest in finite-size corrections and conformal field
theories on more complicated surfaces (higher genus pretzels, ...).

\subsubsection{Dimers on a square lattice with a boundary defect}
In a very recent paper~\cite{w195} fittingly
 dedicated to the 70th birthday
of Michael Fisher who first solved
the dimer problem for the square lattice, W. J. Tzeng and
 F. Y. Wu obtained the dimer generating
function for the square lattice with one corner (or some other
boundary site) of the lattice missing. 
In this work they made use of a bijection between dimer
and spanning tree configurations due to Temperley (and extended
by  Wu in his unpublished 1976 lecture notes as well as
more recently by  Kenyon,
Propp and Wilson).
%establish that the dimer generating function is
%independent of the location of the vacancy, and deduce the closed-form 
%expression for the generating function. 
They also carried out   finite-size analyses which 
lead to a logarithmic correction term
in the large-size expansion for the vacancy problem
with free boundary conditions. They  found a central charge
$\, c\, = \, -2\, $ for the vacancy problem, to be compared with  
$\, c\, = \, -1\, $ when there is no vacancy.
This central charge $\, c\, = \, -2\, $
is in contradiction with the prediction of  the conformal field theory.
 
\subsubsection{Spanning trees}
\label{spanningtrees}
As mentioned above, the problem of spanning trees in graph theory is
intimately related to the dimer problem, and it is not surprising that
  Wu found his way to spanning trees. 
 In 1977 he published a paper~\cite{w58} on the counting of spanning
trees on two-dimensional lattices using the equivalence with a
Potts model. 
Very recently he refined the tools by using a result  in algebraic graph theory
which he and W. J. Tzeng rederived using elementary means.   Tzeng and Wu
 enumerated  spanning trees  for general
$d$-dimensional lattices as well as non-orientable surfaces~\cite{w187}. 
%(Spanning Trees on Hypercubic Lattices and Non-orientable Surfaces).
Applying these results to general graphs and regular lattices,
R. Shrock and F. Y. Wu~\cite{w189} published a 
lengthy paper in which they established
new theorems on spanning trees as well as enumerating spanning trees for a
large number of
regular lattices in the thermodynamic limit.

\subsection{Miscellaneous results on vertex models}
\label{miscel}
In this section we describe an arbitrary choice
of miscellaneous results obtained by F. Y. Wu on vertex models.

\subsubsection{Boundary conditions}
The six-vertex model is known to be 
a boundary condition dependent model. However 
H. J. Brascamp, H. Kunz and  F. Y. Wu~\cite{w40} have
established, for the first time,
that, at sufficiently low temperatures or sufficiently high fields,
the six-vertex model with either periodic or free boundary conditions
are equivalent. 
% The general eight-vertex model without
% fields is not known to be solvable.

\subsubsection{The eight-vertex model in a field}
\label{eight}
A simple result due to F. Y. Wu~\cite{w100} is that a very general staggered eight-vertex
model in the Ising language (as introduced by Kadanoff and Wegner and by 
F. Y. Wu~\cite{w73}) with the special  
 Yang-Lee magnetic field $\, i\pi kT/2$, is equivalent
to Baxter's symmetric eight-vertex model and hence is soluble.
This result is remarkable since  the general eight-vertex model without
this field
is not known to be soluble.
% (a two-dimensional 
%Ising model with crossing and four-spin 
%interactions and a magnetic field $\, i\pi kT/2$).

\subsubsection{The eight-vertex model on the  honeycomb lattice}
\label{honey}

F. Y. Wu has always  been keen 
in providing results for the {\em honeycomb} lattice~\cite{w45,w124}.
One interesting result is that he
 has established the exact equivalence of the eight-vertex model
on the honeycomb lattice
with an Ising model in a nonzero magnetic field~\cite{w45,w124}.
The equivalence also leads to exact analysis of the 
Blume-Emery-Griffiths model for the honeycomb lattice (for details
see section
(\ref{Blume}) below).
 %(eight-vertex model on the honeycomb lattice~\cite{w45},
 % eight-vertex model and Ising model 
%in a nonzero magnetic field on the honeycomb lattice~\cite{w124}).

In pursuit of applications of these results, 
P. Pant and J. H. Barry and  Wu~\cite{w170,w171} 
obtained exact results for 
a model of a ternary polymer mixture which is
equivalent to an eight-vertex model.
A model ternary polymer mixture was considered with bi- and
 tri-functional monomers and a solvent placed on the sites of a honeycomb
lattice. 
  Using the equivalence with an eight-vertex model which further maps the
problem into an Ising model in a nonzero magnetic field, 
an exact analysis of the model was carried out. The phase boundary of the
three-phase equilibrium polymerization regime was determined exactly.

{\em Comment}:
These kinds of results are particularly interesting
 when one realizes that  concepts
and structures corresponding to the 
Yang-Baxter integrability do not exist.  This leads to the natural question:
 % when one 
%considers vertex model on the honeycomb lattices:
{\em how to construct a Yang-Baxter relation
for the honeycomb lattice?}
 
\subsubsection{Exact critical line of a vertex model in 3 dimensions}
 Wu~\cite{w43} has
introduced  a vertex model in three dimensions with real vertex weights, and
determined its exact first-order phase transition line 
by mapping it to an Ising model in a field.  
It also exhibits a critical point.    This is one of the very
few lattice statistical models for which exact results can be deduced
in higher-than-two dimensions.

\section{Spin models: Ising models and other models}
\label{Ising}

We will consider in this section F. Y. Wu's results on {\em spin
models}, mostly {\em Ising} 
models. Due to its importance 
 the Potts models will be treated separately
in section (\ref{Potts}). 

The distinction between vertex models and spin models
(more generally 
 Interaction Round a Face (IRF) models)
 is an important one in lattice statistical
mechanics. However  Wu showed in paper~\cite{w73}
an equivalence between an Ising model with a
vertex model.
%, {\it  J. Phys.} {\bf  A13}, L303-L305 (1980). 
Similarly in 
paper~\cite{w109}  Wu and K. Y. Lin
studied the Ising model 
on the Union Jack lattice showing it to be a
 free-fermion model.
% ({\it J. Phys.} {\bf A20}, 5737-5740 (1987)).
Many of the  free-fermion
results on vertex models of sections (\ref{asymmetric}) 
and (\ref{dimers}) can also be re-styled
as free-fermion Ising models.  

As far as the Union Jack lattice is concerned 
Wu has also obtained the spontaneous magnetization
of the three-spin Ising model~\cite{w48}. It is, in fact, obtained in terms of
the magnetic and  ferroelectric orderings 
of the eight-vertex model, or, equivalently,
the spontaneous magnetization and 
polarization of the eight-vertex model.  It was
 found that the two sublattices
possess different critical exponents.
% namely $\, (\pi\, -\,
%\mu)/4 {\bar \mu}\, $ and  $\, \pi/16{\bar \mu}\, $ 
%with $\, \cos(\mu) \, = \, \pm \tanh(K^{*})$.

An important development in the history of lattice models is the
analysis of
the phase diagram of the Ashkin-Teller
model on the square lattice by F. Y. Wu and K. Y. Lin~\cite{w44}. 
The Ashkin-Teller model is another example of spin models for which the
traditional distinction
of lattice statistical mechanics  between spin and vertex models
is irrelevant. The Ashkin-Teller model can be  seen as two Ising
models coupled together with  four-spin interactions.
Performing a dual transformation on one of the two
  Ising models and interpreting the result as a vertex model one finds
that the Ashkin-Teller model is equivalent to a {\em staggered}
eight-vertex model~\cite{w26}, thus exhibiting two phase transitions. 

Wu's analysis of spin models are not restricted
to two dimensions. For instance,
Barry and Wu have obtained exact results
for a four-spin-interaction  Ising model on the
 three-dimensional pyrochlore lattice~\cite{w122}, and Wu has  also
obtained various results for spin models on Bethe lattice and
Cayley trees~\cite{w50bis,w52}.

Wu also performed  {\em real-space renormalization} studies
for Ising models~\cite{w123}, but, not surprisingly, using
some duality ideas, namely, the duality-decimation
transformation of T. W. Burkhardt. 
 Wu had previously applied the duality-decimation transformation
in order to solve  the two-dimensional Ising model with
nearest-neighbor, next-nearest-neighbor and four-spin interactions
in a pure imaginary field~\cite{w100} (see section {\ref{eight}).  In
 paper~\cite{w123}
 Burkhardt's method which combines a bond-moving and 
duality-decimation transformation is modified 
in order to preserve the free energy in the 
renormalization transformation.

\subsection{Generalized transmissivities for spin models}
\label{transmi}
We will see below that a large number
of F. Y. Wu's work correspond to graph expansions
(see section (\ref{thereview})). 
For spin models with edge interactions this requires the introduction of
certain ``transmissivity'' variables.
Thermal transmissivities are introduced 
 when considering high-temperature expansions
of an edge-interaction spin model or performing  renormalization
analyses. They are also the natural variables 
to use in the decimation of spins in a simple multiplicative way.
%in trying to write, in a simple multiplicative way,
%the decimation of spins. 

Introduce the edge Boltzmann weight 
%of a spin edge model 
$\, W(K_1, \,K_2, \, \cdots K_n;\,\, a, \, b)\, $ 
where $\,K_1, $ \,$K_2,\,
\cdots K_n\, $ denote a set of coupling constants
describing the model, and $\, a$ and $\, b$  
two nearest-neighbor spin states which  can take on $\, q$ values.
Let us assume that the decimation 
procedure yields a  Boltzmann weight of
the {\em same} form:
\begin{eqnarray}
\label{decim}
&&\sum_b \, W(K_1, \,K_2, \, \cdots K_n; \,\, a, \, b) 
\cdot W(K'_1, \,K'_2, \,\cdots K'_n; \,\, b, \, c)\,
\nonumber  \\
&& \qquad \qquad \qquad \qquad = \, \, \lambda \cdot 
W (K''_1, \,K''_2, \,\cdots K''_n; \,\, a, \, c)
% \nonumber 
\end{eqnarray}
Alternatively, one can build a 
$\, q \times q\, $ Boltzmann matrix 
$\, {\bf W}$ with entries $\,  {\bf W}_{i, \, j} \, = \,
W(K_1, \,K_2, \, \cdots K_n; \, i, \, j)  $.
In terms of such matrices the relation  (\ref{decim}) 
becomes  $\, {\bf W} \cdot {\bf W'} \, = \, \, {\bf W''}$.  
The decimation procedures and also the 
high-temperature expansions 
in such models are greatly simplified by introducing a
``transmissivity'' function $\, t_{\alpha}$ such that the  matrix
relation  $\, {\bf W} \cdot {\bf W'} \, = \, \, {\bf W''}$
becomes one or more multiplicative relations of the form:
\begin{eqnarray}
\label{transmit}
 t_{\alpha}({\bf W}) \cdot  t_{\alpha}({\bf W'})
 \, = \, \, \,  t_{\alpha}({\bf W''}), 
\qquad \qquad \alpha \, = \, \,
 1, \cdots , r. \nonumber 
\end{eqnarray}

The simplest example  is the transmissivity
variable for the $\, q$ state standard scalar Potts model,
for which one has $\, t \, = \, \, (e^{K}-1)/(e^{K}\, + \, q\, -1)$. This 
is  the natural expansion variable for the
high-temperature series  of the model (see also  the $\,
f_{i,\, j}$'s in (\ref{fij}) and (\ref{sumrule}) introduced in 
section (\ref{thereview}) below). For the Ising model this reduces to
the $ \tanh(K_{})\, $  variable.
F. Y. Wu et al.~\cite{w140} underlined the fact that 
two quite different situations must be considered.
If the family of Boltzmann matrices $\, {\bf W}$
is a set of {\em commuting} matrices, then
 they can be diagonalized 
simultaneously and the transmissivity
variables are nothing but all the possible ratios
of eigenvalues of the  Boltzmann 
matrices $\, {\bf W}$. If, alternatively, 
the Boltzmann matrices $\, {\bf W}\, $ do not commute, 
then one must perform a simultaneous block-diagonalization
of this family of  Boltzmann matrices,
and, therefore, some of the $\, t_{\alpha}$'s will be
block matrices
%, namely the corresponding blocks, 
from which one can extract
functions $\, \phi_{\alpha}\, $ satisfying $\, \phi_{\alpha}({\bf
W''})\, = \, \, \phi_{\alpha}({\bf W})
\cdot \phi_{\alpha}({\bf W'})$. One obvious 
choice for $\, \phi_{\alpha}\, $ is the
(ratio of)  determinants of these blocks.
A number of non-trivial non-commuting transmissivities
are given in~\cite{w140}.

\subsection{Three-spin interactions: the Baxter-Wu model}
\label{BaxterWu}
Another  important work in the history of  exact solutions of lattice 
statistics is the 
 Baxter-Wu model, which is
an Ising model on the triangular lattice with {\em three-spin}
interactions.  This model was  solved exactly by R. J. Baxter 
and F. Y. Wu in 1973~\cite{w42,w46}.

The three spin surrounding every triangular face
 interact with a {\em three-body}
interaction of strength $\, -J$, so that the Hamiltonian reads:
\begin{eqnarray}
H \, = \,\, -J \cdot \sum \, \sigma_i \, \sigma_j \, \sigma_k 
\end{eqnarray}
Baxter and Wu found that the
 per-site partition function  $\, Z$ has a remarkably simple 
expression:
\begin{eqnarray}
Z \, = \, \, \sqrt{6\,y\,t} \qquad \quad \hbox{with:}
 \qquad \quad t \, = \, \, \sinh(2\,|J|/kT)
\end{eqnarray}
and where $\, y$ is the solution of the algebraic equation:
\begin{eqnarray}
(y-1)^3\, (1+3\,y)\, \, (1+t^2) \cdot t \,\, =
 \,\,\,\, 2\, (1-t)^4 \cdot  y^3
\end{eqnarray}
The partition function has a singular part which behaves as $\, |t-1|^{4/3}$. 

Some  interesting duality properties
of the Baxter-Wu model are
very clearly detailed in~\cite{w42},
and used  to  convert the Baxter-Wu  model 
into  a {\em coloring} problem. This provides 
a very heuristic example showing that   duality
is {\em not} specific to edge-interaction spin  models, but can also be 
introduced with many-body interactions.
% (up to a multiplicative factor). 
In the following we briefly describe how the Baxter-Wu model is transformed
into a coloring problem.

First we introduce a $\, {\bf Z}_2$-Fourier transform
with function $\, g(\lambda, \, \mu )\, $  
which enables us to write simply the Kramers-Wannier duality 
for this three-spin model ($\lambda \, $ and 
$ \, \mu \, $ are Ising
spins) as: 
\begin{eqnarray}
\label{glambda}
&&g(\lambda, \, \mu )\,  = \, \, \, +1 \qquad \hbox{if:} \qquad
 \lambda\, = \, \, +1, \qquad \qquad  
%\hbox{and:} 
\qquad \nonumber \\
&&g(\lambda, \, \mu )\,  = \, \, \, \mu 
\qquad \quad \hbox{if:} \qquad
 \lambda\, = \, \, -1.
\end{eqnarray}
Note that this function is symmetric in 
$\, \lambda \, $ and $ \, \mu \, $, namely,
$\, g(\lambda, \, \mu )\,  = \, \, \, +1 \, $ when $\, 
 \mu \, = \, \, +1\, $ and $\, g(\lambda, \, \mu )\,  = \, \, \,
\lambda\, $ when $\, \mu \, = \, \, -1$.

Returning to the Baxter-Wu model,
each spin $\, \sigma_i\, $ of the triangular lattice
belongs to six triangles around vertex $\, i$,
which form a hexagon with the spin $\, \sigma_i\, $
at the center. 
Let us now consider the close-packing of such 
hexagons.
%  built from six triangles
% around  spins $\, \sigma_i\, $ at the 
% center of these  hexagons. 
The spins $\, \sigma_i\, $ now form
 a (triangular) sublattice of
the initial  triangular lattice.
% (there are three  such sublattices).

Consider next the spins  $\, \sigma_i\, $
%on one of the  sublattices,
 and denote the edge connecting 
nearest-neighboring spins $\, \sigma_k\, $ and $\, \sigma_l\, $
sitting on the hexagon surrounding $\, \sigma_i\, $ by
$\, <kl>\, $.
% denotes one of the six edges of the hexagon),
 Let us introduce
Ising edge variables  $\, \lambda_r$
corresponding to the six edges $\,<kl>\, $  of the hexagon: 
$\, \lambda_r \, = \, \, \sigma_k \cdot \sigma_l$,$\,\  r \, = \, 1, \,
\cdots 6$. The local Boltzmann weight of a hexagonal cell 
around a spin $\, \sigma_i\, $ 
can  be  written as:
\begin{eqnarray}
\label{whex}
W_{\rm hex} \, = \, \, \, {{1 } \over {2}} \cdot 
(1\, + \, \prod_{r \, = \,1}^{6} \, \lambda_r) 
  \cdot \exp\Bigl(K \cdot \sigma_i 
\cdot \sum_{r\, = \, 1}^{6} \, \lambda_r\Bigr) 
\end{eqnarray}
where the factor $\, (1\, + \, \prod\,\lambda_r)  $
  takes into account of the fact that
the Ising edge variables $\, \lambda_r\, $ are not independent but
constrained by the condition 
 $\,\prod \,\lambda_r \,  = \, \,  1 $.
As usual this condition, associated with every hexagon, can be written
by introducing a dummy variable $\, \mu_i$ also
associated with every hexagon:
\begin{eqnarray}
\sum_{\mu_i \, = \,\,  \pm 1} \, \prod_{r \, = \, 
1}^{6}  g(\lambda_r, \, \mu_i) \, \,  = 
\,  \, \,  1\, + \, \prod_{r \, = \,
1}^{6} \, \lambda_r\, , \nonumber 
\end{eqnarray}
enabling us to rewrite (\ref{whex}) as:
\begin{eqnarray}
W_{\rm hex} \,\, \,   = \, \,\,  \,  \sum_{\mu_i \, 
=  \, \pm 1} \, \prod_{r \, = \,
1}^{6} g(\lambda_r, \, \mu_i) \cdot \exp\Bigl(K \cdot \sigma_i 
\cdot \sum_{r\, = \, 1}^{6} \, \lambda_r\Bigr) \nonumber 
\end{eqnarray}
The partition function of the Baxter-Wu model
is now seen as a summation
over all the  (initial) spins $\, \sigma_i\, $ and the
(dummy) spins $\, \mu_i\, $
of a triangular sublattice, and the edge Ising spins $\, \lambda_r$.
Let us focus on one $\, \lambda_r$. The edge $\, r \, = \, <kl>\, $ 
belongs to a hexagon around spin $\, \sigma_i\, $ 
and a neighboring hexagon around another spin, say, $\, \sigma_j$. 
The edge Ising spin $\, \lambda_r$
thus occurs in the Boltzmann factors with a factor:
\begin{eqnarray}
W_{\lambda_r} \, = \, \, \,
 e^{K \cdot (\sigma_i\, + \sigma_j) \cdot \lambda_r}
 \cdot g(\lambda_r, \, \mu_i)
 \cdot g(\lambda_r, \, \mu_j). \nonumber  
\end{eqnarray}
Summing over the  edge Ising spin $\, \lambda_r$
in the partition function 
and using relations (\ref{glambda}),
one thus obtains a factor:
%\begin{eqnarray}
%&&\omega_{ij}\,\,  = \,\,  \sum_{\lambda_r \, = \, \pm 1}
%W_{\lambda_r} \, = \, \, \nonumber  \\
%&& \quad \qquad \qquad \, = \, \,  
%\exp(K\, (\sigma_i\, +\sigma_j)) \, + \, \mu_i\,
%\mu_j \cdot  \exp(-K\, (\sigma_i\, +\sigma_j)) \nonumber 
%\end{eqnarray}
\begin{equation}
\omega_{ij}\,\, = \,\,  \sum_{\lambda_r \, = \, \pm 1}
W_{\lambda_r} \, \, \, = \, \,  
e^{K\, (\sigma_i\, +\sigma_j)} \, + \, \mu_i\,
\mu_j \cdot  e^{-K\, (\sigma_i\, +\sigma_j)} \nonumber 
\end{equation}
between two spins $\, \sigma_i\,$ and $\, \sigma_j\,$ on
the sublattice.  This can be interpreted as the edge weight associated with
a coloring problem, and 
 the Baxter-Wu model is  transformed into a coloring problem.
  
\subsection{The Blume-Emery-Griffiths Model}
\label{Blume}
The Blume-Emery-Griffiths (BEG) model  is a model that 
F. Y. Wu and coworkers  considered 
quite naturally~\cite{w101,w111,w130,w141}, since
%for spin $\, 1$, it is very close to an Ising model
%and even 
it reduces  to an Ising model on the honeycomb
lattice in a special manifold~\cite{w101,w111}.
%(F. Y. Wu,  On  Horiguchi's Solution
% of the Blume-Emery-Griffiths Model, 
% {\it Phys. Lett.} {\bf 116A}, 245-247 (1986),
%and  X. N. Wu and  F. Y. Wu, The
% Blume-Emery-Griffiths Model on the honeycomb
%lattice,  {\it J. Stat. Phys.} {\bf  50}, 41-55 (1988)).

The BEG model 
is defined by the Hamiltonian:
\begin{eqnarray}
\label{BEG}
-\beta \cdot H \, = \, \, \,- J \cdot  \sum_{<i,j>}\, S_i \,
S_j
 - K \cdot  \sum_{<i,j>}\, S_i^2 \, S_j^2 
 - \Delta \sum_i\, S_i^2\, -\, H \cdot  \sum_{i} \; S_i
\end{eqnarray}
where the spins are classical spin-1 spins taking on the
 values $\, S_i \,= \, \ 0, \, \pm 1$.
In the high-temperature expansion the nearest-neighbor Boltzmann factor assumes
the form
\begin{eqnarray}
&&\exp(J\; S_i\;S_j \, + K \; S_i^2\;S_j^2) \,\, = \,\,\,\\ 
&&\qquad \qquad \quad  \, 1\, + \,(e^K\, \sinh J)\, S_i\;S_j \, 
+(e^K\, \cosh J\, -\, 1)\,
  S_i^2\, S_j^2. \nonumber 
\end{eqnarray}
It follows then in the subspace
$\, K \, = \, \, -\ln(\cosh J)$\,, one has the simple relation
%Consider the high-temperature expansion of the partition function
%for the honeycomb lattice.
%It is clear 
%that due to (\ref{identi}) many of the terms  in the expansion will
%vanish.  As a consequence the  high-temperature expansion reduces
%to that of a two-level (Ising) system corresponding to $\, n \, = \, 0\, $ and
%$\, n\, = \, 2$ in (\ref{identi}), and
%One thus immediately sees that %
%the partition function of the BEG model 
%reduces to that of an Ising model
%with the Ising high-temperature variable:
% $\, \tanh(K_1)\, $:
%More generally, one can recall that any system of classical
% $\,q$-state spins, the Potts model included, can be formulated 
%as a spin $\, (q-1)/2$ system.
 \begin{eqnarray}
\exp(J\; S_i\;S_j \, + K \; S_i^2\;S_j^2) \,\, = \,\,\, \, 1\, + \, S_i\;S_j
\; \tanh(J). \nonumber
\end{eqnarray}
and the partition function of
 the BEG model assumes the simpler form:
\begin{eqnarray}
Z_{BEG} \, = \, \, \sum_{S_i \, = \, 0,\; \pm 1}\; \prod_{<i,j>} ( 1\,
+ \, S_i\;S_j\, \tanh J) \;  \prod_{i} \; \exp(-\Delta\; S_i^2\, + \, H \; S_i).\nonumber
\end{eqnarray}
Expanding the products over neighboring pairs, 
 representing each term by a graph
and making use of  the identities
\begin{eqnarray}
\label{identi}
&& \sum_{S_i\, = \,\ 0, \, \pm 1} \, S_i^n \cdot e^{ - \Delta \, S_i^2}
\, = \, \, \, \rho(n)\qquad \quad \hbox {with:}
\qquad \rho(2)\, = \, \, 2 \, e^{ - \Delta} \, ,\qquad
\nonumber \\
&&\qquad \qquad \rho(0)\, = \, \, 2 \, e^{ - \Delta} \, +\, 1 \, ,\qquad 
\qquad \rho(1)\, = \, \, 0,
\end{eqnarray}
 one finds that 
one has eight possible configurations
 at each vertex of the honeycomb lattice, corresponding
%, for the isotropic model,
 to the
following  Boltzmann weights of an (isotropic) eight-vertex model:
\begin{eqnarray}
&&a \, = \, \, 1\, + \,2\; e^{-\Delta} \, \cosh(H), \quad \quad \quad \quad 
b \, = \, 2 \; \sqrt{\tanh(J)} \; e^{-\Delta} \, \sinh(H), 
\nonumber \\
&&c \, = \, 2 \;\tanh(J) \; e^{-\Delta} \, \cosh(H), \quad  \quad  
d \, = \, \, 2 \; (\tanh(J))^{3/2} \; e^{-\Delta} \, \sinh(H)\nonumber
\end{eqnarray}
With these notations one deduces the identity of the partition function of
 the BEG model (\ref{BEG}) with the eight-vertex model on a
 honeycomb lattice (see also sections (\ref{Critgauge}) 
and (\ref{HilbIsinghoney}) below):
\begin{eqnarray}
Z_{BEG} \, = \, \, Z_{8v}(a, \, b, \, c, \, d) \nonumber
\end{eqnarray}
Performing a weak-graph duality transformation 
on this eight-vertex model (see section (\ref{duali}) below)
associated with the $\, 2 \times 2\, $ (gauge) matrix~\cite{w45}:
\begin{eqnarray}
g_1 \, = \, \, g_2 \, = \, \, \,
 {{1} \over {\sqrt{2}}} \cdot 
\pmatrix{ 1 & y \cr -y & 1 }, \nonumber 
\end{eqnarray}
one finds that  the partition function of the eight-vertex model 
$\, Z_{8v}(a, \, b, \, c, \, d)\, $ 
remains invariant under a weak-graph duality transformation~\cite{w5,w111}:
\begin{eqnarray}
\label{weakgra}
&&\tilde{a} \, = \, \, (a\, +3\; y\; b\, +3\; y^2\;c\,
+y^3\;d)/(1+y^2)^{3/2} \, , \,\,\cdots \nonumber \\
&&Z_{8v}(a, \, b, \, c, \, d)\,\, = \, \,\,
 Z_{8v}(\tilde{a}, \, \tilde{b}, \, \tilde{c}, \, \tilde{d})
\end{eqnarray}
The four parameters $\, a, \, b, \, c, \, d\, $ or 
$\, \tilde{a}, \, \tilde{b}, \, \tilde{c}, \, \tilde{d}\, $ 
can be seen, as far as the calculation of the partition function
is concerned, as four {\em homogeneous} parameters. Taking into
account an irrelevant overall factor and 
the irrelevant gauge variable $\, y$ from the weak-graph symmetry
 (\ref{weakgra}), one sees that the partition function of the eight-vertex model 
$\, Z_{8v}(a, \, b, \, c, \, d)\, $ basically depends on
two variables instead of four. Not surprisingly,
Wu found that $\, Z_{8v}(a, \, b, \, c, \, d)\, $
is equivalent to the partition function
of an Ising model 
with nearest-neighbor interactions $K_I$ and a magnetic field $L$:
\begin{eqnarray}
\label{ZIZ8v}
Z_{\rm Ising} \,(L,\,K_I\,)&=&  \, 
\sum_{\sigma} \prod_{<ij>} \exp(K_{I}\; \sigma_i\;\sigma_j) \,
\prod_{i} \exp(L \; \sigma_i) \nonumber \\
\, &=& \,
 Z_{8v}(\tilde{a}, \, \tilde{b}, \, \tilde{c}, \, \tilde{d})\cdot
\Bigl({{2\; \cosh(L) \; \cosh^{3/2}(K_{I})  } \over
{\tilde{a} }}\Bigr)^N 
\end{eqnarray}
where $N$ is the number of lattice sites.
The explicit expressions of $K_I$ and $L$ in terms of the BEG parameters
are   complicated.  But for $H=0$, one has $L=0$
 and
\begin{eqnarray}
\tanh(K_{I}) \, = \, \, 
\, {{2 } \over {2\, + \, e^{\Delta}}} \cdot \tanh(J), \nonumber 
\end{eqnarray}
using which one determines the critical line $K_I=1/\sqrt 3$
in the $J>0$ regime.
The spontaneous magnetization of
 the BEG model for $\, J\, > \, 0\, $ 
and the phase boundary 
of the $\, J\, < \, 0\, $ BEG model
can be similarly determined~\cite{w111}.

The proof of the equivalence of the honeycomb
eight-vertex model with an Ising model in a field as outlined in the above
is quite tedious.  However, a more direct derivation has since been   given
by Wu~\cite{w124}.
 
The eight-vertex model on the honeycomb
lattice  can also be seen to be related to 
 a  lattice-gas grand partition function $\, \Xi^{Kag}\, (z,\, J, \, J_3)$
on the {\em Kagom\'e lattice}~\cite{w120} (see also section (\ref{invduakago}) below),
%The Kagom\'e lattice is the covering lattice of the honeycomb 
%lattice whose sites reside in the triangular faces
%of the Kagom\'e lattice.
where  $\, -J\, $ is the nearest-neighbor 
interaction, $\, -J_3$ the triplet interactions 
 existing among three sites surrounding
a triangular face of the lattice, and  $\, z$ denotes the
fugacity. 
%and the
%corresponding lattice-gas grand partition 
%function $\, \Xi^{Kag}(z,\, J, \, J_3)$ where 
%Regarding a honeycomb 
%lattice edge as being covered by a bond if, and only if, 
%the Kagom\'e lattice site residing on it is occupied
%by a lattice-gas molecule, the  Kagom\'e lattice gas then describes 
%{\em precisely an eight-vertex model on a honeycomb lattice}.
Then one has the equivalence:
\begin{eqnarray}
\label{Xsi}
&&\Xi^{Kag}(z,\, J, \, J_3) \, \, = \, \, \,  Z_{8v}(a, \, b, \, c, \, d) \\
&&a \, = \, \, 1, \quad \quad b\, = \,\, \sqrt{z}, \, \quad \quad 
c \, = \, \, z \cdot e^{J}, \quad\quad 
d \, = \, \,  z^{3/2}\cdot  e^{3\;J\, +J_3}. \nonumber 
\end{eqnarray}
From (\ref{ZIZ8v}) and (\ref{Xsi}), F. Y. Wu and X. N. Wu 
were able to obtain results for the liquid and vapor densities,
showing that an observed anomalous critical behavior
 occurs in the lattice gas only when
 there are nonzero triplet interactions~\cite{w120}.
This analysis has been 
extended to a  lattice gas on the $\, 3-12$ lattice
by J. L. Ting, S. C. Lin and
 F. Y. Wu~\cite{w133}.

Wu's   tricks for the
honeycomb BEG model are not limited to the weak-graph transformation
for the eight-vertex model.  Using a syzygy analysis 
of the invariants under the
O(3) transformation L. H. Gwa and F. Y. Wu have 
obtained an expression for  the critical
variety of the honeycomb BEG model to
an extremely high degree of accuracy~\cite{w141} 
(see section (\ref{HilbO3}) below).

\subsection{Other spin results: disorder points}
\label{disordpoints} 

Let us finally describe, among many results
obtained by F. Y. Wu on spin models, 
one result concerning  disorder points.
Disorder solutions are particularly simple solutions corresponding
to some ``dimensional reduction'' of the model, which
 provides simple exact results for
 models which are generically
quite involved. While this yields severe constraints 
on the phase diagrams, the series expansion and the analyticity
properties of the model,    it does lead to exact solutions
of models which are otherwise nonintegrable.

For example, 
using a decimation approach, Wu~\cite{w95} has deduced the disorder
solution for the triangular Ising model in a nonzero magnetic field. 
 Wu and K. Y. Lin~\cite{w115} have used  a checkerboard Ising 
lattice   to
illustrate that there may exist more
than one disorder points in a given
spin system.
%(Spin model
% exhibiting multiple disorder points, 
%{\it Phys. Lett.}  {\bf 130A},   335-337 (1988)). 
Along the same vine
 N. C. Chao and   Wu~\cite{w96} have explored the validity of the decimation
approach by considering disorder solutions 
   of a general checkerboard Ising model in a field.

\section{Weak-graph dualities and Hilbert's syzygies}
\label{duali}
 
In a pioneering paper F. Y. Wu and
 Y. K. Wang~\cite{w54} introduced a duality transformation 
  for a general spin model which can have chiral interactions.
 This is the first time that chiral spin model has been explicitly considered.
In terms of $\, R$ matrices such as
(\ref{8v}) these transformations are the tensor product of two
similarities:
\begin{eqnarray}
\label{tensor}
R \,\qquad  \longrightarrow \, \qquad  
g_1 \otimes g_2 \cdot R \,\cdot\, g_1^{-1} \otimes g_2^{-1} 
\end{eqnarray}
where $\, g_1\, $ and $\, g_2\, $ are 
two $\, q \times q\, $ matrices,  $\, R$ is a $\, q^2 \times q^2$
matrix
($\, q\, = \, 2$ for the sixteen-vertex model). This symmetry group
is an $\, {\rm sl}(q) \times {\rm sl}(q)$  symmetry group.
The high- and low-temperature duality (\ref{weakgraph})
given  in section (\ref{Liebmonograph}) for the sixteen-vertex model
 is a particular case
of such transformations,   corresponding to $\, g_1 \, $ 
and  $\, g_2 \, $ being two involutions:
\begin{eqnarray}
g_1 \, = \, \, g_2 \, = \, \, \,
 {{1} \over {\sqrt{2}}} \cdot 
\pmatrix{ 1 & i \cr -i & 1 }. \nonumber 
\end{eqnarray}
This  duality relation~\cite{w54,w125}
%(see also subsection (\ref{duali}) and
is now well-known for vertex models.
It corresponds to symmetries of the model 
and can be used, as will be seen in the next two subsections,
to find good variables to
express the critical manifolds of a lattice model, and, hopefully,
to determine their exact expressions  when algebraic.

\subsection{Hilbert's syzygies, gauge-like 
dualities and critical manifolds}
\label{Critgauge}

Let us give some hints as to 
how the gauge-like dualities enable us to deduce
results  on
critical manifolds or varieties.
The main idea is to construct
algebraic invariants under these 
gauge transformations.
% (see for instance~\cite{w125}).
% J. H. H. Perk, F. Y. Wu and X. N. Wu, Algebraic
% invariants of the symmetric
% gauge transformation, {\it J. Phys.} {\bf  A23}, L131-L135 (1990)).

Hilbert has shown that all invariants of a linear transformation are
algebraic and can be expressed in terms of a set of homogeneous
polynomials, the  {\em syzygies}.
 Considering the sixteen-vertex model, the transformation
is O(2) and the fundamental invariants
corresponding to the 
%$\, O(2)\, $ 
O(2) group
have been constructed by
J. H. H. Perk, F. Y. Wu and X. N. Wu 
in~\cite{w125}. Likewise for 3-state vertex models
the transformation is O(3) and the associated invariants
have been constructed by L. H. Gwa and
F. Y. Wu~\cite{w139} (see section (\ref{HilbO3}) below).  
 
%Very little is 
%known on the behavior
%of this general model which encompasses 
%so many lattice models: eight-vertex models, 
%Ising models in nonzero magnetic field, etc.
%A very preliminary and qualitative
%study of  the phase diagram 
%of the sixteen-vertex model 
%have been performed by F. Y. Wu in~\cite{w31}. 
 
\subsection{Hilbert's syzygies and 
the square lattice Ising model in a magnetic
 field}
\label{HilbIsingsquare}
With an algebraic prejudice for critical manifolds, it is very
tempting to conjecture closed algebraic formula
for critical manifolds that will be reproduce known
exact results in various limits.
For instance, closed-form expressions for the critical line
 of the square lattice antiferromagnetic Ising model in a magnetic
 field were proposed\footnote{This is just one example 
in a very long list of wrong algebraic conjectures
for critical manifolds one can find in the literature.} by 
M\"uller-Hartmann and Zittartz. However, 
it has  been shown that the expression is numerically incorrect.

X. N. Wu and F. Y. Wu~\cite{w129}  considered
the square lattice antiferromagnetic Ising model in a magnetic
 field which can be seen as a subcase of the sixteen-vertex model
under the O(2) group.
Introducing the variables:
\begin{eqnarray}
\label{Isingfield}
&&a\, = \, 1, \quad \quad b\, = \, \sqrt{v}\, h, \quad \quad
c\, = \, v, \quad \quad d\, = \, v^{3/2}\, h, 
\quad\quad  e\, = \, v^2 \qquad
\qquad \nonumber \\
&& \quad \hbox{with:} \qquad \qquad
v \, = \, \tanh(J/kT), \qquad \quad  h\, = \,  \tanh(H/kT), \nonumber 
\end{eqnarray}
the five fundamental Hilbert's invariants of O(2) read:
\begin{eqnarray}
\label{I1I4}
&&I_1 \, = \,\,  a\, +2\;c\, +e, \qquad \quad  I_2 \, = \,\,
(a-6\;c+e)^2+16\;(b-d)^2, \qquad \nonumber  \\
&&I_3 \, = \,\, (a-e)^2\,+4\;(b+d)^2, \qquad 
 I_5 \, = \, \,a^2\;d\, -b\;e^2\, -3\;(a-e)\,(b+d)\;c \nonumber\\
&&I_4 \, = \,\, (a-6\;c+e)\cdot ((a-e)^2-4\;(b+d)^2) \, +\,
4\;(a-e)\;(b^2-d^2) \quad\nonumber
\end{eqnarray}
and the critical line proposed by Wu and 
Wu assumes the form: 
\begin{eqnarray}
\label{c1c2c6}
c_1 \cdot I_1^4 \, c_2 \cdot I_1^2\, I_2 \, + \, c_3 \cdot I_1^2\, I_3
+ \, c_4 \cdot I_2^2 \, + \, c_5\cdot I_3^2 \, + c_6 \cdot I_2\, I_3
\, = \, \, 0.\nonumber 
\end{eqnarray}
%From the known zero-field critical point one gets 
%the constraints:
%\begin{eqnarray}
%&&c_6 \, = \, {{1} \over {49}} \cdot \Bigl((234 \sqrt{2} -331)\, c_1
%\, + (98 \sqrt{2}-147) \cdot c_2 \, +(40 \sqrt{2}-57)\cdot c_3 \nonumber \\
%&& \qquad \qquad -(294 \sqrt{2}+539) \cdot c_4\, +(6 \sqrt{2}-11) \cdot c_5 
%\Bigr).
%\end{eqnarray}
They then determined the $\, c_i$'s using the various known
results including a constraint dictated by the known
zero-field critical point, as well as the results of finite-size analyses
which they carried out.  This leads to the values $c_1=1$, $\, c_2=\, -0.044\ 338$,
$c_3=\, 0.362\ 73$, $\, c_4=\, 0.000\ 4938$, 
$\, c_5=\, 0.042\ 779$,
 $\, c_6=\, -0.008\ 9149\, $
The resulting closed form expression
for the critical line 
%of the square-lattice 
% antiferromagnetic Ising model 
reproduces all  known numerical data to a high
degree of accuracy. For instance, 
%with the $\, c_i$'s deduced 
%from a finite-size analysis,
the critical line yields for small magnetic field: $\, T_c \, \simeq
\, \, T_0 \cdot (1\, - \, u \cdot (H/J)^2)\, $ with $\, u \, \simeq \,
0.038\ 022$.  This  is compared to
 the presumably exact value obtained by M. Kaufmann:  $\, u \, \simeq 
0.038\ 0123\ 259 \, \cdots $ 

\subsection{Hilbert's syzygies and the honeycomb lattice
Ising model in a magnetic
 field}
\label{HilbIsinghoney}
Similar  analyses have been carried out 
for  the 
honeycomb lattice by F. Y. Wu, X. N. Wu and H. W. J. Bl\"ote~\cite{w126}.
For the corresponding honeycomb eight-vertex model we have:
\begin{eqnarray}
&&a \, = \, \, 1, \qquad\quad  b \, = \, \, \sqrt{v} \, h , \qquad \quad 
c \, = \, \, v, \qquad \quad d \, = \, v^{3/2} \, h \nonumber \\
&& \quad \hbox{with:}\quad  \qquad \quad  v \, = \, \tanh(J/kT), \quad
\qquad h\, = \,
\tanh(H/kT). \nonumber 
\end{eqnarray}
%The partition function of the eight-vertex model
%on the honeycomb lattice becomes the monomer-dimer
%generating function in the $\, c \, = \, $
%$\, d \, = \, \, 0$
%limit~\cite{w45}. 
Analogous to (\ref{I1I4}), we introduce the following Hilbert's
syzygies:
\begin{eqnarray}
\label{PQP2}
&&P \, = \, \, a^2\, + \, 3 \, a\;c\, + 3\; b\;d\, +d^2, \qquad \quad \quad 
Q \, = \, \, b^2\, - a\;c\, +\; c^2\, - b\; d  \quad   \nonumber   \\
&&P_2 \, = \, 2\, (a^4+d^4)\, -6\; (a^2\;c^2+b^2\;d^2)\, + 12 \;
(a^2\;b^2+c^2\;d^2)
-5 a^2\;d^2  \,\quad  \nonumber \\
&&\quad  \qquad  \qquad \qquad +\, 27\,b^2\;c^2 
+ 36 \, (a\,b\, +c\, d) \, b\, c \, +18\, a\, b\, c\, d. \nonumber 
\end{eqnarray}
The critical line proposed by Wu, Wu and Bl\"ote now
reads~\cite{w119}: 
\begin{eqnarray}
\label{c1P2}
c_1\cdot P_2 \, + c_2\cdot P^2\, + c_3 \cdot P\,Q \, + \, c_4 \cdot
Q^2 \, \,= \, \,\,\,  0. 
\end{eqnarray}
 After an extensive search by mapping with all known exact results, they
proposed the numbers: $\, c_1\, = \, 1, \, $ $\, c_2\, = \, -(4\, + \, 3\;
\sqrt{3})/6, \,  $  $\, c_3\, = \, -(1\, - \, 9\;
\sqrt{3})/8, \,  $ and $\, c_4 \, = \, \,-\, 3\, (3\, - \;
\sqrt{3})/8$. 

The initial slope of this critical frontier for small
$H$ is   $\, -\ln(z_c)\, $ where  $\, z_c \, $
is the critical fugacity of the nearest-neighbor
 exclusion gas.  Their expression leads to the  value
% now agress with plotted in $\, (H, \, T)\,
%$ reads With the previous values of the $\, c_i$'s 
%one gets a closed expression for $\, z_c \, $ 
$\,z_c \, \simeq \,
7.851\ 780\ 04 \, \cdots\,$
which is in very good agreement with the value 
obtained from finite-size analysis,
  namely,  $z_c \, \simeq \,
7.851\ 725\ 175(13)$. The critical line (\ref{c1P2}) is probably not the
exact one but certainly a very accurate approximation.

{\em Comment}: The Hilbert's invariant theory amounts to
considering {\em linear} gauge-like symmetries
of the model and the associated invariants.
% associated with these linear transformations. 
 From the inversion relations
one has further an infinite discrete set of  birational
{\em non-linear}  symmetries, one can couple with these
continuous linear groups. 
In fact all the above analyses can be revisited by combining the gauge
transformation with the {\em infinite discrete symmetries}
generated by the {\em inversion relations} of the 
sixteen- (or simply eight)-vertex models. This
 would lead us to consider 
a unique ``superinvariant'' which
 is, in fact, the {\em modular invariant}
of the elliptic curves parametrizing the sixteen-vertex model.

\subsection{Hilbert's syzygies and the honeycomb BEG model} 
\label{HilbO3}

To apply the syzygy  consideration to the BEG model
which is a 3-state spin model, one needs to consider the O(3)
gauge transformation and its associated invariants.
But the construction of the O(3) invariants is very complicated.
However, the day is saved since there exists a mapping between O(3) and
sl(2), and invariants for the latter have been worked out by
mathematicians a long time ago.  
While the  mathematics to decipher the old results
 is  involved,
L. H. Gwa and F. Y. Wu~\cite{w139} have succeeded in carrying out such an  analysis
and deduced that there are 5 independent invariants for the O(3) group.
They next applied the analysis to the isotropic BEG model on the honeycomb 
lattice~\cite{w141},
and found one of the invariants to vanish identically.
The remaining 4 invariants are then used to determine the
critical variety of the BEG model as in the case of the O(2) gauge.  The
 resulting closed-form 
expression  for the critical variety agrees extremely well with 
numbers obtained from  a finite-size analysis, which they also carried out.
 
\section{Critical manifolds and critical varieties}
\label{crit}

A problem solver like F. Y. Wu 
first tries to find the exact solution of a problem. 
He tries to ``dig out''  problems that can be solved.
However since most of the problems one looks at cannot
 be solved exactly,  one then tries to study  models for which some exact 
results can be ``salvaged". It could be the
critical manifolds which are submanifolds 
along which the models are Yang-Baxter integrable.
In such cases  the 
critical manifolds are, in fact, 
critical {\em varieties}. For other models the critical 
manifolds are {\em algebraic} 
varieties without hidden Yang-Baxter integrability~\cite{w74,w153}. 

For two-dimensional lattice models the situation
is more specific: one can have 
some ``conformal prejudice'' that 
 critical manifolds should be  submanifolds
where the model has a two-dimensional conformal 
(infinite) symmetry yielding some integrability in the 
scaling limit. Therefore, as far
 as critical manifolds are concerned, 
it is crucial to understand the inter-relation between 
1) algebraicity consequence of Yang-Baxter integrability,
2) conformal integrability consequence 
of a two dimensional criticality, and 
3) self-duality.

Many criticality conditions have been obtained, or simply
conjectured, in the literature of lattice models 
in statistical mechanics~\cite{w44,w61,w67,w68,w69,w74,w75},  and all
 these conjectures were {\em algebraic}~\cite{w119}. A straightforward 
situation corresponds to the case where
the model possesses a duality 
symmetry (see section (\ref{duali})) for which
it is always possible to give a {\em linear} representation
of this duality transformation. One can sometimes find 
varieties 
%(hyperplanes)
 which are 
globally invariant under  this symmetry. 

Let us, instead, consider the
 fixed points of the linear duality
transformation, which  belong to 
some algebraic variety
 (hyperplane). If the algebraic variety
separates the phase diagram into two disconnected parts,
and {\em if one assumes that the critical temperature is unique},
one can actually deduce that this algebraic variety
is a critical variety~\cite{w74}. Of course if the algebraic
variety is only globally invariant (and 
not invariant {\em point-by-point} on
the algebraic variety) one cannot draw any conclusion.
%on the localization of the critical (or transition) points.

In fact, for most of the time one is not 
in   a  situation
where a simple self-dual argument allows the determination
of the critical 
%(or transition)
 points.  A good example is the duality transformation (\ref{tensor})
for the   sixteen-vertex model for which  
self-dual arguments 
%for this very general
%model (16 homogeneous parameters)
 are insufficient.
In fact there exists a ``super-invariant'' on this model after
taking into account the gauge (weak-graph) duality 
symmetries
%~\cite{w54} 
(\ref{tensor}) and  the 
%(non-linear) 
inversion relation
symmetries\footnote{Remarkably this consideration even extends the 
$\, sl(2) \otimes sl(2)\, $ weak-graph symmetry group to a 
$\, sl(2) \otimes sl(2)\otimes sl(2)\otimes sl(2)\, $ 
symmetry group.}. But that is another story.

 When the critical varieties are exact, {\em they are
almost always related to some integrability of the model,
 the algebraicity thus being a consequence of
 the integrability}. A paradigm is the standard scalar Potts
model (see section (\ref{thereview})) which
 is integrable at criticality.
% There exist no 
%such algebraic critical varieties for regular
%\footnote{Euclidian 
%lattices, square, triangular, checkerboard, $\, 3-12$ lattice,
%Kagom\'e ... Of course this is no 
%longer true for instance for Bethe lattices, or self-similar lattices.}  
%lattices that are not related to integrability. 
However, we will give below, in the case of the two- and 
three-site interaction Potts model~\cite{w153}, 
an algebraic variety which is the critical
 condition~\cite{w119} but is {\em unrelated} 
to any simple Yang-Baxter-like integrability.

Throughout the years
F. Y. Wu has obtained numerous results of 
critical varieties for two- and three-dimensional
spin models, and many related conjectures as well. 
%All these
% results are useful to better understand 
%the previously mentioned interrelation
%between criticality, algebraicity and/or integrability 
%and/or self-duality.
One can only say that the  seeking 
of critical manifolds and critical varieties
is  a fascinating subject matter by itself for specialists  like  Wu.

%For instance, let us consider a quite non-trivial example
%namely the $\, q$-state Potts models 
%on the triangular lattice with three-spin 
%interaction on the up-pointing triangles~\cite{w153}. A critical
% variety was found by F. Y. Wu on this 
%model based on self-dual arguments~\cite{w74}. 

\subsection{Inversion relations, duality and critical varieties}
\label{invduacrit}

The considerations of 
{\em inversion relations} has been shown to be a
powerful tool for analyzing the phase diagram
 of lattice models and,  particularly, for obtaining 
 critical algebraic manifolds 
in the form of algebraic varieties 
(see (\ref{multiplicativePotts}) in 
section (\ref{commentchecker})). 

Let us consider the standard scalar $\, q$-state 
 Potts model on an anisotropic 
triangular lattice with nearest-neighbor  and three-spin interactions
around up-pointing triangles~\cite{w74,w153} as shown:
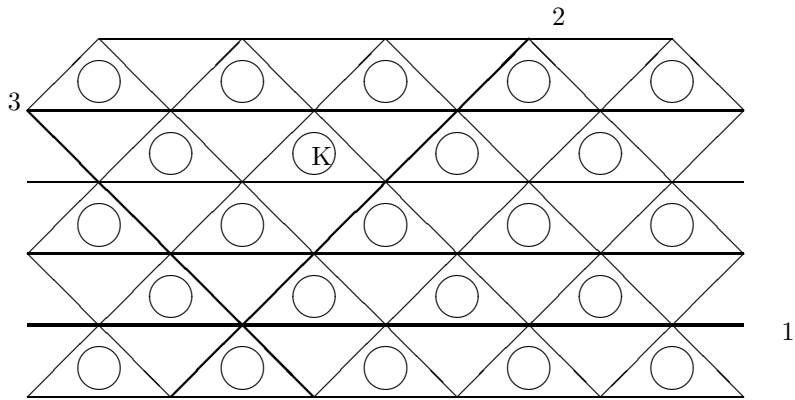
\[
\label{las}
%\bigskip
 \hskip 4cm
\null\hfill
\setlength{\unitlength}{0.0150in}
\begin{picture}(200,200)(72,600)

\thinlines

\put(25,775){\line( 1,0){200}}
\put(0,750){\line( 1,0){250}}
\put(0,725){\line( 1,0){250}}
\put(0,700){\line( 1,0){250}}
\put(0,675){\line( 1,0){250}}
\put(0,650){\line( 1,0){250}}

\multiput(25,775)(50,0){3}{\line(1,-1){125}}
\multiput(0,650)(50,0){3}{\line(1,1){125}}

\put(0,750){\line(1,-1){100}}
\put(0,700){\line(1,-1){50}}

\put(0,700){\line(1,1){75}}
\put(0,750){\line(1,1){25}}

\put(250,750){\line(-1,-1){100}}
\put(250,700){\line(-1,-1){50}}

\put(250,700){\line(-1,1){75}}
\put(250,750){\line(-1,1){25}}

\multiput(25,760)(50,0){5}{\circle{16}}
\multiput(50,735)(50,0){4}{\circle{16}}
\multiput(25,710)(50,0){5}{\circle{16}}
\multiput(50,685)(50,0){4}{\circle{16}}
\multiput(25,660)(50,0){5}{\circle{16}}
\thicklines
\put(0,675){\line( 1,0){250}}
\put(50,650){\line(1,1){125}}
\put(0,750){\line(1,-1){100}}
\put(260,670){\makebox(0,0)[lb]{\raisebox{0pt}[0pt][0pt]{ 1}}}
\put(180,780){\makebox(0,0)[lb]{\raisebox{0pt}[0pt][0pt]{ 2}}}
\put(-10,750){\makebox(0,0)[lb]{\raisebox{0pt}[0pt][0pt]{ 3}}}
\put(96,731){\makebox(0,0)[lb]{\raisebox{0pt}[0pt][0pt]{ K}}}
\end{picture}
\hfill\null
\]
The partition function of the models reads:
\begin{eqnarray}
\label{funcpart}
{\cal Z}\,  \, =\, \, \sum_{\{\sigma_{i}\}} 
\prod_{<i,j>} e^{K_{1}\;\delta_{\sigma_{i},\sigma_{j}}}
\;\prod_{<j,k>} e^{K_{2}\;\delta_{\sigma_{j},\sigma_{k}}}
\;\prod_{<k,l>} e^{K_{3}\;\delta_{\sigma_{k},\sigma_{l}}}
\;\prod_{\Delta} e^{K\;\delta_{\sigma_{i},\sigma_{j}}
 \delta_{\sigma_{j},\sigma_{k}}}. \nonumber 
\end{eqnarray}
Here the summation is taken over all spin configurations,
the first three products denote  edge Boltzmann weights 
%along the three directions of the
%triangular model, 
and the 
last product is over all up-pointing 
triangles.
% of the three-site interaction Boltzmann weights.

A {\em duality}  transformation 
exists for this model~\cite{w74}. We introduce the following notation:
\begin{eqnarray}
\label{xy}
x &=& e^{K},  \qquad x_{i} \, = \, e^{K_{i}}, \quad \quad i= \,1,2,3  \nonumber \\
y &=&\,  x\;x_{1}\;x_{2}\;x_{3}\, -(x_{1}+x_{2}+x_{3})\, +2   
\end{eqnarray}
With the notation (\ref{xy}) the duality
%, denoted $\, D$, 
reads: 
\begin{eqnarray}
\label{dual}
D: \quad
\left\{
\begin{array}{lll}
x_{i} & \longrightarrow & {\displaystyle x_{i}^{*}=1+q\;
\frac{x_{i}-1}{y}}\, , \qquad 
y \,  \longrightarrow \,  {\displaystyle y^{*}=\frac{q^{2}}{y}} 
\\
x & \longrightarrow & {\displaystyle x^{*}= 
\frac{x_{1}^{ }+x_{2}^{ }+x_{3}^{ }
-2+q^{2}/y}{x_{1}^{}\;x_{2}^{}\;x_{3}^{}}} 
\end{array} \right.
\end{eqnarray}
and the partition transforms as
%As far as the partition function per site is concerned:
\begin{eqnarray}
\label{Zdefi}
Z(x_1, x_2, x_3, y) \, \, = \, 
%\, \, \, \, \sum\prod_\Delta e^{-E_{abc}}
%\, \, \sim \, 
(y/q)^N \cdot Z(x_1^*, x_2^*, x_3^*, y^*) 
\end{eqnarray}
where $N$ is the number of sites.

On the basis of this duality Baxter {\em et al.} proposed that the critical
points are located on the {\em algebraic variety}:
\begin{eqnarray}
\label{critBax}
x\;x_{1}\;x_{2}\;x_{3}\, -(x_{1}+x_{2}+x_{3})\, +2 -q 
\,\,  = \, \,  \, 0 
\end{eqnarray}
 which corresponds to the  set of fixed points of $D$.
%the duality (\ref{dual}). 
The critical variety (\ref{critBax}) is not only globally
invariant under (\ref{dual}),  it is 
also {\em point-by-point} invariant, 
namely, every point on
the  variety is invariant. In general when an algebraic
variety is such that every point of the variety is invariant under a
duality symmetry, 
%of the partition function 
it is possible to argue,
subject to some continuity and uniqueness arguments, that the variety
actually corresponds to the criticality variety.
This has been done by Wu and Zia~\cite{w119} 
 for $\, q\, >\, 4\, $ in the ferromagnetic region.
It is important to note that the critical variety (\ref{critBax})
is {\em  not} an algebraic variety on which the model 
becomes {\em Yang-Baxter (star-triangle) integrable}. 
%It is 
% an algebraic variety for which one can actually perform a
%star-triangle transformation for the Potts model 
%for each up-pointing triangle supporting a three-spin interaction,
%thus mapping this two and three site interaction Potts model
%onto a {\em generically non-integrable}
%anisotropic honeycomb standard scalar Potts model.
 This is an
 interesting example of a model where algebraic 
criticality does not
 automatically imply Yang-Baxter integrability.

\vskip .3cm
{\em Comment:} In suitable variables the duality
transformations can be seen as a {\em linear}
 transformation. There are two globally invariant
hyperplanes  under $D$: $\, y \, = +q$ and
$\, y\, = \, -q$. The (ferromagnetic) criticality variety
(\ref{critBax}) corresponds to $\, y \, = +q$. The second hyperplane
$\, y\, = \, -q$ is not a point-by-point invariant 
although it is globally self-dual.   
It is not a locus for critical or transition points.

This illustrates a fundamental question 
one frequently encounters when trying to analyze a lattice model:
 is the critical manifold
an algebraic variety or a transcendental manifold?
It will be seen that a first-order transition 
manifold exists for this model for $\, q \ = \, 3$, and its 
algebraic or  transcendental
status is far from being clear (see~\cite{w158} and 
 (\ref{martin}) in section (\ref{MonteCarlo2})). The
 existence of such very large (nonlinear)
group of (birational) symmetries provides drastic
constraints on the critical manifold and therefore 
the phase diagram.

\vskip .3cm

There exist three inversion relations associated with the three
directions of the triangular lattice
 exist for this model~\cite{w153}.
For instance, the inversion relation which singles out 
direction $\, 1$ (see figure 1) is the
(involutive) rational transformation $\, I_1$: 
\begin{eqnarray}
\label{defI1}
&&I_1:\;\;\; (x, \, x_1, \, x_2, \, x_3) \, \, \longrightarrow \, \, 
\Bigl(
{{ (x\,x_{1}-1)^{2}\,
(x_1\,+q-2)}\over{ (x\,x_{1}^{2}+x\,x_{1}\,(q-3)-q+2\
)\, ( x_{1}-1 )}},\quad \quad \quad \qquad  \,\nonumber \\
&& \quad  \quad \quad 2-q-x_1+{{x_1\,(x-1)} \over
{x_1\,x-1}}, \, \,{\frac { x_{1}-1}{
x_{3}\,(x\,x_{1}-1 )}}, \,\, {\frac { x_{1}-1}{
x_{2}\,(x\,x_{1}-1)}}
\Bigr). 
\end{eqnarray}
These three inversion relations generate a group of symmetries
 which is naturally represented in terms of birational
 transformations in a four dimensional space.
This infinite discrete group of {\em birational} symmetries
 is generically a {\em very large
 one} (as large as a free group). 
The algebraic variety (\ref{critBax})
is remarkable from an algebraic geometry viewpoint:
 it is invariant
under this very large group generated
 by  three involutions (\ref{defI1}).

In this framework of a very large group of symmetries of the model,
 an amazing situation arises: the one for which $\, q$, the number of
states of the Potts model, corresponds to
 {\em Tutte-Beraha numbers} $ \, q \, = \, \, 2 \, +2 \, \cos(2\pi /N)$
where $N$ is an integer.
For these selected numbers of $q$ the group of 
birational transformations is generated by generators 
of finite order: it is seen as a Coxeter group generated by generators and
 relations between the generators. The elements of the group
can be seen as the words one can build from an alphabet of three letters
 $\, A$, $\, B$ and $\, C$ with the constraints $\, A^{N+1}\, = \,
 A$, $\, B^{N+1}\, = \, B$, $\, C^{N+1}\, = \, C$. Since the
 generators $\, A$, $\, B$ and $\, C$ do not commute (and nor do any
 power of $\, A$, $\, B$ and $\, C$) the number of words
of length $\, L$ still grows exponentially 
with $\, L$ (hyperbolic group). 
Among these values of $\, q$, 
%($q \, = \, 2 \, +2 \, \cos(2\pi/N)$), 
two Tutte-Beraha numbers
play a special role: $\, q\, = \, 1$ and $\, q\, = 3$. For
 these two values the
 hyperbolic Coxeter group 
degenerates\footnote{Up to semi-direct products 
by finite groups.} into a  group isomorphic
% Up to semi-direct products by finite groups. 
to $\,\bf{ Z} \times  \bf{Z}$.

For the standard scalar nearest-neighbor Potts model
the Tutte-Beraha numbers correspond to the values of $q$ for which the 
critical exponents of the model are {\em rational} (see (\ref{tutte})
in section (\ref{thereview})). 
%The  Tutte-Beraha numbers 
%$\, q \, = \, \,  2 \, +2 \, cos(2\pi/N)\, $
%actually correspond to values of  $\, u \;\pi$ in (\ref{tutte}) 
%(the shift of the rational spectral parameter
%parametrizing the standard scalar Potts model) ##
%such that  $\, u$ becomes a {\em rational number}.     

\subsection{The exact critical frontier of the Potts model on the 
$\, 3-12$ lattice}
\label{312}
F. Y. Wu {\em et al.} considered a general $\, 3-12$ lattice with two
{\em and three-site} interaction on the triangular cells~\cite{w148}. This  model
 has eleven coupling constants
and includes  the Kagom\'e lattice as a
special case.

In a special parameter subspace of the model, condition 
(\ref{crit312}) below,
an exact critical frontier for this Potts model
on a general $\, 3-12$ lattice  Potts model 
 was determined. The  Kagom\'e lattice limit is unfortunately
not compatible with the required condition 
(\ref{crit312}).

The condition under which they obtained the 
exact critical frontier 
%for this general $\, 3-12$ lattice  Potts model 
reads:
\begin{eqnarray}
\label{crit312}
&& x^2 \, x_1^2 \, x_2^2 \, x_3^2 \,
 - \, x \, x_1 \, x_2 \, x_3 \cdot 
(x_1 \, x_2\, + x_2\, x_3 + \,x_1\, x_3 - \, 1) \,  \\
&&\quad + \, 
( x_1 \,+ x_2 \,+ x_3 \,+ \, q - \, 4)\cdot 
(x_1 \, x_2\, + x_2\, x_3 + \,x_1\, x_3\,+\, 3\, - \, q)\, \nonumber \\
&&\quad  -\,q \,x_1 \, x_2 \, x_3 \, -\,
( x_1^2 \,+ x_2^2 \,+ x_3^2)\, 
 + \, q^2\, -6\; q \, +10\,\,\, = \, \,\, \,\, 0 \nonumber 
\end{eqnarray}
This is nothing but the condition corresponding to 
the star-triangle relation of the Potts model.
%Summing over the configuration of the spin at the center
%of the star, one gets a triangle with 
%nearest neighbor interactions $\, K_1$, $\, K_2$, $\,
%K_3$ and three spin interactions  $\, M$, but of course 
%since the anisotropic star depends on three couplings,
%the four variables $\, x_1$, $\, x_2$, $\,
%x_3$ and  $\, x$ are related: relation (\ref{crit312})
%is actually this relation. 

\vskip .3cm
{\em Comment:} One can show that
condition (\ref{crit312}) {\em is actually invariant under
the inversion relation} (\ref{defI1})
of the previous section (\ref{invduacrit}), and therefore since 
 (\ref{crit312}) is symmetric under the permutations
of  $\, K_1$, $\, K_2$ and $\,
K_3$ under the three inversions generating the very large
group of birational transformations previously 
mentioned in section (\ref{invduacrit}).
More generally, introducing   $\, D_1$, $\, D_2$ and $\, D_3$:
\begin{eqnarray}
\label{d1d2d3}
&&D_1 \, = \, x_1 \,+ x_2 \,+ x_3\, - x \, x_1 \, x_2 \, x_3 \, +
q-2\, , \, \quad \quad 
D_3 \, = \,  x \, x_1 \, x_2 \, x_3 \,
 -  x_1 \, x_2 \, x_3 \quad \nonumber \\
&&D_2 \, = \,\, x_1 \,+ x_2 \,+ x_3\, 
+\, x \, x_1 \, x_2 \, x_3 \, -1\,
 -(x_1 \, x_2\, + x_2\, x_3 + \,x_1\, x_3), \, \quad 
\nonumber 
\end{eqnarray}
one can show  that the algebraic expression 
\begin{eqnarray}
{\bf I}_1(x_1, \,x_2, \, x_3, \, x) \, = \, \,  \,  \, 
{{ D_1\cdot  D_2} \over {D_1\, D_2 \, -q \cdot D_3 }}
\end{eqnarray}
is  invariant under
the three inversion relations and the large group
of birational transformation they generate, the (star-triangle)
condition (\ref{crit312}) corresponding to 
$\, {\bf I}_1(x_1, \,x_2, \, x_3, \, x) \, 
= \, \infty$, namely\footnote{For
 $\, x \, = \, 1$ (no
three-spin interaction, $\, D_3\, = \, 0$), condition
 (\ref{crit312}) factorizes
and one recovers the ferromagnetic 
critical condition (\ref{critBax}) of the
$\, q$-state Potts model on an anisotropic triangular lattice.} 
 $\,  D_1\, D_2 -\, q \, D_3\, = \, 0$.

When $\, x \, = \, 1$, or $\, q\, = \, 1$ 
or $ \, 3$, there are 
additional invariants of the three inversions (\ref{defI1}).
For instance, for $\, x \, = \, 1$, one can build an invariant 
from a covariant we give below 
(see (\ref{mart})). For  $\, q\, = \,
3$, introducing 
\begin{eqnarray}
D_5 \, = \, \, 
x_1\;x_2\;x_3\cdot  \left( x_1^{2} x_2^{2} x_3^{2}{x}^{2}\,
-x_2^{2}\, x \, x_1^{2}\, - x_3^{2}\, x\, x_1^{2}\, -x_2^{2} x_3^{2}
x\,  +x_1^{2}\, +x_2^{2}\, +x_3^{2} \, -1 \right) \nonumber
\end{eqnarray}
one finds that the expression:
\begin{eqnarray}
\label{I2D5}
{\bf I}_2(x_1, \,x_2, \, x_3, \, x) \, = \, \,  \,  \, 
{{ D_1^3\cdot  D_2} \over { 3^5 \cdot D_5}} \nonumber
\end{eqnarray}
is invariant under the three inversions (\ref{defI1}).
One can try to find the  manifold corresponding 
to the first order transition 
(see (\ref{MonteCarlo2}) below) in the form: 
$\, F({\bf I}_1, \, {\bf I}_2) \, = \,\, 0$. It
 still remains an open question to 
determine  whether this variety
is algebraic or transcendental. 
The $\, x \, = \, 1$ limit 
corresponds to $\, {\bf I}_1\, = \, +1$.
The condition $\, {\bf I}_2(x_1, \,x_2, \, x_3, \, x)
 \, = \,1$ yields  
$\, x_1 =\, x_2 =\, x_3 =\,  \, 0.215\ 816\, $ (to be compared with 
$\, 0.226\ 681\, $ from (\ref{mart})
in section (\ref{commentchecker}) below) but still
 different from $\, 0.204\, $ (see  
(\ref{martin2}) in  section (\ref{MonteCarlo2}) below)
which is believed to be the location of the first-order
transition point.

\subsection{The embarrassing Kagom\'e critical manifold}
\label{invduakago}

At the end of the 80's there was a surge of
interest in the Kagom\'e lattice
coming from the theoretical study of high-$T_c$
or strongly interacting fermions in two dimensions
(the 2D Hubbard model, resonating valence bond  (RVB), 
ground state of the Heisenberg model).
% proposed by P. W. Anderson). 
%The Gutzwiller
%product form for the ground-state
%is some kind of product over ``dimers'' 
%which is supposed to up-grade the one-dimensional 
%Bethe ansatz construction (see section (\ref{Hubbard})). 
But the two-dimensional Gutzwiller product RVB
ansatz strategy promoted by P. W. Anderson
for describing strongly interacting fermions
 seemed to fail for regular
lattices (square, triangular, ...). Thus,
because of its ground state entropy and other specific properties,
 the  Kagom\'e lattice seemed to be  the ``last chance'' for the RVB approach. 

Since one can obtain a critical frontier (\ref{crit312})
for the general $\, 3-12$ lattice model, and since 
the $3-12$ model
includes the Kagom\'e lattice as a
special case, it is tempting to try to obtain 
the critical frontier for the Potts Kagom\'e lattice.

%~\cite{w180}
The Kagom\'e Potts critical point was first conjectured by Wu~\cite{w69} as:
\begin{eqnarray}
\label{wuconj}
&&y^6\, -6\; y^4\, +2 \; (2-q) \cdot y^3 \, +3\, (3-2\;q)\cdot y^2 \,
-6\, (q-1) \cdot (q-2) \; y \qquad \qquad  \nonumber \\
&& \qquad \qquad \qquad  -(q-2)\;(q^2-4\;q+2) \, = \,  \, \, 0
\end{eqnarray}
which gives for $\, q \, = 2$ the correct critical point $\,y^4\, -6\; y^2\,
-3\, = \, 0$ and for $\, q\, =0$ gives (also correctly) $\, y\,
= \, 1$. Furthermore, for large $\, q$,  $\, y\, $ behaves like $\,
\sqrt{q}\, $ as it should. 
However in the percolation limit $\, q \,
\rightarrow \, 1$, it gives a percolation threshold $\, p_c\, $ 
for the Kagom\'e lattice of: $\, p_c \, = \, \, 0.524\ 43 \cdots$
which is compared to the best numerical 
estimate\footnote{R.M. Ziff and P.N. Suding,
Determination of the bond percolation threshold
for the Kagom\'e lattice, J. Phys. A 30, 5351 (1997) 
and cond-mat/9707110.}
 obtained by R.M. Ziff and P.N. Suding, namely
 $\, p_c \, = \, \, 0.524\ 405 \ 3\cdots\, $ 
with uncertainty in the
last quoted digit.   Wu's conjecture is thus
 wrong {\em but by less than} $\, 5
\cdot 10^{-5}$. Some very long high-temperature series of I. Jensen,
A. J. Guttmann and I. G. Enting
on the $\, q$-state Potts model on the Kagom\'e lattice
further confirm 
%for specific values of $\, q$ 
that the conjecture is wrong for arbitrary values of $\, q$.
   Nevertheless the Wu conjecture remains to be  an
extraordinary approximation.

It is a bit surprising that no 
exact result on integrability (along some algebraic
 subvariety) or exact expression for the 
critical variety is known for the
standard scalar Potts model on 
the {\em Kagom\'e lattice},
as  generally one expects  the
 integrability on one lattice, say the 
square lattice, implies integrability for most
 of the other  Euclidian lattices.
% (triangular, 
%honeycomb, checkerboard ...).
% This fact can be simply understood  in 
%the framework of vertex models, from the concept 
%of ${\bf Z}$-invariance, which clearly shows
% that integrability exists far beyond specific
% Euclidian lattices as far as Yang-Baxter
% integrability is concerned.
 This is certainly not the case for the Kagom\'e lattice.
 
\section{Potts models }
\label{Potts}

The Potts model encompasses a very large number of problems in
statistical physics and lattice statistics.
The Potts model, which is a generalization of the two-component 
Ising model to $\, q$ components for 
arbitrary $\, q$, has been the subject matter
of intense interest in many fields ranging from condensed matter
to high-energy physics.  
%It includes the ice-vertex and
%bond-percolation models as special cases. 
It is also related 
to coloring problems in graph theory.
%\footnote{If one colors each vertex in a lattice
%with one of $\, q$  colors, in how many ways does one have exactly
%$\, n$ pairs of adjacent vertices colored alike?}.

%For reviews on 
%the Potts model and its relevance see, for example~\cite{w84}.
However, exact results on the Potts 
model have proven to be extremely elusive.
Rigorous results  are limited, and include 
essentially  only a closed-form evaluation of its free energy 
 for  $\, q\, =\, 2$, the Ising model, and 
critical properties 
for the square, triangular and honeycomb lattices~\cite{w66}.
Much less  is known about its  correlation functions.

\subsection{ Wu's review on the Potts model }
\label{thereview}

 F. Y. Wu's 1982 review on the Potts model is very
 well-known~\cite{w84} (see also~\cite{w93}). 
%This
% paper has been receiving more than one hundred citations each year
% ever since it was published. 
It is an exhaustive expository review of most of the results known on the Potts model
up to 1981,  a time when interests on the model began to
mount.  It has
remained  extremely valuable for
anyone wishing to work on the standard scalar Potts model. 
In particular, it explains
 the $\, q \, \rightarrow \, 1\, $ limit
 %of the $\, q$-state Potts model
%corresponds to 
of the percolation problem (see also ~\cite{w60}), 
the  $\, q \, \rightarrow \, 1/2\, $ limit
of the dilute spin
glass problem,
and the  $\, q \, \rightarrow \, 0\, $ limit
of the resistor network problem, the equivalences with the
  Whitney-Tutte polynomial~\cite{w84} (see section
 (\ref{Pottsgraph}) and also~\cite{w53}))
and many other related models are also detailed.
For instance,
the Blume-Capel and the Blume-Emery-Griffiths model 
 (see (\ref{BEG}) in (\ref{Blume})) 
 can also be seen as a Potts model.
More generally it is shown that any system of classical $\,
q$-state spins, the Potts model included, can be formulated 
as a spin $\, (q-1)/2$ system.

However, Wu's review was not written in time to include discussions of the 
 {\em inversion functional 
relations}.
% per site
% of the anisotropic $\, d$-dimensional model.
For the two- and three-dimensional
anisotropic $\, q$-state  Potts models,
the partition functions satisfy, respectively,
the functional relations:
\begin{eqnarray}
\label{invfunc}
Z(e^{K_1}, \,e^{K_2}) \cdot 
 Z(2\, - \, q \, -\, e^{K_1}, \,e^{-K_2}) \, = \, \, 
(e^{K_1}-1) \cdot (1-q\, -e^{K_1}), \qquad \qquad 
\end{eqnarray}
\begin{eqnarray}
\label{invfunccub}
&&Z_{cubic}(e^{K_1}, \,e^{K_2}, \,e^{K_3}) \cdot 
 Z_{cubic}(2\, - \, q \, -\, e^{K_1}, \,e^{-K_2}, \,e^{K_3}) \, \,
 \, \nonumber \\
&&\qquad \qquad \qquad \qquad \qquad  \, = 
\, \, (e^{K_1}-1) \cdot (1-q\, -e^{K_1}).
\end{eqnarray}
%Of course, in the  thermodynamic limit, the 
There are also permutation symmetries like, in 3 dimensions,
% of the $\, d$-dimensional model, 
%$\, K_i, \,$
% \, i\, = \, 1, \cdots d\, $ are on the same footing and
% the partition function per
% site satisfies, for instance,
 $\, Z_{cubic}(e^{K_1},   \,e^{K_2},$ 
$\,e^{K_3}) \, $$= \, Z_{cubic}(e^{K_3}, \,e^{K_1}, \,e^{K_2}) 
 \,$$ = \, Z_{cubic}(e^{K_3}, \,e^{K_2}, \,e^{K_1})\,  $.
 Combining these relations one generates an infinite 
set of discrete symmetries which yield  a canonical rational
parametrization of the Potts model at 
and beyond\footnote{At  $\, T \, = \, T_c\, $ 
one recovers the well-known rational
parametrization
of the model (occurring in (\ref{squarePotts}) see below). 
} $\, T \, = \, T_c$, and shows clearly
the  role played by the  Tutte-Beraha numbers. 
%$\, q \, = \, 2 \, + \, 2\; cos(2\; \pi/N)$ ($\, N$ integer).
%After these results were published, 
These infinite
 sets of discrete symmetries 
impose very severe constraints 
 on the critical manifolds and the integrability
(see sections (\ref{crit}), (\ref{invduacrit})). 
The inversion relation study has subsequently been carried out by
F. Y. Wu {\em et al.}~\cite{w153}. 

%Up to this
% omission of the inversion relation results,
%F. Y. Wu's review was very exhaustive.
%One can find all the results a  Potts model specialist needs. 
Graph theory plays a central role in Wu's work on the 
Potts model. 
The Potts  partition function  can be written as~\cite{w84}: 
\begin{eqnarray}
\label{truc}
 Z \, \, \,  \equiv \,\,  \, Z_G(q,K) \,  \, = \, \, \, 
\sum_{G' \subseteq G} (e^K-1)^{b}  q^{n},  
\end{eqnarray}
where $\, K\, =\, J/kT$, the summation is taken over all subgraphs
$\, G' \subseteq G$, and $\, b\,$ and $ \,  n\,$ 
are, respectively, the number
of edges and clusters, including isolated vertices, of $\, G'$.
The duality relation of the Potts model
can be obtained from a graph-theoretical viewpoint
by using the Euler relation    $\, c\,+\,N = \,b\, + n\,$  
where $c$ is the number of 
independent circuits in the subgraph $G'$, 
  and $\, N$ is the total number of vertices in $G$.
This leads to the duality relation
\begin{eqnarray}
\label{identdual}
Z_G(q, K) \,  \,= \, \, \, v^{|E|}
 \cdot q^{1-N_D} \cdot Z_D(q, K^{*}), \nonumber 
\end{eqnarray}
where $D$ is the graph dual to $G$, and
the dual variable $\, K^{*}$ is given by:
\begin{eqnarray}
\label{dualvar}
(e^K \, -1) \cdot (e^{K^{*}} \, -1)\, = \, \, q.
\end{eqnarray}
The generalization of the duality to multisite interactions is also
given in the review.

A consequence of (\ref{truc})
is that
 one finds
the following  connection with the 
{\em chromatic polynomial} $\, P_G(q)\, $
on $\, G$  by taking the
antiferromagnetic zero-temper-ature 
limit $\, K \, \rightarrow \, -\infty$:
\begin{eqnarray}
\label{chrom}
Z_G(q, \, K \, =\, -\infty) \,  \, = \, \,\, P_G(q). \nonumber
\end{eqnarray}
%For instance, recalling the Whitney rank function:
%\begin{eqnarray}
%\label{whitn}
%W_G(x, \, y) \, = \, \, \, 
%\sum_{G' \sqsubseteq G} \, x^{b} \, 
%\Bigl({{ y} \over {x}} \Bigr)^{c} \nonumber
%\end{eqnarray}
%vertices in $\, G$, 
% one finds the identity:
 %\begin{eqnarray}
%\label{identWhit}
%Z_G(q, K) \,  \, = \, \, q^N \cdot W_G\Bigl({{v} \over {q}}, \,
%v\Bigr), \qquad \hbox{where:} \qquad v \, = \, e^K\, -1. \nonumber
%\end{eqnarray}
%The duality on the partition function of  Potts model now
%follows from a similar relation on the Whitney rank function:
% \quad \quad \hbox{or:}
%\quad \quad 
%e^{K^{*}} \, = \, \, {{ e^K \,+q  -1} \over { e^K \, -1}}.
 
The  (high-  and low-temperature) series expansions
are described from a graph-theoretical viewpoint. For instance, the 
high-temperature expansions are written in the (Domb) form:
\begin{eqnarray}
\label{fij}
Z_G(q,\, K) \, = \, \,
 \sum_{\sigma_i \, = \, 0}^{q-1} \prod_{<ij>} {{q+v} \over {q}} \cdot
 (1\, +\, f_{ij}), \quad f_{ij} \, = \, {{v} \over {q+v}} \cdot
 \Bigl( q \, \delta(\sigma_i, \,\sigma_j)\, -1 
\Bigr) 
\end{eqnarray}
The introduction of these $\, f_{ij}$ variables comes
from the fact that:
\begin{eqnarray}
\label{sumrule}
\sum_{\sigma_j \, = \, 0}^{q-1} \,f_{ij} \, = \, \, 0 
\end{eqnarray}
and, consequently, all subgraphs with vertices of degree 1
give rise to zero contributions. 
The number of subgraphs that occur in the expansion is therefore
greatly reduced.

The location of the critical points of the anisotropic Potts model
on a square, triangular and honeycomb lattice were given
in terms of the variables 
$\, x_r \, = \, (e^{K_r} \, -1)/\sqrt{q} $ 
(see also (\ref{varxr}) below). These
expressions are invariably the various special cases of  Wu's conjecture~\cite{w84}) for
the critical point of the more general checkerboard  lattice, namely,
 \begin{eqnarray}
\label{checker}
&&\sqrt{q} \, + x_1\,_+ \,  x_2\, +x_3\, +x_4 \,  \,  \,  \quad\quad \quad\\
&&\, \quad\quad= \, x_1\;x_2\;x_3 + \,
x_1\;x_2\;x_4 + \, x_1\;x_3\;x_4 + \, x_2\;x_3\;x_4\, + \, \sqrt{q}
\cdot x_1\;x_2\;x_3 \;x_4.  \nonumber
\end{eqnarray}
Using the notation $\, a \, = \, e^{K_1}$, $\, b \, = \, e^{K_2}$, 
 $\, c \, = \, e^{K_3}$,  $\, d \, = \, e^{K_4}$, this 
critical algebraic variety (\ref{checker}) reads:
\begin{eqnarray}
\label{checkerbis}
&&- \left( q-1 \right)  \left( q-3 \right) 
+ \left( a+b+c+d \right)  \left( 2-q \right)
 \\
&& \qquad \qquad \qquad  -(ab+ac+bc+ad+bd+cd) \, 
+abcd \, = \, \,\, 0 \nonumber
\end{eqnarray}
%As far as critical conjectures are concerned, a conjecture
%by Ramshaw was given in~\cite{w84} which happened to be incorrect.
The critical point of a mixed
ferromagnetic-antiferromagnetic square Potts model considered
by Kinzel, Selke and Wu~\cite{w77}:
\begin{eqnarray}
\label{conjwrong}
(e^{K_1} \, -1) \cdot (e^{K_2} \, +1)\, = \, \, -\,q
\end{eqnarray}
was also given.
While this expression coincides with the critical point for $q=2$, it
%These results were, in fact, ruled out in latter 
%publications. They can simply be ruled out 
%because they are not 
is incompatible with the inversion relations
(\ref{invfunc}) for general $q$, and hence is not
a critical variety. 
 This is because the infinite discrete group
of symmetries generated from the inversion relations
of the square Potts model transforms 
 (\ref{conjwrong})
% conjectured to be
%a critical variety of this mixed model,
into an {\em infinite set} of other algebraic varieties,
and hence cannot be  critical. 

Generally,  critical manifolds need to be (globally) invariant 
under this infinite set of transformations (discrete symmetries). 
Actually, for the anisotropic square Potts model,
for instance, one can show that, when $\, q$  is not a Tutte-Beraha
number\footnote{When $\, q$ is a Tutte-Beraha number the generically
infinite discrete group $\,
\Gamma\, $ generated by the inversion relations on the model
becomes a finite group and many algebraic varieties can be invariant
under such a  finite group: a simple way to build 
such algebraic $\,\Gamma$-invariants amounts to 
performing summations over the group $\,\Gamma$.  
}, the {\em only algebraic varieties
compatible with the inversion relation symmetries} (\ref{invfunc}) 
are given by the well-known ferromagnetic condition:
\begin{eqnarray}
\label{wellferro}
\, (e^{K_1} \, -1) \cdot (e^{K_2} \, -1) \, = \,\,  q
\end{eqnarray}
and the antiferromagnetic condition   obtained by 
R. J. Baxter:
 \begin{eqnarray}
\label{antiferro}
(e^{K_1} \, +1) \cdot (e^{K_2} \, +1) \, = \,\, \,4\, - \, q
\end{eqnarray}
(for which the model is exactly soluble). Note that these two varieties 
can be deduced from the conjecture (\ref{checker}) by
taking $\, \, x_1\, = \, x_3$ and $\, x_2\, = \, x_4$.
In this limit the critical condition (\ref{checker})
factors into   conditions (\ref{wellferro})  
and  (\ref{antiferro}).
In fact, it has since been shown that
%some time after the publication
%of F. Y. Wu's review on the Potts model,
the critical condition (\ref{checker})
 corresponds to an integrability condition
of the checkerboard Potts model.
% the partition function per site
%reducing to the two previous ferromagnetic
%and antiferromagnetic critical exact solutions
%for the partition function per site.

 At {\em criticality} the Potts model is {\em exactly solvable}. Let us give the
example of the square lattice. 
The free energy of the isotropic 
Potts model at the {\em ferromagnetic} critical point $\,
T \, = \, T_c$ reads:
\begin{eqnarray}
\label{squarePotts}
&&f(q, \, T_c) \, = \, \, {{1} \over {2}} \, 
+ \, \theta \, + \, 2 \cdot \sum_{n=1}^{\infty}\, {{e^{-n\;\theta}
\cdot \tanh(n\;\theta)} \over {n}}, 
\qquad \qquad \hbox{for:} \qquad q \, >\, 4 \nonumber \\
&&f(q, \, T_c) \, = \, \, 
\ln(2) \, + 4\cdot \ln\Bigl( {{\Gamma(1/4) } \over {2 \,
\Gamma(3/4)}}\Bigr) \qquad  
\qquad \hbox{for:} \qquad q \, = \, 4
 \\
&&f(q, \, T_c) \, = \, \, 
 {{1} \over {2}} \, 
+ \, \int_{-\infty}^{+\infty} \, {{dx} \over {x}} \, \tanh(\mu\; x) \,
{{ \sinh((\pi \, - \mu)\, x) } \over {  \sinh(\pi\; x)}}
\qquad \hbox{for:} \qquad q \, < \, 4 \nonumber
\end{eqnarray}
where the variables $\, \theta$ and $\, \mu$ correspond to the
rational parametrization of the model at $\, T\, = \, T_c$, 
namely $\, \cosh(\theta)\, = \, \, \sqrt{q}/2\, $ or 
 $\, \cos(\mu) \, =\, \, \sqrt{q}/2$.

The exact critical exponents of the standard scalar Potts model are
also given in this review.
  These  critical exponents, which
are {\em rational} when $\, q$ is the
  Tutte-Beraha numbers, are:
%$ \, q \, = \, \,  2 \, +2 \, \cos(2\pi/N)\, $. 
%For the standard 
%scalar nearest neighbor Potts model
%the Tutte-Beraha numbers correspond to the values for which the 
%critical exponents of the model are rational:
\begin{eqnarray}
\label{tutte}
&&\alpha \, = \,\alpha' \, = \, {{2} \over {3}} \cdot {{ 1\;-2\;u }\over
{1\;-\;u}}, \qquad 
\beta \, = \, \, {{ 1+u} \over {12}} , \qquad \gamma \, = \, \gamma'
\, = \, {{ 7-4\;u\, +u^2} \over { 6 \cdot (1-u)}}, \nonumber \\
&&\delta \, = \, {{ (3-u)\cdot (5-u) } \over {1-u^2}}, \qquad 
\nu \, = \, \nu' \, = \, {{ 2-u} \over {3 \cdot (1-u)}}, \qquad 
\eta \, = \, {{ 1-u^2} \over {2 \cdot (2-u)}} \nonumber 
\end{eqnarray}
where the parameter $\, u$ is related to $\, q$ by:
\begin{eqnarray}
 2\, \cos\Big({{\pi \, u} \over {2}}\Big) \, = \, \, \sqrt{q}, 
\qquad \hbox{or:} \qquad \qquad 
2\, + \, 2 \, \cos( \pi \, u)\, = \, q \qquad 
\end{eqnarray}
%The  Tutte-Beraha numbers $\, q \, = \, \,  2 \, +2 \, \cos\,(2\pi/N)\, $
%thus correspond to values of  $\, u\; \pi \, $ 
%(the shift of the rational spectral parameter
%parameterizing the standard scalar Potts model) 
%such that  $\, u$ becomes a {\em rational number}.     
These  results   play a key role
in the emergence of the conformal theory.
%explain why the two-dimensional Potts model
%plays such an important role in the theory
%of phase transitions and critical phenomena. 
%Along this topics more work certainly 
%remains to be done
%for more involved Potts models~\cite{w89}.
% (F. Y. Wu 
%and H. E. Stanley,
%Universality in Potts models with 
% two- and three-site interactions).

\subsection{Comments on the checkerboard Potts model}
\label{commentchecker}
The Wu conjecture (\ref{checker}) for the 
 criticality condition
of the $\, q$-state  checkerboard
 has since been confirmed from
%after the publication
%of this review the exact result 
%for the partition function per site of the critical checkerboard 
%Potts model was deduced from 
an inversion relation analysis.
To discuss the inversion relation we  introduce   variables $\,u,\, v,\, w,\,
z,\, t\,$ defined by:
 \begin{eqnarray}
\label{new}
&&{{ e^{K_1} -1} \over {e^{K_1} \,+q-1}} \, = \, \, 
t \cdot {{ u-t} \over { 1-t^3 \, u}}, \, \qquad 
{{ e^{K_2} -1} \over {e^{K_2} \,+q-1}} \, = \, \, 
t \cdot {{ v-t} \over { 1-t^3 \, v}}, \,\,    \cdots \qquad \\
&& \quad \hbox{or:} \quad 
e^{K_1} \, = \, a  \, = \,\, {\frac {u \, - \, t^3}{t \cdot ( 1\, -u\,
t ) }}, \quad 
e^{K_2} \, = \, b  \, = \,\, {\frac {v \, - \, t^3}{t \cdot ( 1\, -v\,
t ) }}, \,  \,  \cdots \qquad \nonumber \\
&& \qquad \hbox{with:} \qquad \qquad t \, + t^{-1} \, = \,
\, \sqrt{q}\nonumber
\end{eqnarray}
In these variables the criticality 
condition  (\ref{checker})
reads:
\begin{eqnarray}
\label{cricheck}
u\, v\, w\, z \, \, \, = \, \, \, 1.
\end{eqnarray}
Using the  inversion trick\footnote{
With the well-suited variables (\ref{new})
the inversion trick naturally  yields exact formulae
as products over an infinite discrete group, in this case Eulerian
products (\ref{multiplicativePotts}).},
the partition function of the checkerboard model at {\em criticality}
can be written in a multiplicative form
%Instead of the free energy of the isotropic 
%Potts model (\ref{squarePotts}), let us give 
%the partition function per site of the full checkerboard 
%Potts model, {\em at criticality}, which can be given
%as a multiplicative 
%closed formula this time 
\begin{eqnarray}
\label{multiplicativePotts}
&&Z(u, \, v, \, w, \, z)_{[u\, v\, w\, z \, \, =  \, 1]} \, \,   \, \, \, \\
&&=\ {{q} \over {t^2}} \cdot  {{ F(u) \cdot F(1/u)} \over {1\, -\, t\, u}} 
\cdot  {{ F(v) \cdot F(1/v)} \over {1\, -\, t\, v}} 
\cdot  {{ F(w) \cdot F(1/w)} \over {1\, -\, t\, w}} 
\cdot  {{ F(z) \cdot F(1/z)} \over {1\, -\, t\, z}} \nonumber \\
&&\qquad \qquad  \hbox{where:} \qquad  
\qquad  F(u) \, = \, \, 
\prod_{n=1}^{\infty} {{ 1\, - \,t^{4\;n-1} \,u}
\over  { 1\, - \,t^{4\;n+1} \,u}} .  \nonumber 
\end{eqnarray}
This formula is
reminiscent of (\ref{multiplicative}) for
the six-vertex model (see section (\ref{Liebmonograph})),
as it should since it
also follows from the Potts and six-vertex model  correspondence (see section
(\ref{correspond}) below).

Note that, in the anisotropic case $\, w\, = \, u \, $ and $\, z
\, = \, v$,  the critical condition (\ref{cricheck})  
%of the checkerboard Potts model $\, u\, v\, w\, z \,  \, = \, \,1\, $ 
factorizes into $\, u\, v \, = \, + 1$
and $\, u\, v \, = \, - 1$ which are 
respectively the aforementioned ferromagnetic
and  antiferromagnetic
 criticalities (and integrability).
 
The checkerboard model reduces to an honeycomb model,
and hence a triangular model by taking the dual, if one of the
four interactions vanishes.
 Therefore it is useful to re-examine the triangular lattice
limit of the checkerboard variety.
P. Martin {\em et al.} have established two varieties for the triangular Potts model:
a ferromagnetic variety which is precisely (\ref{critBax}) with $x=1$, and an
antiferromagnetic variety
\begin{eqnarray}
\label{mart}
(q-2)^2\, -2\, + \,( a+b+c\, +abc)  \,( q-2) + 2\,(ac+\,ab+\,bc)
\, = \, \, 0. \qquad \qquad  
\end{eqnarray}
These two algebraic varieties can be written, respectively, as
 $\,u\, v\, w\, = \, + t$ and  
$\,u\, v\, w\, = \, - t$, which can be deduced by taking,
respectively,  $\,d=1\,$ and    $\, d \, = \,
-(q-2)/2$ in (\ref{checkerbis}).

%As far as antiferromagnetic algebraic varieties are concerned,
%it can be shown, for instance in the case
%of the anisotropic triangular
%Potts model (P. Martin {\em et al.}) that 
%the ferromagnetic critical variety (which can be written $\,
%u\, v\, w\, = \, + t$, see also (\ref{critBax})):
%\begin{eqnarray}
%\label{ferro}
%  2-q \, -(a+b+c) \, +a\,b\,c \, = \, \, 0,\, 
%\end{eqnarray}
%together with the (antiferromagnetic) algebraic variety:
%\begin{eqnarray}
%\label{mart}
%&&(q-2)^2\, -2\, + \,( a+b+c\, +abc)  \,( q-2) + 2\,(ac+\,ab+\,bc)
%\, = \, \, 0, \qquad \qquad  \\
%&&\qquad \hbox{or:} \qquad \qquad \qquad 
%u\, v\, w\, = \, - t \nonumber 
%\end{eqnarray}
%are, for generic $\, q$ (namely
%when $\, q$ is not a Tutte-Beraha number) 
%{\em the only two algebraic varieties compatible} with 
%the inversion relation symmetries. This (antiferromagnetic) 
% algebraic variety (\ref{mart}) corresponds to the limit $\, d \, = \,
%-(q-2)/2$ (namely $\, e^{K_4^{*}}\, = -1$ or
% $\, z \, = \, -1/t$) 
%  of (\ref{checkerbis}). 
 For $q=3$ the antiferromagnetic 
 algebraic variety yields an 
algebraic critical point very close to the first-order transition
point for the three-state Potts model~\cite{w86,w136}.

 Along the  well-known ferromagnetic variety
(\ref{dualvar}) which for the isotropic model reads
$\, (e^{K} \, -1) \cdot (e^{K^*} \, -1) \, = \, q$,
%corresponds to a self-dual condition $\, K_2 \, = \, K_1^{*}\, $
%with  duality . Along this line,
it is interest to point out the existence of  some ``hidden'' duality
$\, K \, \rightarrow \, K^{\dag} \, $
such that the antiferromagnetic condition (\ref{antiferro})
reads $\, K_2 \, = \, \, K_1^{\dag}$. This hidden duality
is the involution:
\begin{eqnarray}
\label{hidden}
e^{K^{\dag}} \, = \, \, 
-\, {{ e^{K}+\, q-3  } \over {e^{K} +1}}.
\end{eqnarray}
Note that the dualities (\ref{dualvar})
and  (\ref{hidden}) commute, and their product gives the
involution:
\begin{eqnarray}
\label{newnew}
e^{K} \, \,\quad  \longrightarrow \,\, \,\quad \quad   e^{{  K^{*}  }^\dag} \,\, 
 = \,\,  \, -\, {\frac {(q-2)\cdot e^{K}\, +2}{2\,e^{K}\, +(q-2)}} 
\end{eqnarray}
The ferromagnetic critical
variety of the anisotropic triangular Potts model (\ref{critBax})
now transforms into the anti-ferromagnetic critical
variety (\ref{mart}) under (\ref{newnew}).
One also finds that the ferromagnetic critical
variety of the anisotropic square-lattice  Potts model (\ref{wellferro}),
 and its anti-ferromagnetic critical
variety (\ref{antiferro}), are {\em both} 
invariant under transformations (\ref{dualvar}),
(\ref{hidden}) and thus (\ref{newnew}).
More generally, for the checkerboard model
one finds that (\ref{checkerbis}) 
is invariant under both  (\ref{dualvar}),
(\ref{hidden}) and thus (\ref{newnew}).
These results can be very simply seen in the
variables $\, u$, $\, v$, $\, w$, $\, z$.  For example, 
%for the inversion relations. With these variables
%one finds that transformation 
(\ref{newnew}) is simply
$\, u \, \rightarrow \, \, -u\, $.
% and one immediately obtains
%these results from the very simple product forms of the
% critical varieties.
 Involution like the ``hidden duality''
(\ref{hidden}), or like (\ref{newnew}), do not yield
simple functional equations on the partition function like
the Kramers-Wannier duality does (see (\ref{identdual})
and (\ref{dualvar})). They are not  symmetries
of the model, but, rather,  ``symmetries of the 
second kind'':  symmetries of the symmetries.

\subsection{Equivalence of the 
$\,q$-state Potts model with a  six-vertex model}
\label{correspond}

Recall that the partition function of the standard scalar
$\, q$-state Potts model  can readily
be written as a  Whitney-Tutte polynomial.
In a further step Temperley and Lieb used operator
methods to show that, for the square lattice,
the Whitney-Tutte polynomial is, in turn, equivalent to
a staggered ice-type vertex model. 
 
In a classic 
paper~\cite{w53}  by  R. J. Baxter, S. B. Kelland and  F. Y. Wu,
this equivalence is {\em rederived} from a graphical approach (see
 (\ref{Temper}) in section (\ref{Pottsgraph})),
% concerns the graphical construction of the 
%equivalence of the partition function 
%of the Potts model (or  Whitney polynomial), 
%with an {\em ice-type model}~\cite{w53}. 
which is easier than the
algebraic method of Temperley and Lieb
 and  is applicable to {\em arbitrary}  planar graph. 
%(square, triangular, etc ...). 
%In~\cite{w53}
%R. J. Baxter, S.B. Kelland and  F. Y. Wu re-derived 
%this equivalence by a graphical method, which is simple
%to apply to {\em any planar lattice}. It was shown, in particular,
Particular, the equivalence was extended to
triangular or honeycomb Potts models
 and a staggered six-vertex model
on the Kagom\'e lattice. 

 The equivalences are as follows  (page 404 in~\cite{w53}).
For the square, triangular and honeycomb lattices, 
the equivalent ice-type vertex model has the Boltzmann weights:
\begin{eqnarray}
\label{correspondboltz}
(\omega_1, \, \, \omega_2, \, \, \cdots , \, \, \omega_6)\,\, = \, \,\, 
(1, \, \, 1, \, \, x_r,  \,\, x_r, \, \, A_r, \, \, B_r)
\end{eqnarray}
where $\, (A_r, \, B_r)\, $ are respectively, for the square,
triangular and honeycomb lattice:
\begin{eqnarray}
\Bigl({{1}\over {s}} \, + \, x_r\cdot s, \,
{{x_r}\over {s}} \, + \, s  \Bigr), \quad 
 \Bigl({{1}\over {t}} \, + \, x_r\cdot t^2, \,
{{x_r}\over {t^2}} \, + \, t  \Bigr), \quad 
  \Bigl({{1}\over {t^2}} \, + \, x_r\cdot t, \,
{{x_r}\over {t}} \, + \, t^2  \Bigr) \nonumber 
\end{eqnarray}
with:
\begin{eqnarray}
\label{varxr}
s\, = \, e^{\theta}, \qquad t\, = \, e^{\theta/3}, \qquad
2\,\cosh(\theta) \, = \, \sqrt{q}, \qquad 
x_r \, = \, {{e^{K_r} \, -1} \over {\sqrt{q}}}
\end{eqnarray}
These mappings  play an important role 
in the Lee-Yang theorem to be discussed latter (see 
section (\ref{HintermanKunzWu})).

\subsection{Miscellaneous Potts model  phase diagrams}
\label{miscphase}

\subsubsection{Potts model with competing interactions}
Kinzel, Selke and Wu~\cite{w77} have studied a square lattice Potts model
with competing interactions alluded to earlier in section (\ref{thereview}).
A similar model in 3 dimensions with  next-nearest-neighbor 
competing interactions has been studied by 
 J. R. Banavar and   Wu
 using the mean-field theory and
Monte-Carlo simulations~\cite{w92}. A rich
phase diagram was found and they
 established positively that the behavior of the
 4-state three dimensional Potts model is
  mean-field-like.
 
\subsubsection{First-order transition in the antiferromagnetic Potts model}
 \label{MonteCarlo1}
  F. Y. Wu {\em et al.} have analyzed specifically
the three-state triangular Potts model and 
considered the (tricky) tricritical behavior of 
 this  model~\cite{w158}. 

Recalling the results of section (\ref{invduacrit}), and in particular 
the special role played by $\, q\, =3$ for the two- and three-site
interaction Potts model (\ref{funcpart}), 
Monte-Carlo calculations of the $q=3$ isotropic limit of
the model have been performed~\cite{w137}. 
These studies confirmed the existence of an antiferromagnetic critical
point (in addition to the well-known ferromagnetic one) probably
corresponding to a first-order transition
occurring near the  variety 
 $\, x\, =\, 1$ isotropic region.
%antiferromagnetic critical point is definitely different~\cite{w137}.
%This negative result does not
% rule out the existence of an  extension of
%this variety to some $x \ne 1$ domain as a critical variety in
%some region of the parameter space which could depend on the value of
%$\, q$.

For the triangular  Potts model
with two-spin interactions, Monte-Carlo calculations~\cite{w158}
confirm the localization $\, a \, \simeq 0.204 \, \pm \, .003$
for a {\em first-order} transition point. This 
transition point is confirmed to be different from 
{\em the algebraic antiferromagnetic
 point localized} at $\, a \, = \, \, 0.22665\, $
derived below (see (\ref{martin})
below).

The antiferromagnetic transition point
$\, a \, \simeq 0.204 \, \pm \ .003\, $
can be interpreted as to
belong to some singular 
manifold in the parameter space of the model
 with two- and three-spin interactions
%(a curve in the ($K_1\, = \, K_2\, 
%= \, K_3$, $\,\, K$) plane). 
This singular 
manifold corresponds to a first-order transition
frontier. Recalling discussions in section (\ref{invduacrit}), 
the question of the algebraic or transcendental status
of this first-order transition frontier still remains open.

\subsubsection{Chiral Potts models}
%: Monte-Carlo simulations}
\label{MonteCarlo2}

  F. Y. Wu {\em et al.} have analyzed a 
particular two-dimensional chiral Potts
model, namely, the  3-state chiral Potts
in order to understand the relation 
between the (higher genus) integrability 
 and  criticality conditions~\cite{w136}.
On the checkerboard model the higher genus integrability
of the 3-state chiral Potts model
 is restricted to the following algebraic variety:
\begin{eqnarray}
\label{P1234Q1234}
&&3\cdot (Q_1\,P_2\,P_3\,P_4\, +Q_2\,P_1\,P_3\,P_4\,
+Q_3\,P_1\,P_2\,P_4\, +Q_4\,P_1\,P_2\,P_3) \, \\
&& \quad -(P_1\,Q_2\,Q_3\,Q_4\, +P_2\,Q_1\,Q_3\,Q_4\,
+P_3\,Q_1\,Q_2\,Q_4\, 
+P_4\,Q_1\,Q_2\,Q_3)\, = \, \, \, 0 \nonumber 
\end{eqnarray}
where:
\begin{eqnarray}
&&P_i \, = \, \, f_i\, - 3\, h_i, \quad 
\quad Q_i \, = \, \, f_i\, -2\;g_i\;
+3\;h_i, \quad \quad
 h_i \, = \, a_i^2\;b_i^2\;c_i^2 \quad \quad \quad  \,  \\
&& f_i \, = \, \, a_i \;b_i \;c_i 
\cdot (a_i^3\, + \,b_i^3\, + \,
c_i^3), \quad \quad \quad g_i \, = \, \, 
a_i^3\,b_i^3\,  + \,b_i^3\,c_i^3\,  + \,
c_i^3 \, a_i^3 \nonumber 
\end{eqnarray}
where $\, a_i$, $\, b_i$, $\, c_i$ denote 
the three possible values of 
the four edge Boltzmann weights
 $\, w_i(\sigma_k-\sigma_l)$ of the checkerboard lattice: 
\begin{eqnarray}
a_i \, = \, \, w_i(0), \, \, \,
b_i \, = \, \, w_i(1)\, = \, \,w_i(-2), \, \, \,
c_i \, = \, \, w_i(2)\, = \, \,w_i(-1), \, \, \,
\, i \, = \, 1, \, 2, \, 3, \, 4
\nonumber 
\end{eqnarray}
Recalling the conformal theory prejudice
(criticality in two dimensions versus  integrability),
one can also wonder if an integrability
condition like (\ref{P1234Q1234})
could correspond to a critical 
subvariety of the phase diagram.
Let us consider the standard scalar limit 
%($\, w(1) \, = \,w(2)\, = \, b \, = \, c$) 
of this model. The 
higher genus integrability
condition (\ref{P1234Q1234}) reduces to
the critical condition (\ref{checker}) or (\ref{checkerbis}) 
of the standard scalar
Potts model on the checkerboard lattice
for $\, q\, =3$:
\begin{eqnarray}
\label{critcheck}
a b c d \,  - (a b+ a d + a c + b c + b d + c d)\, 
- (a + b + c  + d) \, = \, \, \, 0
\end{eqnarray}
together with another algebraic variety: 
\begin{eqnarray}
\label{another}
&&a b c d \, + \, 2 \cdot (a c d + b c d + a b d + a b c)
+\, a b + b d + c d + a c + a d + b c\,\nonumber \\
&& \qquad \qquad  \qquad\qquad -(a+b+c+d)\,  -2 \,  \, =  \, \, \, 0 
\end{eqnarray}
With the variables (\ref{new}) taken for $\, q\, = \, \, 3$, the
critical condition (\ref{critcheck}) 
reads $\,\,  u\, \,v\, w\, z \, = \, \, +1\, $ and 
the algebraic condition
(\ref{another}) reads: $\,  u v w z \, = \, \, -1$.
Considering the similarity of (\ref{another}),
namely $\,  u v w z \, = \, \, -1$,
with  (\ref{critcheck}), namely 
 $\,\,  u v w z \, = \, \, +1$, 
it is tempting
to imagine that (\ref{another}) could also be
in some domain of the
parameter space $\, a, \,b, \, c, \, d$, 
a critical variety.

In the  isotropic triangular 
%limit
%of this checkerboard chiral Potts model, and in a second step
and  standard scalar limit 
%if the triangular model ($\, w(1) \, = \,
%w(2)\, = \, b \, = \, c$). The 
the higher genus integrability
condition (\ref{P1234Q1234}) factorizes into two conditions.  One is
the ferromagnetic critical condition of the standard
scalar Potts model (see also (\ref{critBax})) 
$\,  2-q\,-a-b-c\,+a\,b\,c \, = \, \, 0\, $ (or 
$\, u\, v\, w\, = \,  t$) with $\, q\, = \, 3$, 
the other one is:
\begin{eqnarray}
\label{martin2}
&&( q-2)^2\, -2\, + \,( a+b+c\, +abc)  \,( q-2) + 2\,(ac+\,ab+\,bc)
\, = \, \, 0, \nonumber \\
&& \qquad \qquad \qquad \hbox{or:} \qquad \qquad 
u\, v\, w\, = \, - t
\end{eqnarray}
with $q=3$.
In the isotropic limit and for $\, q\, = \, 3$,
 these two algebraic varieties give respectively:
\begin{eqnarray}
\label{martin}
a^3\, -3\, a \, -1 \, = \, \,0
\qquad \hbox{and:} \qquad  
a^3\,+6\, a^2\,+3\, a\, -1 \, = \, \, 0
\end{eqnarray}
yielding the ferromagnetic 
critical point $\, a \, = \, \,
1.8793\,$ and {\em an antiferromagnetic} transition
 point at  $\, a \, = \, \, 0.22665$.
This antiferromagnetic point must be compared with the
antiferromagnetic critical and {\em first-order}
transition point $\, a \, \simeq 0.204 \, \pm \, 0.003$
obtained from series analysis 
by I. G. Enting and F. Y. Wu~\cite{w86}.
%F. Y. Wu {\em et al.} 
% Monte-Carlo calculations on the 
%three-state chiral Potts model on  triangular 
%lattices of various sizes up to 128. 

\subsection{Zeros of partition functions of Potts models}
\label{Zeros}

  In 1952 Yang and Lee  introduced the concept 
of considering the zeros
of the grand partition function of statistical mechanical systems, a
consideration that has since opened new avenues to the study of
phase transitions.  While Yang and Lee considered the zeros in
the complex fugacity plane, or equivalently the complex magnetic
field plane in the case of spin systems, Fisher
in 1964 called attention to the relevance
of the zeros of the canonical partition function in the
complex temperature plane.
Generally speaking, there exist several 
different kinds of exact results on 
lattice models in statistical mechanics.
Ideally, one would like to obtain the
exact, closed-form, expressions
of thermodynamic quantities such as the per-site free
energy, the surface tension,
spontaneous magnetization,
and correlation functions. A knowledge 
of these exact expressions leads to a complete description of the system
including the phase boundary (critical frontier)
and the location of zeros of the partition
function. 
However, exact evaluations of physical quantities are not always
possible.  In such cases one can sometimes determine
the critical frontier
from properties such as the duality and the inversion  relations, or
 analyzes analyticity properties of the free energy by locating
the zeros of the partition function.
But other than in the case of
some special  one-dimensional model, exact results on the
 zeros have been confined mostly to the Ising model.

\subsubsection{Fugacity variable for checkerboard Potts model,
staggering field, Lee-Yang theorem, duality}
\label{HintermanKunzWu}

Let us consider a $\, q$-state Potts model
on a square, or triangular, lattice,
with no magnetic field. Since one does not
have a  magnetic field one does not expect 
a Lee-Yang theorem to exist. Actually {\em this is not true}:
there is a ``hidden'' field for the standard
scalar $\, q$-state Potts model !

%  {\it J. Stat. Phys.} {\bf 19}, 623-632
% (1978))  
A. Hinterman, H. Kunz and 
 F. Y. Wu~\cite{w66} combined the equivalence
of  the Potts model with a staggered six-vertex model,
together with  Lee-Yang circle theorem
%requires Ising variables, and thus actually 
%functions for vertex model (generalization
%of the Lee-Yang theorem 
due to Suzuki and Fisher (see also~\cite{w27})
to deduce the critical variety of the Potts model.
In this consideration
%to get a Lee-Yang theorem for the $\, q$-state Potts model
%on a  anisotropic square model
%or a (anisotropic) triangular lattice, 
a fugacity variable $\, z$ \, is
 associated with the staggering field
that occurs in the aforementioned correspondence.
%Recalling the (partition function) correspondence
%between the Potts model and staggered six-vertex models
%(see section (\ref{correspondboltz})
 In terms of the variables
 $\, (A_r, \, B_r)\, $ given in (\ref{correspond}),
%A. Hinterman, H. Kunz and  F. Y. Wu 
the fugacity variable $ \, z$ reads respectively 
for the  square 
and  triangular lattices:
\begin{eqnarray}
\label{z4z6}
 z^4 \, = \, \, {{A_1\,A_2 } \over {B_1\,B_2 }}, \qquad \qquad 
z^6 \, = \, \,  {{A_1\,A_2 \,A_3 } \over { B_1\,B_2\,B_3 }} .
\end{eqnarray}
A condition on   the Suzuki-Fisher 
extension of the Lee-Yang theorem  and 
%are that 
%some variables have to be real ($c_r\, $ 
%and $\, x_r\, $ see~\cite{w66})
%which requires 
for real temperatures
requires that we must have  $\, q\,>\,4$.

All these results can be generalized to the
 $\, q$-state Potts model
on the checkerboard lattice.
Not surprisingly,  (\ref{z4z6}) is 
generalized into:
\begin{eqnarray}
\label{z8}
z^8 \, = \, \,\,   {{A_1\,A_2 \,A_3 \,A_4 } 
\over { B_1\,B_2\,B_3 \,B_4}} \nonumber 
\end{eqnarray}
Note that the duality (\ref{dualvar}) of the Potts model 
{\em has a very simple representation
in terms of these fugacity variables} $\, z$: $\, z \, \rightarrow \, 
1/z$. 

The zeros of the partition function of the checkerboard
model will later be seen to
lie on $\, | z| \, = \, 1$, as it should 
from the Lee-Yang theorem which applies when $\, q \, > 4$. 
Remarkably, it was also seen later
by other authors, that the $\, | z| \, = \, 1$ condition
can be extended to $\, q \, < \, 4$
and that one recovers  the well-known Fisher's circles
for the Ising model !!

The fugacity variable $\, z$ is a fundamental 
variable. It corresponds to
a crucial combination of variables encapsulating the 
action of the infinite discrete group
generated by the inversion relations. 
The criticality 
condition correspond to $\, z\, = \, 1$ 
and  $\, z\, = \, -1$ seem also to
 play some role (see (\ref{another})
for $\, q \, = \, 3$ in section (\ref{MonteCarlo2})), but not a critical
or transition point role. 
%Actually recalling the inversion relation
In terms of variables $\, u, \, v, \, w, \, z$ of section
(\ref{MonteCarlo2}), the fugacity variable $\, z$ is  
 simply the product $\, u\,v\,w\, z$. 

%These results are also in agreement with 
%the fact that the Kramers-Wannier duality 
%on the Potts model becomes $\, z \, \rightarrow \, 1/z$.
 
\subsubsection{Zeros for the square lattice: A graph-theoretical
viewpoint}
\label{zerossquare}

Following  F. Y. Wu's approach let us  consider
 the $\, q$-state Potts model 
on the square
lattice from a graph-theoretical viewpoint
with the partition function (\ref{truc}).
Introducing the variable 
\begin{equation}
 x\, =\, \, (e^K-1)/\sqrt q, 
\end{equation}
the partition function  (\ref{truc}) can be written as a polynomial in $x$ as:
\begin{eqnarray}
\label{Zcsum}
Z  \, \equiv \, \,  P_G(q,x) \,  =\, \,
  \,\, \sum_{b=0}^E c_b(q) x^b, 
\quad \, \,  \hbox{  where } \quad \quad 
 c_b(q)\, =\, \, q^{b/2}\: \sum_{G`\subseteq G} \, q^{n},
\nonumber  
\end{eqnarray}
where the second  summation is taken over all
 $\, G' \subseteq G\, $ for a fixed $\, b$. 
 Then, the duality relation (\ref{identdual}) can be rewritten as a duality relation
for the polynomial $\, P_G$~\cite{w54}
\begin{equation}
\label{e5}
 P_G(q,x) \, = \, \, \,\,
 q^{N-1-E/2}\, x^E \cdot  P_D(q, x^{-1}).
\nonumber
\end{equation}
  In the case of the square lattice for which $\, D$ is identical to $\, G$ in the
thermodynamic limit regardless of boundary conditions, 
 (\ref{e5}) implies  that the system is critical at $\, 
x_c\, =\,\,  1$. For finite self-dual lattices
 relation (\ref{e5}) gives an example of a self-dual
polynomial\footnote{A polynomial $\, P(x)$ in $\, x$ is self-dual
 if it is proportional to $\, P(1/x)$. Self-dual polynomials occur
naturally in lattice models in statistical physics
 and in restricted partitions of an
integer in number theory (see section (\ref{3Denumer})
 below).}~\cite{w186}.

 To describe the density of zeros
on the Lee-Yang circle, we introduce
an angle $\, \theta\, $ associated with the location 
of the zero on the unit circle.
For small $\, \theta$
we have 
 $\, \,  g(\theta) \, = \,$
$  a |\theta|^{1-\alpha(q)}\, \, $  for $\,
q\leq \, 4$ and $\, \,  g(\theta) \, = \,$
$ \epsilon (q)\, $ when $\,  q\, >\, 4$.
This leads to the specific heat singularity
$ \, |t|^{-\alpha(q)}\, $ for
$\, q\, \leq \,  4$ and a jump discontinuity 
of  $\, \epsilon(q)\, $ in $\, U$ for $\, q\, >\, 4$.
This is the known critical behavior of the Potts model~\cite{w84}.

The zeros of the  partition
function of the $\, q$-state Potts model on the square
lattice  have been evaluated
numerically~\cite{w162}. On the basis
of these numerical results, it was conjectured~\cite{w162} that,
for both finite self-dual lattices and for lattices  with free
or periodic boundary conditions in the thermodynamic limit,
the zeros in the $\, {\rm Re} (x) \, > \, 0 \,$ region of the
complex $\, x \, $ plane {\em are
located on the unit circle}
$|x|\, =\, 1$.

\subsubsection{Zeros for two- and three-spin
 interactions on triangular lattice}
\label{zerostriang}

We now return to  the 
$\, q$-state Potts model on the triangular lattice with
two- and three-site interactions in 
alternate triangular faces~\cite{w74} (section (\ref{invduacrit})).
%It is noteworthy  that, in terms of the variables
%\begin{eqnarray}
%\label{wyq1w}
%w\, =\, y/q\, =\, 1/w', \hskip 1cm u_i\, =\, 
%\, (x_i-1)/\sqrt w\, =\, u_i', 
%\end{eqnarray}
%the duality relation (\ref{Zdefi}) (see section 
%(\ref{invduacrit}) above) takes the simpler form: 
%\begin{eqnarray}
%\tilde Z(u_1,u_2,u_3,w) \, \, \sim \, \, \, w^N \cdot \tilde Z(u_1,
%u_2, u_3, 1/w) \nonumber
%\end{eqnarray}
% for $\, q \ge 2$ in which the variables $\, u_i\, $ are invariant.
The partition function is:
\begin{eqnarray}
\label{GWG}
Z(x,x_1,x_2,x_3) \,\, &=& \sum_GW(G)\,\, , \qquad  \hbox{where:} \nonumber \\
  W(G) \, &=& \,  \,
\prod _\Delta(1+v\delta_{abc})
(1+v_1\delta_{bc})(1+v_2\delta_{ca})(1+v_3\delta_{ab}) \nonumber \\
\qquad  {\rm and} &&\qquad \quad 
  v\, =\, \, e^K-1, \qquad \quad v_i 
\, =\, \, e^{K_i} -1, \nonumber
\end{eqnarray}
and the product is taken over all up-pointing triangles.
It is convenient to represent terms in the expansion of the partition function
by graphs $G$ in which the up-pointing 
triangular faces are either occupied by a solid triangle with a fugacity
$\, v$, or unoccupied.

We next evaluate the weight $\, W(G)\, $ 
associated with the graph $\, G$.
It is clear that each solid triangle contributes
 a factor $\, v$ to $\, W(G)$,
and each bond a factor $\, v_i$.
In addition, by including the associated bond factors, 
each solid triangle 
contributes   an
additional factor $\, (1+v_1)(1+v_2)(1+v_3) \, =$
$ \, e^{K_1+K_2+K_3}$.  
Consider next the $\, q\, $ dependence of $\, W(G)$.
For the graph  representing
$N$ isolated points we have simply $\, W(G)\, = \, q^N$.  
For other graphs, each 
triangle reduces the factor $\, q^N\, $ by
 $\, q^2$, and each bond by $q$. 
But whenever the triangles and bonds close up to 
form a circuit,\footnote{Here, we
use the term {\it circuit} in the topological sense that solid triangles
can be regarded as stars  having three branches, each of which
can be connected to other triangles and bonds.}
this restores a factor $\, q$ due to
 the overlapping of one lattice site summation.
Thus we have:
\begin{eqnarray}
\label{Zv1v2v3}
Z\, = \,\,  \, \, q^N \sum_G 
\biggl[ {{v}\over {q^2} } e^{K_1+K_2+K_3}\biggr] ^{m}
\biggl[{{v_1}\over q}\biggr]^{b_1}
\biggl[{{v_2}\over q}\biggr]^{b_2}
\biggl[{{v_3}\over q}\biggr]^{b_3} q^{c} 
\end{eqnarray}
where the summation is over the $2^N$ graphs $G$,
$\, m\, $ is the number of solid 
triangles, $\, b_i\, $ the number of 
bonds with weight $\, v_i$,
and $\, c(G)\, $  the number of independent circuits in $G$.
The expression (\ref{Zv1v2v3}) generates 
the high-temperature expansion of the partition function. 

In the case 
of pure three-site interactions, (\ref{Zv1v2v3}) reduces to
\begin{eqnarray}
\label{wqqcG}
 Z\, =\,\, \,  q^N\sum_G \, (w/q)^{m(G)}\,  q^{c(G)},
 \quad  \quad 
\hbox{where } \quad \quad   w\, =\,\,  (e^K-1)/q.
 \quad \quad \nonumber
\end{eqnarray}
%For pure two-site interactions and $\, q\, \geq\,  4$,
%the critical variety  can also be obtained by applying the Lee-Yang circle
%theorem in a vertex model formulation of the model,
%or  by applying an inversion relation  argument.  
For pure three-site interactions, the partition
function is self-dual and the critical 
variety assumes the simple form
$\, w\, =\, 1 $.
% or $\,  e^K-1\, =\, q$. 
%For finite lattices (a twisted $\, 27$-sites triangular 
%lattice and a 36-sites lattice with standard
% periodic boundary conditions) an exact enumeration
%gave the partition function as a polynomial for the pure three-site
%interaction model. For instance 
%for the twisted $\, 27$-sites triangular 
%lattice:
%\begin{eqnarray}
%&&Z(q, \, w) \, = \, q^{27} \, +\,  27\, q^{26}\, + \, 351\, q^{25} \,
%w^2 \, + (2898\;  q^{24}\, +\, 27\; q^{25})\, w^3\, \nonumber \\
%&& + \, \cdots  \,  
%\,  + (2898\;  q^{25}\, +\, 27\; q^{26})\, w^{24}\, + \, 351\;
%q^{26}\, w^{25} \,
%+ \, 27\, q^{27}\,w^{26}\, + \, q^{28}\,w^{27}
%\nonumber
%\end{eqnarray}
On the basis of a reciprocal symmetry 
 and  numerical results, Wu {\em et al.}~\cite{w162}
 conjectured that zeros
for the three-site Potts model (in up-pointing triangles)
lie in the thermodynamic limit
   on the {\em unit  circle} 
$ \, \vert w\vert\,  = \, 1 \, $, as well as
a  line segment on the real negative axis.
%\begin{eqnarray}
%\, - \alpha(q)\, \, \leq \,\,  w \, \, \leq \,\,
%  -1/\alpha(q). \nonumber
%\end{eqnarray}
 %This conjecture holds
%for $\, q\, =\, 1\, $ and 2, and that it reproduces 
%the known critical point for general $\, q\, $
%including  $\, q\, =1$, corresponding to
% site percolation.

\subsubsection{Density of Fisher zeros for the Ising model}

One trademark of F. Y. Wu's research is that he often looks at old
problems and finds new life or new solution that others have not
previously seen.  A good example of a new look at an old problem is
the density of Fisher zeros for the Ising model, which is the $q=2$ Potts model.

In 1964 Fisher pointed out that in
the thermodynamic limit the zeros of the Ising partition function for
a square lattice lie on two circles, now known as the Fisher circles, in
the complex $\tanh K$ plane,  where $K$ is the nearest-neighbor interaction. 
  However, Fisher had not made the argument rigorous and, furthermore,
no one has bothered to look into the distribution of the zeros
on the circles,
except at small angles which dictates the Ising critical behavior.

Both of these two deficiencies have been rectified by W. T. Lu and F. Y. Wu.
First, by considering zeros of the Fisher zeros for the Ising model
on finite self-dual lattices, Lu and Wu~\cite{w179} established 
rigorously that, indeed, the Fisher zeros approach two circles in the
thermodynamic limit.  In a subsequent paper published in 2000~\cite{w191}, they 
deduced 
the close-form expression for the density of the Fisher zeros for many regular
two-dimensional lattices, thus completing the story of the Fisher zeros 
 some 25 years after it was proposed!

\subsection{Duality relation for Potts correlation functions }
\label{correl}

\subsubsection{Correlation dualities}
\label{cordua}
Duality considerations are not  often applied  to  
 correlation functions~\cite{w127}. But
F. Y. Wu initiated a new method for generating duality relations
for correlation functions of the Potts model
 on  planar graphs. 
%The 
%method extends previously known results, by
%allowing the consideration of the correlation function
%for arbitrarily placed vertices on the graph. 
In a pioneering paper~\cite{w172}
he obtained duality relations for 2- and 3-point correlation functions,
for  spins residing on the boundary of a lattice.
The consideration was soon extended to 
$n$-point correlations~\cite{w176,w178} and to
the case where the spins reside on 
2 or more faces in the interior of the lattice~\cite{w194}.
A
graph-theoretical formulation of the results in terms of
rooted Tutte polynomials (see section (\ref{Tutte}) below) was
 also given~\cite{w175,w176}. In addition,
C. King and Wu~\cite{w194} showed that
generally it is linear combinations of correlation functions,
not the individual correlations, that are related by dualities. 
%The method is illustrated in several non-trivial cases, and the
%relation to earlier results is explained. 

%F. Y. Wu and coworkers obtained new  sum rule
%identities for  $\,n$-point correlation functions
%of the  Potts on {\em planar lattices}
%using duality relations extensively as well as graph theoretical
% arguments~\cite{w175,w176}. 
%It was shown that certain sum rule identities  exist which
%relate  correlation functions for $\, n$ Potts spins  on the boundary of
%a planar lattice  for $\, n\, \geq \, 4$.
%  Explicit expressions of the identities are obtained for $\, n\, =\, 4,5$.
%It was also shown  that the
% identities provide the missing link needed for
%a complete determination of  the duality
%relation for the  $\, n$-point correlation function.
% The $\, n\, =\, 4\, $ duality relation 
%is obtained explicitly. More generally we deduce the
%number of correlation identities for any  $\, n$ as well as an 
%inversion relations and a conjecture as to the general form of
%the duality relation.
%Let us just give one of these relations
% for $\, n\, =\, 4$, for example:
%\begin{eqnarray}
%\label{1111}
%&& {q^2 \over {C}} \cdot  Z_{1111} \, = \, \, 
% Z^*_{1111}+p_1(Z^*_{2111}+Z^*_{1211}+Z^*_{1121}+Z^*_{1112}) 
%\nonumber \\
%&& \qquad \qquad +p_1( Z^*_{1212}+p_2Z^*_{1213}\, 
%+p_2Z^*_{2131} +p_2p_3Z^*_{1234}) +p_1(Z^*_{1122}+Z^*_{1221})
% \nonumber \\
%&& \qquad \qquad +p_1\, p_2 \, (Z^*_{1123}\, 
%+Z^*_{2113}+Z^*_{2311}+Z^*_{1231} ). \nonumber
%\end{eqnarray}
%where $\, p_m\, =\, q-m,\, \,  m\, =\, 1,\, 2,\, 3$.

\subsubsection{Correlation function as rooted Tutte Polynomials}
\label{Tutte}
 
As previously mentioned,  the Potts partition function is
also the Whitney-Tutte polynomial, or the Tutte polynomial for brief, 
considered in graph theory.
In one further step,
F. Y. Wu, C. King and W. L. Lu 
formulated the Potts correlation function as
 a rooted Tutte polynomial~\cite{w182}.
 
In graph theory  
 a vertex is rooted 
if it is colored with a prescribed (fixed) color and
a graph is rooted if it contains a rooted vertex.
Interpreting the color of a site (vertex) as spin states, then the
 Potts correlation functions 
for which the spin states of  given sites are fixed can naturally be formulated 
as rooted-Tutte polynomials.
This is the basis of their graph-theoretical formulation of the
Potts correlation function, from which  duality relations of the
Potts correlations become transparent and can be analyzed~\cite{w182}.

%More generally, the rooted Tutte polynomial arises in statistical physics
%as the correlation function of the Potts model. 
%It has been seen in the previous 
%section (\ref{correl}) that F. Y. Wu established a
%duality relation for the Potts correlation 
%function for {\em planar} graphs~\cite{w178}.

\subsection{Potts model and graph theory  }
\label{Pottsgraph}

F. Y. Wu has written several review papers dedicated to the
 analysis of the Potts model from a  graph 
theoretical viewpoint~\cite{w53,w66bis,w112,w116bis}.
%(Potts model and graph theory, {\it J. Stat.   Phys.}
%     {\bf 52}, 99-112  (1988),  Potts model and graph theory, in Progress in
%Statistical Mechanics, Ed C. K. Hu, World Scientific Singapore, (1988)).
The most important one is the
classic paper~\cite{w53} written in collaboration with Baxter and Kelland
alluded to above, which
 %(Equivalence of the Potts model 
% or Whitney polynomial with the ice-type model: A new derivation) written 
%with  R. J. Baxter and S. B. Kelland, which
 gives the graphical construction of the 
equivalence of the partition function 
of the Potts model
%, or {\em Whitney polynomial}, 
with an ice-type model~\cite{w53}. This derivation  simplifies
%and is easier to follow than 
the algebraic method of Temperley and Lieb
which is based on the Temperley-Lieb algebra\footnote{A matrix representation 
of the $\, U_{i,i \pm 1}\, $ 
is for instance the 
$\, q^n \times q^n$ matrices with entries 
$\, q^{-1/2} \, \prod_{j=1;\;j\ne \;i}^{n}\delta(\sigma_j,\,
\sigma_j')\,\,\, $ and $\,\,\, q^{1/2} \, \delta(\sigma_i,\, \sigma_{i+1}) \, 
\prod_{j=1}^{n} \delta(\sigma_j,\, \sigma_j')$.
}:
% Let us recall briefly that
%this method introduces the  {\em Lieb-Temperley algebra}:
\begin{eqnarray}
\label{Temper}
&&U_{i,i+1}^2 \, = \, \, \sqrt{q} \cdot U_{i,i+1}, \nonumber \\
&&U_{i,i-1}\cdot U_{i,i+1}\cdot U_{i,i-1}\, = \, \,U_{i,i-1}, \qquad \quad 
U_{i,i+1}\cdot U_{i,i-1}\cdot U_{i,i+1}\, = \, \,U_{i,i+1},   \nonumber \\
&& U_{i,i+1}\cdot U_{j,j+1} \, = \, \,
  U_{j,j+1}\cdot  U_{i,i+1}\quad
\quad 
\hbox{if }\quad  \quad  |i-j| > 3, \nonumber
\end{eqnarray}
and applies to an arbitrary planar graph.  
This then opens the door
for analyzing the triangular and honeycomb Potts models.

Another important graphical analysis of a Potts model is the
  joint work
with J. H. H. Perk~\cite{w98,w99} 
%(Non-intersecting string model and graphical
%approach:  Equivalence with a Potts model; Graphical
% approach to the non-intersecting 
%string  model:  Star-triangle equation, inversion
% relation and exact solution), 
on the  non-intersecting string (NIS)
model of Stroganov and Schultz, the close-packed loop model.
The NIS model formulated by Perk and Wu in~\cite{w98} 
turns out to be nothing but  the bracket polynomial introduced
by L. H. Kauffman in his state-model formulation of knot invariants.
This fact
 offers a most natural approach to knot invariants from a
statistical mechanical viewpoint~\cite{w147}.

\section{Other miscellaneous topics}
%\section{Graph theory and enumerative combinatorics}
\label{Enumer}

F. Y. Wu has worked on a diverse array of topics in mathematics 
and mathematical physics.   
In this section we present a random choice of topics that 
are not included above.
 
%A large number of F. Y. Wu's papers correspond
% to graph-theoretical
%methods in lattice statistical mechanics. On this topic
%one can read several review papers by 
%F. Y. Wu~\cite{w66bis,w116bis} 
%(Graph Theory in Statistical Physics, in
%Studies in Foundations of Combinatorics, 
%Potts model and graph theory).

%F. Y. Wu obtained many now well-known results
%in graph theory or enumerative combinatorics.
%One should certainly mention, closely related
%to the Potts model, many results
%on percolation~\cite{w59,w60}
% (H.  Kunz and  F. Y. Wu,  
%Site percolation as a Potts model, 
%{\it J. Phys.} {\bf C11}, L1-L4 (1978),
%F. Y. Wu,  Percolation and the Potts model,  
%{\it J. Stat. Phys.} {\bf 18}, 115-123 (1978)),
%and more specifically graph problems like the aforementioned
%counting of the number of {\em spanning trees}~\cite{w58,w189}. 

\subsection{Topics in graph theory}
F. Y. Wu is fond of graphs and has made many contributions to graph theory.
A fine example is the
 aforementioned introduction of the rooted Tutte polynomial.
 Even when  Wu does not obtain 
new results, he  tries to provide
simpler derivations, or find new 
consequences of known results. A 
good example is his work on random graphs~\cite{w87}.

Random graphs is a topic well-known to graph theorists after the work by 
 Erd\"os and Renyi who introduced the problem in 1960.
In the simplest formulation each pair of points of a set
of $N$ can be connected (by a bond, say) with
a probability $\alpha/N$, where $\alpha$ is a constant.
One then asks  questions such as what is the mean 
cluster size and the probability $P(\alpha)$
that the
set becomes fragmented, namely not connected, etc.

Using a Potts model formulation, Wu~\cite{w88} reproduced
the Erd\"os-Renyi result in just a few steps, showing
 that a transition occurs
at $\alpha_c=1$ and computed the critical exponents as well as
the  mean cluster size at criticality.
This work has drawn considerable attention from graph theorists.
 Wu has also applied the result to evaluate the reliability 
probability
of a communication network~\cite{w87}.

As another example of Wu's work in graph theory, one can mention
his paper on the Temperley-Nagle identity
for graph embeddings~\cite{w65} where   
he provides a simple derivation of the Temperley-Nagle
identity
\begin{eqnarray}
\sum_{G} \; x^{v}\; y^{l} \, \, = \, \, \, \, 
(1+x)^N \, \sum_{L} \; (y-1)^{l} \, \Bigl( {{x} \over {1+x}} \Bigr)^{v}
\nonumber
\end{eqnarray}
where  $G\, $ denotes  section graphs of the original 
graph of $\, v\, $ vertices and $\, l\, $ lines, and $\, L$ is  a line 
set of $\, G\, $ containing $\, l\, $ lines 
 covering $\, v\, $ vertices.  
Using a weak-graph expansion he also deduced 
a sum-rule relation connecting the lattice constants
of weak and strong embeddings. 
%He also  
%implication of the Temperley-Nagle
%identity, namely that it yields the fluctuation 
%of the number of bonds (resp. sites) in a site (resp. bond)
%percolation problem. 

\subsection{The vicious neighbor problem}
%Among many, F. Y. Wu also obtained results 
%on the vicious neighbor problem~\cite{w106}, on some
%restricted random walks on graphs~\cite{w181},
% on some lattice animal problem~\cite{w163} related with
% three-dimensional enumerations 
%in relation to some arithmetic problems 
%(restricted partition of an integer~\cite{w166}). 
%Let us sketch briefly some of these results. 
 \label{viciouspar}
 
Consider $\, N$ points randomly distributed in a bounded
$\, d$-dimensional space. At a given instance, each point
destroys his nearest neighbor (vicious neighbors)
with a probability $\, p$. What
 is the probability $\,P_N(p)$ that a given point
will survive in the $\, N \, \longrightarrow \, \infty\, $ limit?

The $d=2,\, p=1$ version of this problem was first posed by the 
Brandeis mathematician R. Abilock in
American Mathematical Monthly in 1967,
 which remained unsolved for almost two decades.
In 1986 the {\em Omni} magazine posted a prize
for its correct solution, and R. Tao and Wu claimed the prize
by publishing the solution for general $d$ and $p$ in 1987~\cite{w106}.

The idea of their solution is very simple.  In $d$ dimensions
a given point can be killed by at most a finite number $n_d$ of other points.
In two dimensions, for example, the number is 
$n_2=5$  (theoretically 
a point can also be killed by 6 other points but the phase
space for that to happen has a zero measure).  Therefore, one  computes the
volumes of the phase space for a point to be killed  by 1, 2, $\cdots$,
$n_d$ neighbors, and uses the inclusion-exclusion principle to write
the probability in question as an alternate series in $p$, whose highest power 
is $n_d$.  However, the evaluation of the volumes of the phase space
is tedious requiring special techniques.
 
For $d=1$   the result is quite simple and one has
$ \, P_{\infty}(p)\,= 1-p+p^2/2 $.  
For $d=2$ the result is:
\begin{equation}
\label{vin}
P_{\infty}(p)\,=1-p+ 0.316\ 3335 \, p^2
-0.032\ 9390\,  p^3+0.000\ 6575\, p^4 -0.000\ 0010 \, p^5
\end{equation}
where the coefficient of each term is evaluated from 
integrals which can in principle be computed to any numerical accuracy.
The coefficient of the last term, for example, is obtained by
combining two 8-fold integrations.

For $\,p=1\,$, (\ref{vin}) yields $\,P_{\infty}(1)=0.284\ 051 ...$,
a solution which   claimed the {\em Omni} prize.
Tao and Wu also carried out Monte Carlo simulations to obtain
the solution for $d=3,4,5$.

 %\begin{eqnarray}
%\label{vicious}
%I(1, \, 1, \, 0) \, = \, \, {{2\; \pi} \over {3}} \cdot
%\int_{\pi/3}^{\pi/2} \, d\theta_1 \, 
%\int_{\pi/3}^{\pi/2} \, d\theta_2 \, 
%\int_{2\, \cos(\theta_1)}^{1/cos(2\, \theta_1)} 
%\,t_2^3\cdot  dt_2 \, 
%\int_{2\, \cos(\theta_2)}^{1/cos(2\, \theta_2)} \,
% {{ t_3} \over {V_3}} \cdot  dt_3 \, \nonumber
%\end{eqnarray}
%where $\, V_3\, $ and, more generally, expressions like  $\, V_n$
%read:
%\begin{eqnarray}
%\label{Vnzn}
%&&V_n\, = \, \, z_1 \, + \, t_2^2 \, z_2 \, 
%+ \, \,  t_2^2 \,t_3^2 \, z_3 \,
%\cdots \, + \, \,(t_2^2\, t_3^3 \cdots  t_n^2) \, z_n, \qquad
%\hbox{where } \nonumber \\
%&&\qquad z_n \, = \, \pi \, -\alpha_n \, +\, {{1} \over
%{2}} \cdot \sin(2\; \alpha_n) \, -\, \beta_{n-1} \,  +\, {{1} \over
%{2}} \cdot \sin(2\; \beta_{n-1})\nonumber
%\end{eqnarray}
 
\subsection{Counting partitions: from Potts to three-dimensional enumeration
and beyond}
\label{3Denumer}

F. Y. Wu {\em et al.}~\cite{w163} considered a directed  lattice animal
problem on the $\, d$-dimensional hypercubic lattice, and established its
equivalence first with the infinite-range Potts model and,
in a second step, with the enumeration of $\, (d-1)$-dimensional
restricted partitions of an integer. The directed compact lattice
animal problem was solved exactly in two and three dimensions, using
known results in number theory. They found that the number of
lattice animals of $n$ sites grows as:
\begin{eqnarray}
\label{expo}
 \, \exp( c \cdot n^{(d-1)/d}).  \nonumber 
\end{eqnarray}
Furthermore, the infinite-state Potts model solution leads to a
conjectured limiting form for the generating function of restricted
partitions for $\, d >\, 3$, which is a long-standing  
unsolved problem in number theory.

Let us denote by $ \, A_n(L_1, \, L_2, \, \cdots\,  L_d)$ 
the number of $n$-site animals that can grow on an
$L_1 \times L_2 \times \cdots\times  L_d$ 
 lattice.
 %\begin{eqnarray}
%%&n \, = \, \, \sum_{n_1\, = \, L_1}^{L_1}
 %\sum_{n_2\, = \, L_2}^{L_2}
%\cdots \sum_{n_d\, = \, L_d}^{L_d} \, m(n_1, \,n_2, 
%\,\cdots n_{d-1}),\nonumber  \\
%&&\qquad \qquad \qquad  \qquad \hbox{with:} \quad \quad \qquad 
%m(n_1, \,n_2, \,\cdots \, n_{d-1} \, > \, 0 \nonumber 
%\end{eqnarray}
F. Y. Wu {\em et al.} showed that $A_n$ is precisely the number of 
$(d-1)$-dimensional restricted partitions
of the integer $n$ into non-negative parts
to units of a hypercube of size  $L_1 \times L_2 \times \cdots\times  L_{d-1}$,
with the size of each part at most $L_d$.

Define the generating function
 \begin{eqnarray}
G(L_1, \, L_2, \, L_3, \, \cdots\,  L_d; \, t) 
\,  \,= \, \, \, 
1 \, + \, 
\sum_{n\, = \, 1}^{L_1 \cdot L_2\cdot  L_3 \cdots L_d}
\; A_n(L_1, \, L_2, \, \cdots\,  L_d) \cdot t^n, \nonumber 
\end{eqnarray}
which is of interest in number theory.
F. Y. Wu {\em et al.} showed that $G$ 
 is precisely the  partition function of a Potts model 
on the $d$-dimensional lattice in the infinite-state limit,
  provided one identifies $\, t $
 with $\, x^d$ where  $\, x
\, = \, (e^K \, -1)/q^{1/d}$.
This then connects the Potts model with the theory of partitions in number theory.

For $\, d\, = \, 2\, $ 
the generating function corresponding to
 the square lattice $(L_1, \, L_2)\, $ reads\footnote{Let us recall
the classic analysis
 due to Rademacher which yields 
the celebrated Hardy-Ramanujan asymptotic result:
 $\, A_n \,\, \simeq \,\, 1/(4\;n\;\sqrt{3}) \cdot
\exp(\pi \sqrt{ 2\;n/3})$. One has the following 
asymptotic behavior 
for  $\, A_n(L_1, \, L_2)\, $ 
when $\, L_1 \, = \, L_2\, $:
$\, A_n(L, \, L)\, \simeq \, \,  \,
 (\sqrt{3}/(2\, \pi \, n)) 
\cdot 2^{2 \, \sqrt{2n}}$.
}:
\begin{eqnarray}
\label{gauss}
G(L_1, \, L_2;\, t) \, = \, \, {{ (t)_{L_1\, + \, L_2} }
 \over { (t)_{L_1} \cdot  (t)_{L_2} 
}}.
\end{eqnarray}
where $\, (t)_p \, = \, \,  \prod_{q \, = \, 1}^{p}\; (1-t^q)$.
%Note that, despite its appearance, all :zeros 
%in the denominator cancel: 
$ \, G(L_1, \, L_2) \,$ 
is a polynomial in $\, t$ also known as the
Gaussian polynomial or the ``$q$-coefficient''. 
%\quad \quad  n \, < \, \min(L_1, \, L_2) 
%\nonumber \\
%%\end{eqnarray}
%giving the asymptotic result:
%\begin{eqnarray}
%\label{An}
% A_n(L_1, \, L_2)\, \,&  \simeq &\,\, 
% \exp\bigl(c\, ({{ L_1}/L_2)  } \sqrt{n}\, \bigr),
%\quad \quad \quad  n \,
%\simeq \, \, {{L_1\cdot L_2} \over {2}}
% \nonumber 
%\end{eqnarray}
%with 
%\begin{eqnarray}
%c(u)\, = \, \, \sqrt{2} \cdot 
%\Bigl(\sqrt{1/u} \cdot \ln(1+u) \, + \, \sqrt{u} \cdot
%\ln(1+1/u)\Bigr). \nonumber 
%\end{eqnarray}
%One has also he following asymptotic behavior 
%of  $\, A_n(L_1, \, L_2)\, $ 
%when $\, L_1 \, = \, L_2\, $:
%\begin{eqnarray}
%A_n(L, \, L)\, \simeq \, \,  \,
%\Bigl( {{\sqrt{3} } \over {2\, \pi \, n}} \Bigr) 
%\cdot 2^{2 \, \sqrt{2n}} \nonumber 
%\end{eqnarray}
For $\, d\, = \, 3\, $ 
the generating function 
%corresponding to
 %the cubic lattice of size $\, L_1 \times L_2 \times L_3\, $ 
reads:
\begin{eqnarray}
\label{gauss3}
G(L_1, \, L_2, \, L_3;\, t) \, = \, \,\, {{ 
[t]_{L_1+ \, L_2+ \, L_3\, } \cdot [t]_{L_1\, }
\cdot [t]_{L_2\, }\cdot [t]_{L_3\, }
} \over { [t]_{L_1\,+ \, L_2\, }\cdot
[t]_{L_2\,+ \, L_3\, }\cdot [t]_{L_3\, \, +L_1 \, }
}},
\end{eqnarray}
where $\,[t]_L \, $ denotes:
\begin{eqnarray}
\label{tL}
[t]_L \, = \, \, \prod_{p \, = \, 1}^{L-1}\; (t)_p \, ,\quad L\,  >\,  1,
 \quad \quad \quad
(t)_p \, = \, \,  \prod_{q \, = \, 1}^{p}\; (1-t^q). 
\end{eqnarray}

For $d=4$
$\, G(L_1, \, L_2, \, L_3, \, L_4;\, t)\,$ is 
 the generating function of restricted solid
partitions of a positive integer into parts on a
 $\, L_1 \times  L_2 \times  L_3\, $ cubic lattice, 
with each part no greater than $\, L_4$.  The evaluation of a closed-form
expression for $G$ in this case has remained an unsolved problem for almost
a century.

The expression which straightforwardly generalizes (\ref{gauss}) and
(\ref{gauss3}) would be:
\begin{eqnarray}
\label{gauss4}
&&G_{\rm straight}(L_1, \, L_2, \, L_3, \, L_4; t) \, = 
\, \, {{ N_G(t)} \over {D_G(t)}},
\quad \quad \quad\hbox{with:}  \\
&&  N_G(t)\, = \, \,  
     \{t\}_{L_1+ \, L_2+ \, L_3\,+ \, L_4} 
     \cdot \{t\}_{L_1\,+ \, L_2 } \cdot
     \{t\}_{L_1\,+ \, L_3 } 
      \cdot \{t\}_{L_1\,+ \, L_4 } \nonumber  \\
     &&\qquad \qquad 
     \cdot \{t\}_{L_2\,+ \, L_3 }\cdot \{t\}_{L_2\,+ \, L_4 } \cdot
     \{t\}_{L_3\,+ \, L_4 } \nonumber  \\
     &&D_G(t) \, = \, \,   \{t\}_{L_1\,+ \, L_2\,+ \, L_3 }\cdot
     \{t\}_{L_2\,+ \, L_3\,+ \, L_4 }\cdot \{t\}_{L_1\,  +L_3 \,+ \, L_4 } 
     \cdot \{t\}_{L_1\,  +L_2 \,+ \, L_4 }\nonumber  \\
     &&\qquad \qquad 
     \cdot \{t\}_{L_1} \cdot \{t\}_{L_2} \cdot \{t\}_{L_3} \cdot \{t\}_{L_4}
     \nonumber 
    \end{eqnarray}
   where:
  \begin{eqnarray}
  \label{tLL}
  \{t\}_L \, = \, \, \prod_{p \, = \, 1}^{L-1}\; [t]_p \, ,\quad L \,  > \, 2
  \end{eqnarray}
%[[t]]_{L_1+ \, L_2+ \, L_3\,+ \, L_4} 
%\cdot [[t]]_{L_1\,+ \, L_2 } \cdot
%[[t]]_{L_1\,+ \, L_3 } 
% \cdot [[t]]_{L_1\,+ \, L_4 } \nonumber  \\
%&&\qquad \qquad 
%\cdot [[t]]_{L_2\,+ \, L_3 }\cdot [[t]]_{L_2\,+ \, L_4 } \cdot
%[[t]]_{L_3\,+ \, L_4 } \nonumber  \\
%&&D_G(t) \, = \, \,   [[t]]_{L_1\,+ \, L_2\,+ \, L_3 }\cdot
%[[t]]_{L_2\,+ \, L_3\,+ \, L_4 }\cdot [[t]]_{L_1\,  +L_3 \,+ \, L_4 } 
%\cdot [[t]]_{L_1\,  +L_2 \,+ \, L_4 }\nonumber  \\
%&&\qquad \qquad 
%\cdot [[t]]_{L_1} \cdot [[t]]_{L_2} \cdot [[t]]_{L_3} \cdot [[t]]_{L_4}
%\nonumber 
%\end{eqnarray}
%where:
%\begin{eqnarray}
%\label{tLL}
%[[t]]_L \, = \, \, \prod_{p \, = \, 1}^{L-1}\; [t]_p \, ,\quad L \,  > \, 2
%\end{eqnarray}
%The explicit expression of $\,  G(L_1, \, L_2, \, L_3, \, L_4;\, t)\,$
%is not known for general  $\, L_1, \, L_2, \, L_3, \, L_4$. However
%for  $\, L_1\, =  \, L_2\,  = \, L_3 \, = \, 2\, $
But the explicit expression of $G(2, \, 2, \, 2, \, L_4;\, t)$ obtained by Major P. A.
MacMahon in 
%(P. A. MacMahon, {\em Combinatorial Analysis, Vol. 2}, Cambridge,
1916 is
\begin{equation}
G(2, \, 2, \, 2, \, L_4;\, t) = \, G_{\rm straight}(2, \, 2, \, 2, \, L_4;\, t)
 + C(2, \, 2, \, 2, \, L_4;\, t) \nonumber
\end{equation}
where
\begin{eqnarray}
\label{MacMahon}
&&G(2, \, 2, \, 2, \, L_4;\, t)\, = \, \,\,  \, 
\sum_{i=0}^{4} g_i \cdot {{ (t)_{L_4+8-i} } \over {(t)_{8}
\cdot (t)_{L_4-i}  }}, \quad
 \quad \quad \hbox{with:} \quad \quad
\quad \nonumber \\
&&g_0\, = \, \, 1, \quad \quad \quad g_1 \, = \, \, 2\, t^2 \cdot
(1+t+t^2+t^3+t^4)\, + t^4 , \quad \quad \quad  \nonumber \\
&& g_2 \, = \, \, t^5 \cdot (1+3\, t\, +\,4\,
t^2\, +\, 8 \, t^3\,  +\,4\, t^4\,+3\, t^5\, +\, t^6) \nonumber \\
&& g_3\, = \, \, 2\, t^{10} \cdot (1+t+t^2+t^3+t^4) \, +t^{12}  ,
\quad \quad  \quad  g_4\, = \, \, t^{16},\quad \quad \quad   \nonumber
\end{eqnarray}
%with $\, (t)_L\, $ given by (\ref{tL}).
and
%
%The exact expression $\, G(2, \, 2, \, 2, \, L_4)$
%does not coincide with the straightforward generalization
%$\,  G_{\rm straight}\, $ given in (\ref{gauss4}) 
%to which one must introduce an
%additive correction $\,C(2, \, 2, \, 2,
% \, L_4; \, t)\,$ assuming the nice form:
\begin{eqnarray}
C(2, \, 2, \, 2, \, P;\, t)\, = \, \, 
-\Bigl({{ t^6\cdot (t+1)^2 \cdot (t^4-2\;t^3+t^2-2\;t+1) } \over 
{ t^2+t+1}}
\Bigr) \cdot \Bigl(
{{ (t)_{P+6} } \over { (t)_8 \, (t)_{P-2}}}
\Bigr) \nonumber 
\end{eqnarray}

%A fruitful approach to many enumeration problems is to look at 
%the :zeros of the generating functions, especially in these cases where
%Eulerian products like (\ref{tL}) are floating around. It is clear 
%that all the $\, L_1\cdot L_2 \cdot L_3 \cdot L_4 \, + \,
%1\, $ :zeros of the straightforward generalization
%$\,  G_{\rm straight}(L_1, \, L_2, \, L_3, \, L_4; t)\, $
%are on the unit circle $\, |t| \, = \, 1$.
H. Y. Huang and F. Y. Wu~\cite{w169}
decided to look into the zeros of  
the generating function
 $\, G(2, \, 2, \, 2, \, L_4;\, t)$
for various increasing values of $\, L_4$.
They found that the zeros
  are not exactly on the unit circle but seem to converge to
the  unit circle as $\, L_4$ increases.
This indicates that a multiplicative correction 
$\,C_{\rm mult}(2, \, 2, \, 2, \,L_4;\, t)\,=\, $
$\,G(L_1, \, L_2, \, L_3, \, L_4; t)/G_{\rm straight}(L_1,
 \, L_2, \, L_3, \, L_4; t) $
would not have any simple Eulerian
product form as in (\ref{gauss4}) and
(\ref{tLL}):
\begin{eqnarray}
\label{forms}
C_{\rm mult}(2, \, 2, \, 2, \,L_4;\, t)\,= \,\,\, 
\prod_{n=1} (1-t^n)^{\alpha_n} 
\end{eqnarray}
where $\,\alpha_n$ are positive integers, since these product forms
(\ref{forms}) would necessarily yield zeros on the unit circle. 
H. Y. Huang and F. Y. Wu 
conjectured however, on the basis on their numerical
results, that the zeros tends to be on the unit circle in
the limit when any one of $\, 
 L_1, \, L_2, \, L_3, \, 
L_4 \rightarrow \, \infty$.

\subsubsection{Directed percolation and random walk problems}
\label{Restric}

F. Y. Wu and H. E. Stanley~\cite{w85} have considered a
directed percolation problem on square and triangular lattices
in which the occupation probability
is unity along one spatial direction. They
formulated the problem as a random walk, and evaluated in closed-form the
percolation probability, or the arriving probability of a walker.   
To this date this solution stands as the only exactly solved
model of directed percolation.

In another random walk problem  Wu and H. Kunz~\cite{w181}
  considered {\em restricted random walks} on graphs, which
 keep track of the number of immediate reversal steps, 
 by using a transfer matrix formulation. A closed-form 
expression was obtained for the number of
$\, n$-step walks with $\, r$ immediate reversals for any graph. In the
 case of graphs of a uniform valence, they established  a
probabilistic meaning of the formulation, and deduced explicit expressions for the
 generating function in terms of the eigenvalues of the adjacency matrix.

\section{Knot theory}
\label{knot}

The connection between knot theory and statistical mechanics
was probably first discovered by Jones. His derivation
of the V. Jones polynomial reflects the resemblance of the von Neumann
algebra he uses with the Lieb-Temperley 
algebra occurring in the Potts model (see section (\ref{Pottsgraph})).
 The direct connection 
came to light when L. Kaufman produced a simple 
derivation of the Jones polynomial using the very diagrammatic  
formulation of the non-intersecting string (NIS) model of J. H. H. Perk and
F. Y. Wu~\cite{w98,w99}. Soon thereafter Jones worked out a derivation 
of the Homfly polynomial using a vertex-model approach.
The connection between knot theory and 
lattice statistical mechanics
was further extended by Jones to include spin and IRF models.
 
F. Y. Wu has written several 
 papers on the connection
 between knot theory and statistical mechanics~\cite{w143,w144,w147}
including the comprehensive review~\cite{w143}.
	In hindsight, knot invariants arose naturally 
in statistical mechanics even before the connection with
solvable models was discovered.   
In their joint paper~\cite{w98}, for example,
 J. H. H. Perk and  F. Y. Wu
described a version of an NIS model which is 
  precisely the
bracket polynomial of L. Kauffman.   Similarly,
%This was clear, even before Kauffman,
%when one considers the non-intersecting-string model (NIS) 
the $q$-color  NIS model studied by J.H.H. Perk and C. Schultz
%(J.H.H. Perk and C. Schultz, Non-intersecting-string model
%and graphical approach: Equivalence with a Potts model, 
%J. Stat. Phys. (1986) {\bf 42}, 727-742)  
is a 
$\, q \, (q-1)$ vertex model which generates the Homfly polynomial.
Here we briefly describe the latter connection.

The $q$-color NIS model 
has vertex weights ($\delta_{abcd}\, = 1$
if $\, a \, = \, b \, = \, c \, = \, d$ and zero otherwise): 
\begin{eqnarray}
\label{NIS}
&&w(a,\, b, \, c, \, d) \, \,  = \,  \\
&& \quad (W(u)-S(u)-T(u)) \cdot \delta_{abcd}\, + \, S(u)
\cdot \delta_{ab} \, \delta_{cd}\, + \, T(u)
\cdot \delta_{ac} \, \delta_{bd}, \, \nonumber \\
&& \hbox{where:} \qquad \quad
 W(u) \, = \, \, \sinh(u)\, = \, \sinh(\eta \, +u), \,\quad 
 S(u) \, = \, \, \sinh(u), \quad \nonumber \\
&&\quad T(u) \, = \, \, \sinh(\eta \,-\, u), 
\quad \quad \quad q \, = \, e^{\eta}\, +\, e^{-\eta}, \nonumber 
\end{eqnarray}
 and the {\em Homfly polynomial} is a {\em two variable} knot invariant 
polynomial, discovered after Jones' work, by Freyd {\em et al.}
The Homfly polynomial knot invariant 
has since been re-derived and analyzed 
by Jones using the Hecke algebra of the 
braid group. It can also be constructed  from 
the Perk-Schultz NIS model. Actually the partition function
$\, Z(q, \, e^{\eta})\, $ of the NIS model is a
knot invariant related to the Homfly polynomial $\, P(t, z)$:
\begin{eqnarray}
\label{relHomf}
 P(t, z) \, = \, \, {{ \sinh(\eta) } \over {\sinh( q\, \eta)}}
\nonumber
 \cdot Z(q, \, e^{\eta})
\end{eqnarray}

In the infinite rapidity limit this model leads
 to the Jones polynomial.
The Boltzmann weight (\ref{NIS}) of the non-intersecting string model
becomes:
\begin{eqnarray}
w(a,\, b, \, c, \, d) \, = \, 
-\;e^{\pm 2 \; \eta} \delta_{a,b}\, \delta_{c,d} \, 
+ \, e^{\pm  \; \eta} \delta_{b,d} \qquad \hbox{with:} \qquad 
q \, = \, \, e^{ \eta} \, + \,  e^{ -\eta}
\end{eqnarray}
The Jones polynomial $\, V(t)\, $ is then obtained from the Homfly
polynomial $\, P(t, z)$ by taking $\, z \, = \, \sqrt{t} \,
-1/\sqrt{t}$.

F.Y. Wu discussed  many knot invariants in his  review~\cite{w147}:
the Alexander-Conway polynomial,
the Jones polynomial, the  Homfly polynomial,  
the Kaufman polynomial and  
the Akutsu-Wadati polynomial, etc. The 
Alexander-Conway polynomial can be
 obtained from the Homfly polynomial by setting  $\, t\, =\, 1$ 
in the Homfly polynomial $\, P(t,\, z)$. 
%The Jones polynomial 
%can be obtained from the Homfly  polynomial setting by $\, z\,
%=\, \sqrt{t} \, -1/\sqrt{t}$ 
%in the Homfly polynomial $\, P(t,\, z)$.
The Akutsu-Wadati polynomial is an example
 of a {\em new} knot invariant 
derived from exactly solvable models in statistical mechanics.
%(Y. Akutsu and M. Wadati, Exactly solvable
% models and new link polynomials, 1987). 

%shown {\em directly} in~\cite{w147} 
%(page 1126), that the Potts model leads to 
%the Jones polynomial. One key step
%is  that the Skein relation is satisfied provided:
%\begin{eqnarray}
%\label{key}
%&&{{1} \over {t}} \cdot \Bigl( 
%{{1} \over {\alpha}} \cdot W_{-}(a, \, b) \, - t \cdot W_{+}(a, \, b) 
%\Bigr) \, = \, \, \, \sqrt{t} \, - \, {{ 1} \over {\sqrt t}} \nonumber \\
%&&\quad \hbox{where:} \qquad 
% W_{\pm} (a, \, b) \, = \, \, A_{\pm}  \cdot e^{K_{\pm} \delta_{a, \,
%b}}, \qquad \alpha \, = \, -t^{-3/4}, \qquad \hbox{and:} \nonumber \\
%&&\quad 
%q \, = \, t\, + 2 \, + {{1} \over
%{t}}, \quad A_{\pm} \, =\, \, t^{\pm\; 1/4}, \quad 
%t \equiv -e^{-K_{+}}\, \, \, \quad \hbox{or:}\,\quad 
%\, \,  -e^{-K_{-}} \nonumber
%\end{eqnarray}
As our final example of F. Y. Wu's versatility, he and P. Pant and C. King~\cite{w154},
have obtained a new knot invariant using 
  the exactly solvable chiral 
Potts model and a generalized Gaussian summation identity. Starting 
from a general formulation of link
     invariants using edge-interaction spin models, they establish the 
uniqueness of the invariant for self-dual models. They applied the
 formulation to the self-dual chiral Potts
     model, and obtain a {\em link invariant} in the form of a lattice sum
 defined by a matrix associated with the link diagram. A generalized 
Gaussian summation identity was then used
     to carry out this lattice sum, enabling them to cast the invariant 
into a tractable form. The resulting expression for the link invariant
 was characterized by roots of unity and does
     not appear to belong to the usual quantum group family of
 invariants. 

Finally, Pant and Wu~\cite{w174} 
 have  derived  a link invariant 
associated with  the Izergin-Korepin model.

\vskip .3cm

\section{Conclusion}
\label{concl}

It would not be fair to summarize
 F. Y. Wu's contributions by a quick conclusion such as:
he wrote several important monographs on
 vertex models, on the Potts model and on knot theory,
obtained many important results, in particular
the Lieb-Wu solution of the Hubbard model, the Fan-Wu free-fermion
vertex model, the solution of the Baxter-Wu model, 
and many other results on dimers or free-fermion models,
3D dimers, $\, d$-dimensional
free-fermion models, Potts models, Ising and
 vertex models, using a large set of
 tools including analytic calculations, expansions,
 series analysis, Monte-Carlo, ..., with a particular emphasis on 
 graph-theoretical methods.  

Most of the work of F. Y. Wu could be said to 
correspond to exact results in lattice statistical mechanics,
 or in mathematics, with  particular emphasis on {\em graph theory}
 and enumerative combinatorics.
 We have tried to give here some hints as to
F. Y. Wu's very large ``graph'' of concepts,
results, tools, models, with many ``intellectual loops''.
We have not tried to provide an exhaustive description
of F. Y. Wu's contributions but, rather, only to provide a few
{\em comments} on some of his results, emphasizing
the fruitful cross-fertilizations between
the various domains 
of mathematical physics and mathematics, and also 
to show the motivation and relevance of
 these results, tools, concepts
and methods.  

Beyond a post-modern accountant's evaluation and from a research viewpoint,
 one must say that the important and numerous
results F. Y. Wu has obtained are not due to
  publish-or-perish productivity pressure, but, on the contrary, are 
the natural consequence of the pleasure of a scientist who loves to play  
with concepts and mathematical objects
(dimers, graphs with particular boundary conditions,
 dualities, Potts models, series expansions with 
transmissivities, ...)
and who has a strong desire to reach 
ambitious goals such as obtaining
 new results in three dimensions,
 new results for non-critical Potts models, or even for the Ising model
in a magnetic field. 

A scientist does not become as productive as F. Y. Wu in response to 
external pressure but, on the contrary, only
by  forces being in harmony with his 
 deep personal motivations. This is the only way to be 
as efficient and productive as F. Y. Wu
and, as the famous French mathematician Jean Dieudonn\'e once
wrote, to work efficiently, ``pour l'honneur de l'esprit
humain''.
% \footnote{In French in the text.}''. 

%The simple listing of Professor Wu's 
%results and contributions,
%and the inter-relations between these results and the associated
%concepts and tools, is by itself a  ``challenge 
%of enumerative combinatorics''. 

\vskip .3cm

{\bf Acknowledgments:} I would like to thank  Professor Chin-Kun Hu
for giving me the opportunity to participate in this conference in 
honor of  Professor F-Y. Wu. 
I would also like to thank Fred for so many passionate discussions
on lattice models throughout the years. I would like to thank
A. J. Guttmann for many valuable comments on this paper,
and J-C Angl\`es d'Auriac, S. Boukraa 
for careful readings of the manuscript.

\vskip 1cm

% \section{Appendix A: XXX }

\end{document}